\documentclass[aps, prx, twocolumn, superscriptaddress, longbibliography]{revtex4-2}

\usepackage{epsfig,amsmath,amssymb,color,amsfonts,physics}
\usepackage[protrusion=true,expansion=true]{microtype}
\usepackage[bookmarks=true,colorlinks=true,linkcolor=blue, citecolor=blue, urlcolor=blue]{hyperref}
\usepackage{dsfont}
\usepackage{wrapfig}
\usepackage{tikz}
\usepackage{physics}
\usepackage{mathrsfs}
\usepackage[export]{adjustbox}
\usepackage{comment}
\usepackage[french,english]{babel}
\usepackage{tikz}
\usepackage{soul}
\usepackage{leftidx}
\usepackage[T1]{fontenc}

\begin{document}

\title{Generalised BBGKY hierarchy for near-integrable dynamics}
\author{Leonardo Biagetti}
\affiliation{Laboratoire de Physique Th\'eorique et Mod\'elisation, CNRS UMR 8089, CY Cergy Paris Universit\'e, 95302 Cergy-Pontoise Cedex, France}

\author{Maciej Łebek}
\affiliation{Faculty of Physics, University of Warsaw, Pasteura 5, 02-093 Warsaw, Poland}

\author{Miłosz Panfil}
\affiliation{Faculty of Physics, University of Warsaw, Pasteura 5, 02-093 Warsaw, Poland}

\author{Jacopo De Nardis}
 \affiliation{Laboratoire de Physique Th\'eorique et Mod\'elisation, CNRS UMR 8089, CY Cergy Paris Universit\'e, 95302 Cergy-Pontoise Cedex, France}

\begin{abstract}
We study quantum and classical many-body Hamiltonian systems that combine integrable contact interactions with generic long-range two-body potentials. Starting from an ansatz for the state at time $t$, which we call the \textit{correlated fluid-cell ensemble}, we show that the dynamics of local observables at macroscopic times and length scales can be cast into a generalized Bogoliubov--Born--Green--Kirkwood--Yvon (gBBGKY) hierarchy formulated in terms of the quasiparticle densities of the underlying integrable model and their correlations. 
We derive this hierarchy and validate these predictions against microscopic molecular-dynamics simulations, finding perfect agreement.

At late times, the one-particle distribution relaxes via a Boltzmann-type scattering integral encoding the interplay between integrable contact processes and long-range collisions, whereas higher-point correlations remain strongly non-thermal on thermalization time scales, indicative of a form of incomplete or \emph{generalised thermalisation}. Focusing on long-range dipolar quantum gases, where the relevant matrix elements can be obtained explicitly, we show that our collision integral reduces exactly to the Fermi golden rule result and provide a complete theoretical account of the experimental observations of Tang \textit{et al.} (Phys.\ Rev.\ X \textbf{8}, 021030 (2018)). More broadly, our framework extends the BBGKY program to regimes with strong local interactions, and applies to a wide class of experimentally relevant systems, from one-dimensional dipolar cold-atom gases to Lennard--Jones molecular fluids.
\end{abstract}

\maketitle
\section{Introduction}\label{s:introduction}
The study of non-equilibrium many-body physics continues to reveal ever more intricate phenomena arising from complex interactions. Classical theories of chaos, turbulence, and active dynamics remain vibrant research areas, and in recent years, quantum many-body systems have also entered the non-equilibrium arena \cite{calabrese2016introduction,Polkovnikov2011,Cheneau2012,Huse2014}. In low spatial dimensions, integrable models—distinguished by their stable quasiparticles—form a key class of dynamical systems. These quasiparticles govern transport properties, entanglement spreading \cite{PhysRevA.78.010306,Alba2018SciPost,alba2019entanglement}, anomalous scrambling \cite{PhysRevB.100.115150,PhysRevB.104.104307}, extended coherence times \cite{Fendley_2016,Pasnoori2023,PRXQuantum.5.020323}, and exotic hydrodynamic behavior \cite{castro2016emergent,bertini2016transport,PhysRevD.85.085029,ruggiero2020quantum,Lucas2018,Malvania2021,schemmer2019generalized,Scheie2021,moller2021,cataldini2022,Li2023,bulchandani2018bethe,Bulchandani2020,piroli2017transport}. Typical examples of interacting integrable systems include bosons, spins, or fermions with two-body contact interactions, such as the Lieb--Liniger gas \cite{Lieb1963,Lieb1963a} and the Fermi--Hubbard chain \cite{Essler2005}. This naturally raises the question: how do quasiparticles and the overall dynamics evolve when the Hamiltonian includes additional, longer-range interactions? Integrability breaking has attracted considerable interest \cite{Mallayya2019,PhysRevLett.130.247101,PhysRevE.102.022201,PhysRevB.101.180302,Bastianello2021,PhysRevB.107.184312,10.21468/SciPostPhys.11.2.037,2402.12979,PhysRevLett.115.180601,Bertini2016IB,PhysRevB.89.165104,PhysRevB.103.L060302,PhysRevResearch.5.043019,lebek2024aa,PhysRevB.99.054520,Bastianello2019,GdV2022,Durnin2021,Panfil2023,Lebek2025b,PhysRevLett.125.040604,delacretaz2022thermalization,miron2019derivation,chen2014nonintegrability,zhao2018fourier,lepri2020too}, prompted by experiments \cite{Gring2012,PhysRevLett.126.090602,PhysRevX.8.021030,2412.14153} that report anomalously slow thermalization in quasi-one-dimensional settings, a challenge shared across molecular dynamics, plasma physics, and astrophysics \cite{Lepri2005,Mareschal1988,Eltohfa2024,Parameshwaran2025,Benettin2023,Eldridge1963,Dawson1964,Rouet1991,Fouvry2019,Fouvry2020,2504.18754,PhysRevE.108.054108,Rybicki1971,Reidl1988,Milanovi1998,Tsuchiya1996,Miller1996}. 

Significant progress has been achieved for systems whose unperturbed integrable limit is free—where matrix elements of local operators are typically known, allowing a Boltzmann scattering approach via Fermi’s golden rule. However, for genuinely interacting integrable models, matrix elements depend nontrivially on both the state and the model, and existing methods accurately capture only late-time behavior, leaving short-time dynamics—often of primary experimental interest—largely unexplored. 

Despite advances in specific contexts \cite{Durnin2021,Panfil2023,Lebek2024}, a general framework is still lacking to compute the dynamics of any integrable model perturbed by weak two-body or higher-order potentials. Developing such a framework would be both theoretically valuable and directly relevant to cold-atom experiments \cite{PhysRevX.8.021030}, where the precise form of interparticle potentials can be uncertain. 

In kinetic-theory texts, the Bogoliubov–Born–Green–Kirkwood–Yvon (BBGKY) hierarchy is a classic tool \cite{Balescu_1975,Kirkwood1947,Kirkwood1946,Yvon1935,Harris2004,Bogoliubov1947,2107.10872}, typically applied to free particles with weak or long-range interactions (as in plasma and astrophysical systems), truncating at two- or three-point correlations to describe thermalization and dissipation.

Yet, to date, BBGKY framework has not been implemented in cases where the unperturbed Hamiltonian is an interacting integrable model with well-defined quasiparticles.

Here, we derive a generalization of BBGKY hierarchy for the macroscopic time and scale evolution of systems
where the unperturbed (classical or quantum) theory features arbitrarily strong integrable contact interactions.

Specifically, we consider a hard-core repulsion $V_a^{\rm int}(r)$ of radius $a$ supplemented by a long-range field $\varphi(r)$, leading to the two-body potential (see Fig.~\ref{fig:1})
\begin{equation}\label{eq:potential}
   U(r) = V_a^{\rm int}(r) +  V(r),
   \quad V(r)=\tfrac{V_0}{\xi}\,\varphi(r/\xi),
\end{equation}
where $V_0>0$ ($<0$) describes repulsive (attractive) long-range interactions acting over the scale $\xi$. Crucially, we assume a strict separation of length scales. We require the range of the perturbation $\xi$ to be much larger than the typical microscopic correlation lengths characterizing the state of the unperturbed integrable model.

Notable examples include atomic dipolar systems \cite{Chomaz2023,Chen2023,Chomaz2023,PhysRevX.8.021030,PhysRevA.95.043606,PhysRevB.92.115107,PhysRevA.109.043316,PhysRevB.95.245111,PhysRevX.6.041039,Li2023,Recati2023,2501.08179,Yang2024}, whose Hamiltonian in terms of the bosonic field $\psi(x)$ can be written as
\begin{align}\label{eq:lll0}
H = & \int \dd x \Big( \frac{1}{2m} |\partial_x \psi(x) |^2 + a |\psi(x)|^4  - \mu |\psi(x)|^2 \Big) \nonumber \\
& + \int \dd x \dd y  \  V(x-x') |\psi(x)|^2 |\psi(x')|^2,
\end{align}
in which a local contact interaction is again supplemented by a long-range component. Our approach provides a fully predictive theory for a wide class of near-integrable dynamics in both classical and quantum regimes, encompassing bosonic, fermionic, and spin-chain models. In particular, it unifies the description of prethermalization and full thermalization dynamics.

\section{Background and Summary of the Results}

In this work, we focus on the broad class of integrable systems that are characterized by stable quasiparticle excitations. Similarly to free systems, these excitations provide a complete description of local properties. In particular, a generic thermodynamic state is specified by the \textit{quasiparticle momentum distribution}, or one-particle (quasi)momenta distribution, denoted by \(\rho_\theta(x)\), where we use \(\theta\) as a generic parametrization of the quasiparticle momenta \(k(\theta)\). Under the unperturbed time evolution, the spatial integral of \(\rho_\theta(x)\) is conserved.

\begin{figure}[t!]
\includegraphics[width=1.00\linewidth]{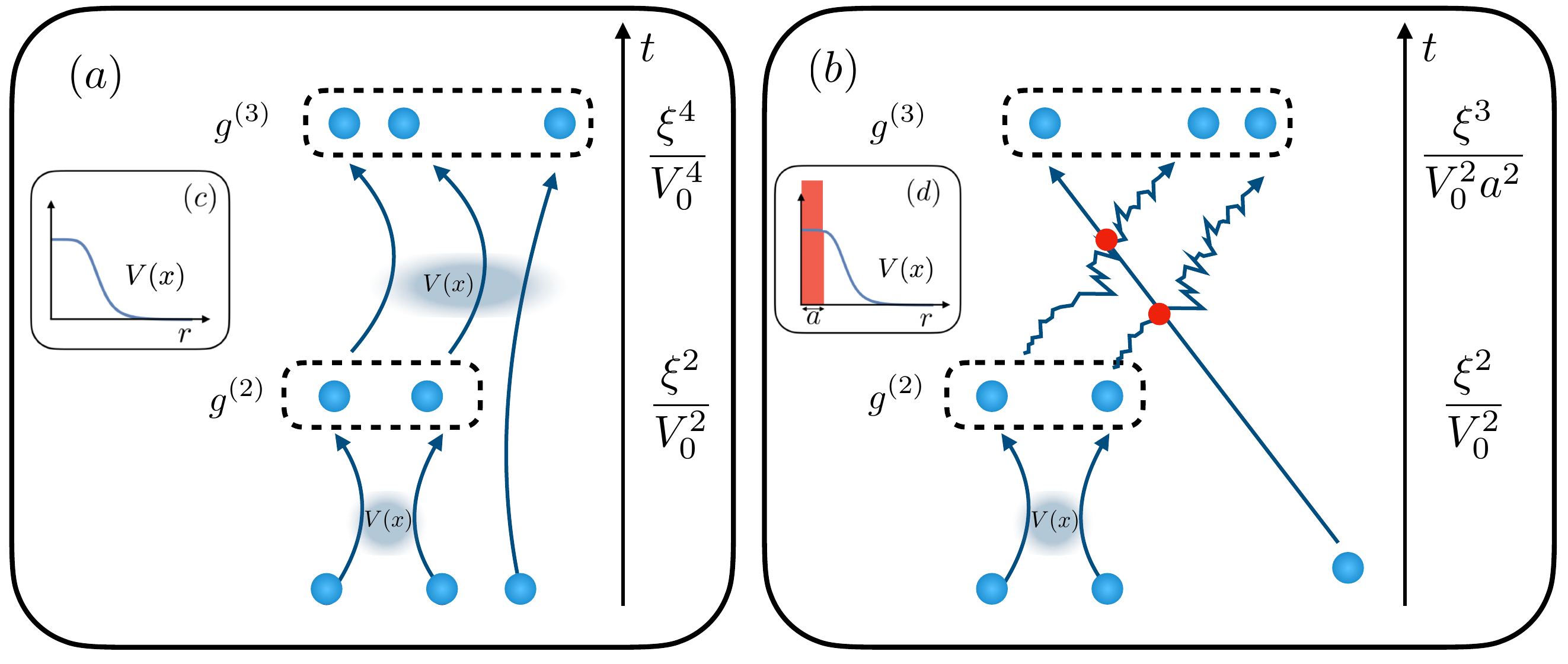}
\caption{
Schematic evolution of (a) BBGKY and (b) gBBGKY for the two-point functions \(g^{(2)}\), the two \textit{insets} show, respectively, (c) the long-range potential for free particles and (d) the combination of long-range and short-range interactions in the interacting integrable case. In (a), two particles drawn from the momentum distribution \(\rho(\theta)\) scatter via \(V(r)\) and become correlated on the time scale \(\xi^2/V_0^2\), this is the \textit{prethermalisation} stage where correlations build up but no extensive mixing occurs, they then scatter again and generate three-point correlations \(g^{(3)}\), which control the \textit{thermalisation} stage on time scales \((\xi/V_0)^4\). In (b), with integrable local interactions in homogeneous systems, the two-point correlation is generated by the large-scale interaction \(V(r)\), while the three-point function is produced by a different and faster mechanism. Particles propagate and scatter through local interactions, shown as red circles, providing an effective diffusive bath for the two-point functions, and leading to the thermalisation of one and two-point functions, namely \textit{generalised thermalisation}, on time scales \(\xi^3/(V_0 a)^2\).
}
\label{fig:1}
\end{figure}

Such systems arise in both classical and quantum physics \cite{Bernard2016,Scheie2021,Doyon2025,2406.17569,Vasseur2016,Kerr2023, Alba_2021, ESSLER2023127572,PhysRevA.108.013324, Suret2024,Bastianello2025}, typically in \(1+1\) dimensions, where, in certain cases, scattering processes factorize. Generic perturbations, such as couplings between different one-dimensional channels or the addition of long-range interactions, break integrability. In this situation quasiparticles become unstable, and most distributions \(\rho_\theta(x)\) are no longer stationary. Under weak integrability breaking, typically only thermal distributions remain stationary, as first recognized by Boltzmann.

The evolution from a generic out-of-equilibrium quasiparticle distribution toward a thermal state is the subject of \emph{prethermalization} and kinetic theory. These frameworks have applications ranging from high-energy collisions and plasma physics to, more recently, quantum many-body dynamics in ultracold atomic systems \cite{Kinoshita2006,PhysRevX.8.021030,Gring2012,Schreiber2015,Kao2021,Morong2022,Birnkammer2022,Le2023}. A common approach models this evolution through a collision integral derived from Fermi’s golden rule \cite{Durnin2021,Panfil2023,Mallayya2019,2508.00254}, the perturbation induces transitions between quasiparticle states with rates determined by the corresponding matrix elements, or by classical cross sections. This approximation, however, neglects correlations that build up dynamically, and the computation of the relevant matrix elements is generally difficult. The BBGKY hierarchy provides a more comprehensive framework, as it allows one to track correlations up to arbitrary order. It also enables a systematic derivation of the collision integral without requiring explicit knowledge of all transition amplitudes, and it goes beyond the late-time approximation inherent in the Boltzmann approach \cite{Janssen2003,Garnier2012}.

\medskip

Concretely, we consider a model of particles in the continuum, the extension to lattice systems is immediate, with Hamiltonian, quantum or classical,
\begin{equation}
\label{eq:HPerturb}
    \hat{H} = \hat{H}_a^{\rm int} + \frac{1}{2} \int \!\!\dd{x} \dd{x'} \; V(x - x') \,\hat{q}_0(x)\,\hat{q}_0(x') \,,
\end{equation}
where \(\hat{q}_0(x)\) is the local particle density operator, and \(\hat{H}_a^{\rm int}\) is an integrable Hamiltonian containing standard kinetic terms together with an integrable interaction \(\hat{V}_a^{\rm int}\). For example, in the Lieb--Liniger model, as in \eqref{eq:lll0}, the integrable interaction is \(V^{\rm int}(x - x') = c\,\delta(x - x')\), with \(c=0\) and \(c\to\infty\) corresponding to free and hard-core bosons, effectively free fermions, respectively. We denote by \(a\) the strength of the integrable interaction. In the limit \(a \to 0\), the integrable potential vanishes and the model reduces to free particle propagation,
\begin{equation}
   \lim_{a \to 0} \hat{H}_a^{\rm int} = \hat H^{\rm free} = \int {\dd} x \ \frac{(\hat{q}_1(x))^2}{2m} \,,
\end{equation}
with \(\hat{q}_1(x)\) the local momentum density.
The integrable Hamiltonian \(\hat{H}_a^{\rm int}\) possesses an extensive family of conserved charges \(q_i(x)\), with \(i\in\mathbb{N}\). 
The gBBGKY theory describes the time evolution of the expectation values of these densities and of their connected correlations.

\begin{figure}[h!]
\includegraphics[width=0.80\linewidth]{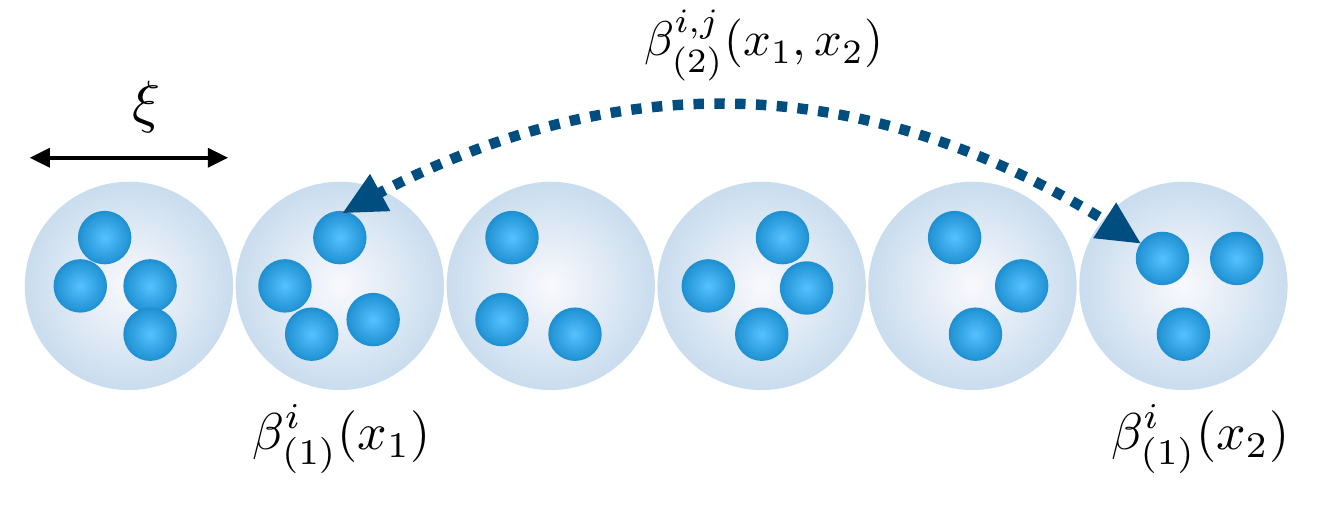}
\caption{
Schematic representation of the correlated fluid cells ensemble. Fluid cells of size \(\xi\) can be correlated over long distances by the external potential \(V\), or by the interplay of contact interactions and ballistic dynamics in the system. The corresponding density matrix is introduced in Eq.~\eqref{eq: longrange_gge_state}.
}
\label{fig:corr}
\end{figure}

To describe the system at finite macroscopic time \(t\), we introduce the following ansatz. The state %at any macroscopic time \(t\)
is given by the \textit{correlated fluid cells ensemble}, which can be formulated either as a density matrix or as a phase-space density, for a precise definition see Eq.~\eqref{eq: longrange_gge_state}. This state can be viewed as an embedding of a GGE, \(\rho\propto \exp(-\sum_i \int\dd{x}\beta^i(x) \hat q_i(x))\), determined by a set of generalized temperatures \(\beta^i(x)\) associated with the local conserved quantities \(\hat{q}_i(x)\). The correlated fluid cells ensemble additionally involves higher-rank Lagrange parameters \(\beta_{(n)}^{i_1,\ldots,i_n}(x_1,\ldots,x_n)\), which encode information about long-range correlations in the system. When these higher-order terms are nonzero, the state is not simply a product of independent local stationary GGE cells. It is instead a collection of correlated fluid cells as illustrated in Fig.~\ref{fig:corr}.
\\
The reason to expect this form is as follows. The hydrodynamic postulates assert that, for macroscopic times \(t\gg \tau_{\rm mic}\), much larger than the microscopic time \(\tau_{\rm mic}\),  a generic interacting system relaxes locally giving rise to quasiparticle description. Fast degrees of freedom associated with non-conserved modes decay, and the evolution becomes confined to a reduced manifold of states. Since the slow modes are precisely the conserved densities and their products, we include them here in full generality.
\\
 
Crucially, the correlated fluid cells ensemble correctly describes the out-of-equilibrium state of pure integrable systems when observed on macroscopic scales, i.e. much larger than the homogeneous GGE connected correlation functions typical decay length $\xi_0$ \cite{Hubner2024}.\\ 
Our framework fundamentally relies on the assumption that the integrability-breaking potential $V(x-x')$ acts strictly on macroscopic scales $\xi \gg \xi_0$.
Due to this scale separation,  the integrability-breaking potential does not corrupt the microscopic local structure of the integrable state. Instead, its effect is purely hydrodynamic: it smoothly modifies the local inverse temperatures $\beta^i(x)$ and the higher-rank parameters $\beta_{(n)}^{i_1,\ldots,i_n}(x_1,\ldots,x_n)$ over time. This clear separation of scales is precisely what justifies our core assumption: under the condition $\xi \gg \xi_0$, the macroscopic state of the system remains validly described by the correlated fluid cells ensemble throughout its time evolution.
A similar logic can be applied to standard BBGKY hierarchy. The only difference lies in the fact that $\tau_{\rm mic} \to 0$ as there is no transient phase needed for local relaxation to GGE and emergence of quasiparticles. Thus, in the usual setting, where one starts from states without correlations, there is little practical need to make the structure of~%\eqref{eq: longrange_gge_state}
the correlated ensemble explicit. Our formulation therefore \textit{generalizes BBGKY to situations in which the stationary states of the unperturbed Hamiltonian already displays nontrivial local correlations} (in contrast to \(H^{\rm free}\), where local correlations are absent and the hierarchy simplifies substantially). We will thus refer to the conventional hierarchy as the \emph{standard BBGKY}, and to the present extension as the \emph{generalized BBGKY} (gBBGKY). 
However, it is crucial to emphasize a fundamental conceptual distinction between the two frameworks: while the standard BBGKY hierarchy is an exact microscopic identity valid across all length and time scales prior to any truncation, the gBBGKY is inherently a macroscopic effective theory that fundamentally relies on coarse-graining and the assumption of local generalized equilibrium at large scales.

In principle, the full hierarchy of multi-point correlations encodes complete marcoscopic information about the state, but, to obtain a tractable description, it must be truncated at some finite order.
Truncating the hierarchy at the level of three-point functions yields the explicit evolution equations in Eqs.~\eqref{eq: evo_eq_explicit_deltas_1}–\eqref{eq: evo_eq_explicit_deltas_3}. Equivalently, these equations can be rewritten in terms of quasiparticle densities \(\rho_\theta(x,t)\) and their correlations by the replacement \(i \to \theta\) and by identifying the local densities, or operator or phase-space densities, as \(\langle \hat q_i(x,t) \rangle \to \rho_\theta(x,t)\), which also recovers standard textbook results in the limit \(a\to0\).

In this work we show that \textit{the presence of local interactions strongly affects the hierarchy}. The main new feature is the emergence of delta-function jumps in the evolution equations for multi-point correlations, which appear as discontinuities in spatial derivatives at coinciding points. These jumps encode ballistic, \textit{non-linear}, namely interacting, modes, and they persist even when integrability is weakly broken, since ballistic transport remains relevant at intermediate times, thereby driving the evolution of lower-order correlations. This has important physical consequences, see Fig.~\ref{fig:1}. For \(a=0\), correlations are generated solely by scattering mediated by the perturbing potential \(V(x)\). In one dimension, two-body scattering alone cannot thermalize the system, so three-point correlations must be retained, leading to thermalization times scaling as \(V_0^{-4}\), a phenomenon known as \emph{kinetic blocking}. For \(a\neq0\), local integrable interactions allow quasiparticles to scatter via \(V\) and then interact again locally with a ballistically propagating particle. These processes generate three-point functions with coincident-point jumps, which in turn induce dissipative, diffusive contributions to the dynamics of the two-point correlations, and lead to effective thermalisation. As a result, thermalization of one and two-point functions, which we refer to as incomplete or \textit{generalised thermalisation}, is reached at times of order \((a\,V_0)^{-2}\). Higher-point functions, such as three-point functions, remain far from thermal on these time scales, preserving remarkable \textit{long-range correlations}, as a direct consequence of the interplay between local contact interactions and long-range interactions. Recent preliminary experiments \cite{PhysRevX.8.021030} have observed this scaling, and our gBBGKY framework places it on firmer theoretical ground. Moreover, we show that truncation at third order reproduces the dynamics of one and two-point functions \textit{very precisely and at all times}, as we demonstrate by comparison with exact numerical simulations.
Moreover, we show that truncation at third order reproduces the dynamics of one and two-point functions throughout the dynamically relevant relaxation process, in excellent agreement with exact numerical simulations.
\\
\\

Before proceeding to the detailed derivation, we explicitly outline the physical boundaries, required inputs, and inherent limitations of the gBBGKY framework:
\begin{itemize}
\item Model Constraints: The practical application of this framework restricts the class of unperturbed Hamiltonians. The integrable model must admit a well-defined macroscopic description in terms of Generalized Hydrodynamics (GHD). This requires a known spectrum of stable quasiparticle excitations, established microscopic scattering shifts, and a valid Thermodynamic Bethe Ansatz (TBA) to rigorously define the local root densities.

\item Macroscopic Effective Theory (Length Scales): The gBBGKY approach is a macroscopic effective theory built upon the correlated fluid cells ensemble. It relies on a strict scale separation, $\xi \gg \xi_0$, where $\xi$ is the range of the integrability-breaking perturbation and $\xi_0$ represents the typical microscopic correlation length of the unperturbed model. Consequently, our equations do not resolve multi-point correlations at spatial separations comparable to or smaller than the fluid cell size.

\item Initial Conditions and Quenches (Time Scales): Because the framework fundamentally assumes local equilibrium, it cannot resolve the highly complex, microscopic dephasing transient immediately following a sudden integrable quantum quench, when the system is not in GGE. The macroscopic gBBGKY equations apply only after an initial local relaxation time ($\tau_{\rm mic}$). Furthermore, the framework requires the post-quench quasiparticle rapidity distribution to be known and provided as an external input to initialize the time evolution.
\end{itemize}

After the derivation of the gBBGKY hierarchy, we illustrate its predictive power by studying the non-equilibrium dynamics in the system of hard rods and in dipolar quantum gases. The theoretical predictions will be benchmarked against the numerical simulations of the former and experimental thermalisation rates measured in the latter.

\subsection*{Organisation of the paper}

The paper is organized as follows. In Section~\ref{s:gBBGKY} we derive the gBBGKY hierarchy, an infinite set of coupled equations for correlation functions of charges and currents. We then show how to truncate this hierarchy consistently, obtaining a closed set of equations for connected charge correlators. This requires neglecting correlations beyond a given order, and assuming a separation of scales between local integrable interactions and the long-range integrability-breaking potential. We also show that the generalized hierarchy reduces to the standard one when the integrable interactions vanish. The resulting equations are later solved numerically and validated against numerical simulations.

To clarify the mechanisms behind generalised thermalization, in Section~\ref{s:Landau} we derive a kinetic equation, dubbed the generalized Landau equation, from which the thermalization time scale quoted above follows. This derivation assumes a separation of time scales between the fast dynamics of correlations and the slow evolution of the one-body function.

In the next two sections we apply the gBBGKY and generalized Landau equations to two systems.
First, in Section~\ref{s:hard-rods} the predictions of the preceding sections are tested against exact numerical simulations for classical hard rods, hard spheres, with long-range interactions, showing perfect agreement. Second, in Section~\ref{s:dipolar} we apply the framework to dipolar cold-atomic gases trapped in an array of quasi-one-dimensional tubes. This requires an extension of the methods to multicomponent systems. It also provides an opportunity to compare the generalized Landau equation, derived here under explicit assumptions, with a heuristic approach where the scattering integral is obtained from Fermi’s golden rule. We find exact agreement between the two approaches in the regime where both apply.

In the final Section~\ref{s:conclusions}, we summarize our findings and outline future directions. More technical aspects are relegated to the appendices. In the accompanying Supplementary Materials we provide derivations of the generalised Landau equation and of the gBBGKY hierarchy for multicomponent systems.

\section{The \MakeLowercase{g}BBGKY hierarchy}\label{s:gBBGKY}

We start by introducing the notation needed to set up the generalised BBGKY hierarchy. We first work with densities of conserved quantities, and later move to particles or quasiparticle occupation functions, as is customary in standard formulations of BBGKY.

We denote by \(\hat{q}_i(x)\) the density of the conserved charge \(\hat{Q}_i\), which commutes with the integrable Hamiltonian, \([\hat{H}_a^{\rm int},\hat{Q}_i]=0\), noting that a suitable gauge must be chosen since these densities are defined up to a total derivative. The first standard conserved quantities are particle number \(\hat{Q}_0 = \hat{N}\), momentum \(\hat{Q}_1 = \hat{P}\), and unperturbed energy \(\hat{Q}_2 = \hat{H}_a^{\rm int}\), to which we refer as the \textit{kinetic energy of quasiparticles}. Throughout the derivation we focus on the quantum formulation, yet all steps can be translated straightforwardly to classical models by replacing the commutator with the Poisson bracket, \({\rm i} [\cdot,\cdot ] \to \{ \cdot,\cdot \}\). We will often refer to the case of free classical particles, where the equations are simplest and essentially all steps are exact, and we also assume Galilean invariance, although violations can be incorporated without difficulty.
We set \(\hbar=1\) and start from the Heisenberg equation for the charge densities,
\begin{equation}
    \partial_t \hat{q}_i = {\rm i} [\hat{H},\hat{q}_i].
\end{equation}
We decompose the commutator and introduce currents \(\hat{j}_i\) and generalized currents \(\hat{j}_{i,k}\), defined by
\begin{equation}
    {\rm i}[\hat{H}_a^{\rm int},\hat{q}_i]+\partial_x \hat{j}_i=0, \qquad {\rm i}[\hat{Q}_i,\hat{q}_k]+\partial_x \hat{j}_{i,k}=0\,,
\end{equation}
where the density does not generate any dynamics, \(\hat{j}_{0,i}=0\).
To simplify the Heisenberg equation, we use locality of the charges,
\begin{equation}
    {\rm i}[\hat{q}_i(x),\hat{q}_0(x')] =  \hat{j}_{i,0}(x) \delta'(x-x').
\end{equation}
Using this relation for commutators between local charges, the quantum generalisation of the classical Liouville equation reads \cite{doyon2017note,durnin2021diffusive}
\begin{align}
\label{eq:chargedyn}
   &\partial_t \hat q_{i_1}(x_1)  +\partial_{x_1}  \hat j_{i_1}(x_1)  =-\frac{1}{2}\int {\rm d}x_2 V'(x_1-x_2) \times\\
   & \nonumber \times\Big[ \hat q_0(x_2)\hat j_{i_1,0}(x_1)+\hat j_{i_1,0}(x_1)\hat q_0(x_2) \Big] \,.
\end{align}
Similar equations to \eqref{eq:chargedyn} can be written for multi-point operators involving arbitrary products of charge densities. Taking expectation values then generates the hierarchy of dynamical equations
\begin{equation}
\begin{aligned}
    &\partial_t \langle q_{i_1}(x_1) \ldots q_{i_n}(x_n) \rangle = \\
    &-\sum_{k=1}^n \partial_{x_k} \langle q_{i_1}(x_1) \ldots j_{i_k}(x_k) \ldots q_{i_n}(x_n) \rangle-\sum_{k=1}^n\int {\rm d}y \times\\
    &V'(x_k-y)\langle q_{i_1}(x_1) \ldots q_0(y)j_{i_k,0}(x_k) \ldots q_{i_n}(x_n) \rangle.
\end{aligned}
\end{equation}
To isolate the relevant parts of correlators we move to connected correlation functions. For \(n=2\) we set \(\langle o_1 o_2 \rangle^c:= \langle o_1 o_2 \rangle - \langle o_1 \rangle \langle o_2 \rangle\),
whereas for \(n=3\) we define
\begin{equation}
\begin{aligned}
    &\langle o_1 o_2 o_3\rangle ^c :=\langle o_1 o_2 o_3 \rangle -\langle o_1  o_2 \rangle^c \langle o_3 \rangle-\\
    &\langle o_1  o_3 \rangle^c \langle o_2 \rangle-\langle o_2  o_3 \rangle^c \langle o_1 \rangle - \langle o_1 \rangle \langle o_2 \rangle \langle o_3 \rangle,
\end{aligned}
\end{equation}
and for \(n>3\) we proceed analogously. The first three layers of the hierarchy can then be written as
\begin{widetext}
    \begin{equation}
    \begin{split}
        \label{eq: evo_eq_implicit_deltas_1}
        \partial_t \langle q_{i_1}(x_1) \rangle +\partial_{x_1} \langle j_{i_1}(x_1) \rangle&= - \int {\rm d}x_2V'(x_1-x_2)\langle j_{i_1,0}(x_1) \rangle \langle q_0(x_2) \rangle - \int {\rm d}x_2 V'(x_1-x_2) \langle j_{i_1,0}(x_1) q_0(x_2) \rangle^c,
        \end{split}
    \end{equation}
    \begin{equation}
    \label{eq: evo_eq_implicit_deltas_2}
    \begin{aligned}
        &\partial_t \langle q_{i_1}(x_1) q_{i_2}(x_2)\rangle^c + \left[\partial_{x_1} \langle j_{i_1}(x_1) q_{i_2}(x_2) \rangle^c\right]_{(1,2)}=\\&= - \Big[\int {\rm d}x_3 V'(x_1-x_3) \big( \langle j_{i_1,0}(x_1) \rangle \langle q_{i_2}(x_2) q_0(x_3) \rangle^c + \langle j_{i_1,0}(x_1) q_{i_2}(x_2)\rangle^c \langle q_0(x_3) \rangle+ \langle j_{i_1,0}(x_1)q_{i_2}(x_2)q_0(x_3) \rangle^c \big)\Big]_{(1,2)},
    \end{aligned}
    \end{equation}
    \begin{equation}
    \label{eq: evo_eq_implicit_deltas_3}
    \begin{aligned}
        &\partial_t \langle q_{i_1}(x_1) q_{i_2}(x_2) q_{i_3}(x_3) \rangle^c + \left[\partial_{x_1} \langle j_{i_1}(x_1) q_{i_2}(x_2) q_{i_3}(x_3) \rangle^c\right]_{(1,2,3)}=\\
        &-\bigg[\int {\rm d}x_4 V'(x_1-x_4) \Big( \langle j_{i_1,0}(x_1) \rangle \langle q_{i_2}(x_2)q_{i_3}(x_3)q_0(x_4)\rangle^c+ \langle j_{i_1,0}(x_1) q_{i_2}(x_2) q_{i_3}(x_3) \rangle^c \langle q_0(x_4) \rangle + \\
        &+ \langle j_{i_1,0}(x_1)q_{i_2}(x_2) \rangle^c \langle q_{i_3}(x_3) q_0(x_4) \rangle^c+\langle j_{i_1,0}(x_1) q_{i_3}(x_3) \rangle^c\langle q_{i_2}(x_2) q_0(x_4)\rangle^c + \langle j_{i_1,0}(x_1) q_{i_2}(x_2) q_{i_3}(x_3)q_0(x_4)\rangle^c \Big) \bigg]_{(1,2,3)},
    \end{aligned}
    \end{equation}
\end{widetext}
% where for one-point functions we wrote $\langle o \rangle \equiv o$.
Here the symbol $[\ldots]_{(i_1,\ldots,i_n)}$ denotes the sum of the expression inside the square brackets over all \textit{cyclic} permutations of the indices \((i_1,\ldots,i_n)\).

\begin{figure}
    \centering
    \includegraphics[scale=0.5]{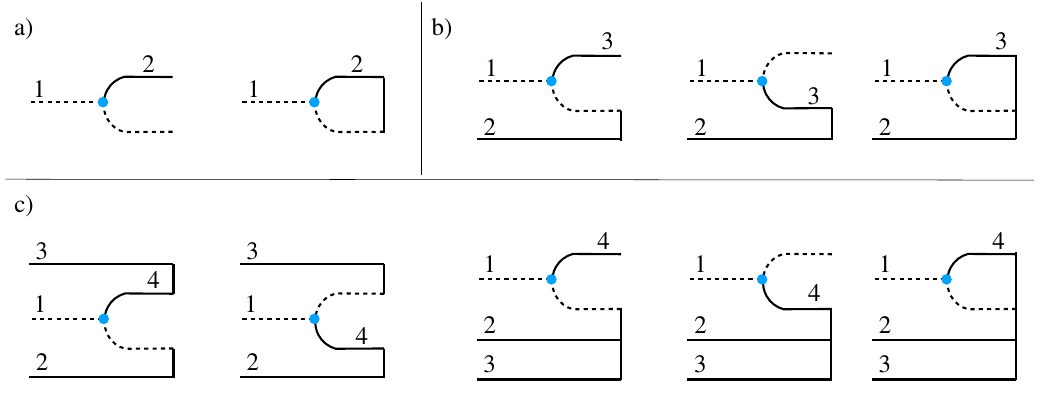}
    \caption{
Diagrams representing the right-hand sides of the hierarchy equations \eqref{eq: evo_eq_implicit_deltas_1}, \eqref{eq: evo_eq_implicit_deltas_2} and \eqref{eq: evo_eq_implicit_deltas_3}. Dashed lines represent the current operator \(j_{i_1,0}(x_1)\), while solid lines represent charge operators \(q_{i_k}(x_k)\), with \(k\) indicated by the number above the line. Vertical lines represent connected correlations between operators. Each diagram contains a single vertex associated with the potential \(V(x-x')\), which couples a current to an additional operator \(q_0\). To translate a diagram into an expression, one takes products of connected correlation functions according to the vertical lines, multiplies by \(V(x_1-x_{m+1})\), where \(m+1\) labels the extra charge operator, and integrates over \(x_{m+1}\). To draw diagrams at the \(m\)-th level, one considers one dashed line from which an extra solid line emanates, together with \(m-1\) additional solid lines, and then enumerates all possible connections, or partitions, between the \(m+1\) lines. The case with no connections is excluded, as it corresponds to the disconnected part of the correlator.
}
    \label{fig:diagrams}
\end{figure}

When the integrable model reduces to a free theory, this hierarchy is equivalent to the standard BBGKY hierarchy, as discussed later in Section~\ref{ssec:standrd_BBGKY}. A key difference is that the present formulation is expressed in terms of correlations of conserved charges and currents, whereas the standard BBGKY hierarchy is usually written directly for quasiparticle distributions. Nevertheless, the familiar structures are readily recognized. Truncating the hierarchy at the level of expectation values, namely setting the two-point functions to zero, yields the Vlasov equation \cite{Vlasov1968,Braun1977,Chavanis1996}. Retaining two-body correlations instead leads, at large times, to either the Landau or the Boltzmann equation, depending on the truncation procedure. Here we keep three layers of the hierarchy because, due to kinetic blocking, the first two layers are not sufficient to go beyond the mean-field regime \cite{Fouvry2019,Fouvry2020}. Finally, as in the standard BBGKY case \cite{Balescu_1975}, the hierarchy can be represented diagrammatically. In Fig.~\ref{fig:diagrams} we show the diagrams corresponding to the first three layers written above.

The hierarchy of equations \eqref{eq: evo_eq_implicit_deltas_1}, \eqref{eq: evo_eq_implicit_deltas_2} and \eqref{eq: evo_eq_implicit_deltas_3} is not closed. First, as in the standard BBGKY hierarchy, the equation for an \(n\)-point function involves the \((n+1)\)-point function. Closing the system therefore requires assuming that connected correlation functions beyond a given order are negligible. Second, the hierarchy contains correlators involving currents, of the form \(\langle \ldots q_i j_k q_{i'} \ldots \rangle^c\) and \(\langle \ldots q_i q_0 j_{k,0} q_{i'} \ldots \rangle^c\). To obtain a consistent truncation scheme, these correlators must be expressed in terms of correlators involving charge densities only. In a hierarchy built on a non-interacting theory, current operators are linear functionals of the charge densities, so this step is straightforward. In the present setting, it requires identifying the relevant structure of correlations in integrable theories.

To achieve this, two ingredients are needed, the ensemble with respect to which expectation values are computed, and the associated fluid-cell averaging procedure. These steps are discussed in full generality below, and later specialized to homogeneous systems.

Before proceeding, we summarize the assumptions made throughout the manuscript:
\begin{itemize}
    \item At large space-time distances the connected correlators scale as \(\langle q_{i_1}(x_1)\ldots q_{i_n}(x_n)\rangle^c\sim1/\xi^{n-1}\), where \(\xi\) is the characteristic range of the long-range potential, and is much larger than the range \(a\) of the integrable interaction. This is discussed in Sec.~\ref{sec: corr_fluid_cell_ens}.
    \item The state of the system is described by the correlated fluid cells ensemble \eqref{eq: longrange_gge_state}. It encodes the expectation values of local charge densities and the correlations between them, as discussed in Sec.~\ref{sec: corr_fluid_cell_ens}. Using this ensemble one also obtains equations for correlation functions involving currents, see Sec.~\ref{sec:currents}.
    \item The resulting partial differential equations, obtained by imposing the previous points in Eqs.~\eqref{eq: evo_eq_implicit_deltas_1}, \eqref{eq: evo_eq_implicit_deltas_2} and \eqref{eq: evo_eq_implicit_deltas_3}, are expanded at large \(\xi\), up to and including order \(\xi^{-3}\), yielding the final hierarchy reported in Sec.~\ref{sec:finalhyear}.
\end{itemize}

\subsection{The correlated fluid cells ensemble}
\label{sec: corr_fluid_cell_ens}

% Until this point, the derivation does not contain any approximation. In order to be able to treat the aforementioned complex charge-current correlators, we now perform a macroscopic rescaling of space $x\to \xi x$ and time $t\to \xi t$ and use the hydrodynamic assumptions to reduce the complexity of the problem.
To truncate the hierarchy and obtain a closed description of the system at time~$t$, correlations must be retained only up to a chosen maximal order. This mirrors standard hydrodynamics, where a mesoscopic length~$\ell$ separates microscopic fluctuations from macroscopic, observable fields. In the homogeneous setting considered here, the natural choice for this scale is the interaction length~$\xi$ of the potential. Although additional length scales may emerge in inhomogeneous systems, we assume that, in the present case, \(\xi\) is the only relevant spatial scale.

We postulate that, at macroscopic times, the system relaxes to the state described by the \textit{correlated fluid cells density matrix}
\begin{align}
\label{eq: longrange_gge_state}
        &\rho  = {\rm exp} \Big( - \int \dd x_1 \beta_{(1)}^i(x_1) q_i(x_1) \\&\nonumber - \int\limits_{x_1\neq x_2}\hspace{-0.2cm} {\dd \vec{x}}{} \,  \beta_{(2)}^{i,j}(x_1,x_2) \, \delta q_{i}(x_1)\delta q_{j}(x_2)\\&\nonumber - \int\limits_{x_1 \neq x_2\neq x_3 \neq x_1} \hspace{-0.7cm} {\dd \vec{x}}{} \,  \beta_{(3)}^{i,j,k}(x_1,x_2,x_3) \delta q_{i}(x_1)\delta q_{j}(x_2)\delta q_{k}(x_3)
        \\&\nonumber+ \ldots \Big)\,.
\end{align}
In the following, the ensemble average \(\langle\ldots\rangle\) is always taken with respect to density matrix \eqref{eq: longrange_gge_state}. The first term corresponds to a family of \textit{local} GGE states, one for each uncorrelated fluid cell at position \(x\). Accordingly, we denote by \(\langle\ldots\rangle_1\) the average with respect to the zeroth order of the correlated fluid cells ensemble, namely the local GGE average given by eq. \eqref{eq: longrange_gge_state} with $\beta_{(n>1)}=0$. In \eqref{eq: longrange_gge_state} we also introduced fluctuations around the local GGE,
\(\delta q_i(x)=q_i(x)-\langle q_i(x)\rangle_{1}\).

For the states defined by \eqref{eq: longrange_gge_state}, the entropy is maximized independently in each cell, with additional constraints imposed by conservation laws. Setting all higher terms to zero amounts to neglecting nonlocal constraints, which implies that only short-range, thermal-like correlations are retained, decaying on distances much smaller than the scale \(\xi\). As in the standard BBGKY hierarchy, the goal is instead to keep higher-order correlations in the description of the state. The state \eqref{eq: longrange_gge_state} therefore maximizes the entropy under constraints given not only by fixed one-point functions, but also by correlations of arbitrary order. As we will show in this section, the nonlocal generalized temperatures \(\beta_{(n)}^{i_1,\ldots,i_n}(x_1,\ldots,x_n)\) are directly related to the connected \(n\)-point functions \(\langle q_{i_1}(x_1)\cdots q_{i_n}(x_n)\rangle^c\), and, in particular, they display the same scaling with the length scale \(\xi\).

For generic states, computing expectation values with respect to ensemble~\eqref{eq: longrange_gge_state} is a difficult task, and, in close analogy with (g)BBGKY, it calls for a perturbative scheme. Here a natural truncation arises from the assumed separation of length scales. A key hypothesis for hydrodynamic states is that connected correlation functions satisfy a large-deviation scaling
\begin{equation}
\label{eq:scaling_xi_corr}
\bigl\langle q_{i_1}(x_1)\cdots q_{i_n}(x_n)\bigr\rangle^c \sim \frac{1}{\xi^{n-1}},
\end{equation}
which also implies \(\beta_{(n)}^{i_1,\ldots,i_n}(x_1,\ldots,x_n)\sim  {\xi^{-n+1}}\).
Equivalently, as the order \(n\) increases, the magnitude of the connected correlator decreases by a factor \(1/\xi\) for each additional field. This ensures that higher-order correlations become progressively smaller and can be neglected beyond a finite truncation order. This scaling is central for defining a controlled truncation scheme, both for the gBBGKY hierarchy and for the correlated fluid cells density matrix. Under the scaling \eqref{eq:scaling_xi_corr}, the density matrix \eqref{eq: longrange_gge_state} can be evaluated perturbatively in the parameter \(\xi^{-1}\). As an example, we compute the two-point function using the state \eqref{eq: longrange_gge_state}, expanded to first order in \(\xi^{-1}\),
\begin{align}
\label{eq: def_2pointfunc_delta_1}
    &\langle q_{i_1}(x_1)q_{i_2}(x_2) \rangle ^c = \langle q_{i_1}(x_1)q_{i_2}(x_2) \rangle ^c_1+\\&\nonumber\int\limits_{x_3\neq x_4}\hspace{-0.3cm} \dd x_3\dd x_4   \beta_{(2)}^{i_3,i_4}(x_3,x_4)  \langle\delta q_{i_3}(x_3)\delta q_{i_4}(x_4)q_{i_1}(x_1)q_{i_2}(x_2)\rangle_1
    \\&\nonumber+O(\xi^{-2})\,.
\end{align}
Using the cumulant expansion for four-point functions and the locality of the GGE we obtain
\begin{align}
\label{eq: def_2pointfunc_delta_2}
    \langle q_{i_1}(x_1)q_{i_2}(x_2) \rangle ^c &= \delta(x_1-x_2)C_{i_1,i_2}+\\&\nonumber+2\beta_{(2)}^{i_3,i_4}(x_1,x_2) C_{i_1,i_3}C_{i_2,i_4}+O(\xi^{-2})\,,
\end{align}
where we used \(\langle q_{i_1}(x_1)q_{i_2}(x_2) \rangle ^c_1=\delta(x_1-x_2)C_{i_1,i_2}\), and where \(C_{i_1,i_2}\) denotes the local charge-charge susceptibility matrix.
From \eqref{eq: def_2pointfunc_delta_2} we define the long-range charge-charge correlation \(g^{(2)}_{i_1,i_2}(x_1,x_2)\) through
\begin{equation}
    \label{eq:def_g2_beta2}
    g^{(2)}_{i_1,i_2}(x_1,x_2)=2\beta_{(2)}^{i_3,i_4}(x_1,x_2) C_{i_1,i_3}C_{i_2,i_4}+O(\xi^{-2})\,,
\end{equation}
so that
\begin{equation}
    \label{eq: def_2pointfunc_delta}
    \langle q_{i_1}(x_1)q_{i_2}(x_2) \rangle ^c = \delta(x_1-x_2)C_{i_1,i_2}(x_1)+g^{(2)}_{i_1,i_2}(x_1,x_2)\,.
\end{equation}
Similarly, as shown in Appendix \ref{app:corrfuns}, for the three-point function we find
\begin{equation}
    \begin{split}
    \label{eq: def_3pointfunc_delta}
        \langle &q_{i_1}q_{i_2}q_{i_3}\rangle ^c = \delta(x_1-x_2)\delta(x_1-x_3)C_{i_1,i_2,i_3}(x_1)+\\&+[\delta(x_1-x_2)C_{i_1,i_2,j}C^{j,k}g^{(2)}_{k,i_3}(x_1,x_3)]_{(1,2,3)}\\&+g^{(3)}_{i_1,i_2,i_3}(x_1,x_2,x_3)\,,
    \end{split}
\end{equation}
with
\begin{multline}
     g^{(3)}_{i_1,i_2,i_3}(x_1,x_2,x_3) =\\= 3\beta_{(3)}^{j_1,j_2,j_3}(y,y')C_{i_1,j_1}C_{i_2,j_2}C_{i_3,j_3}+O(\xi^{-3})\,,
\end{multline}
and where \(C_{i_1,i_2,i_3}\) is the local GGE three-point function, and \(C^{i,j}\) is the inverse covariance matrix, \(C_{i,j}C^{j,k}=\delta_{i,k}\). We emphasize that the matrices \(C_{i_1,i_2}\) and \(C_{i_1,i_2,i_3}\) are known functions of the expectation values of the conserved charge densities \(\langle q_i\rangle_1\).
It is also important to note that the functions \(g^{(2)}(x_1,x_2)\), \(g^{(3)}(x_1,x_2,x_3)\), and higher orders, as well as the generalized temperatures \(\beta_{(n)}\), are smooth everywhere except at coinciding points, \(x_i=x_j\), where discontinuities may appear but no divergences occur. For this reason, to avoid ill-defined expressions, in Eq.~\eqref{eq: longrange_gge_state} the integrals are defined over configurations with non-coinciding points.

\subsection{The expression for the currents in the ensemble}
\label{sec:currents}

Importantly, once the state \eqref{eq: longrange_gge_state} is assumed, the averages of all local, or quasi-local, observables can be expressed through cumulant expansions in the charge densities.

In this section we use the state \eqref{eq: longrange_gge_state} to derive the cumulant expansion for the average current
\begin{align}
\label{eq: def_2pointfunc_delta_1}
    &\langle j_{i_1}(x_1) \rangle = \langle j_{i_1}(x_1)\rangle _1+\\&\nonumber+\int\limits_{x_2\neq x_3} \dd x_2\dd x_3   \beta_{(2)}^{i_2,i_3}(x_2,x_3)  \langle q_{i_2}(x_2)q_{i_3}(x_3)j_{i_1}(x_1)\rangle^c_1
    \\&\nonumber+O(\xi^{-2})\,.
\end{align}
We now observe that, due to the locality of the GGE average, one has
\begin{gather}
    \label{eq: def_2pointfunc_delta_delta_cond}
    \langle q_{i_2}(x_2)q_{i_3}(x_3)j_{i_1}(x_1)\rangle^c_1=0\,,
    \\\nonumber{\rm for:}\,(|x_2-x_1|+|x_3-x_1|>\varepsilon)
\end{gather}
where \(\varepsilon\) denotes a typical microscopic length scale of the model. Since the generalized temperature \(\beta_{(2)}\) is in general discontinuous at coinciding points, due to fast microscopic-scale fluctuations encoded in the local GGE, this contribution must be treated with care. 

Condition~\eqref{eq: def_2pointfunc_delta_delta_cond} effectively localizes the integral on the r.h.s. of Eq.~\eqref{eq: def_2pointfunc_delta_1} to a neighborhood of size $\varepsilon$ around the discontinuity. Within this region, the correlation function decomposes into its local parity-time ($\mathcal{PT}$) symmetric and antisymmetric components, $\beta^{i_1,i_2}_{(2)} = \beta^{i_1,i_2}_{(2),\rm sym} + \beta^{i_1,i_2}_{(2),\rm anti}$.Crucially, because the microscopic jump is inherently antisymmetric, the singularity is entirely confined to $\beta^{i_1,i_2}_{(2),\rm anti}$. This implies that the symmetric component, defined locally as 
\begin{multline}
    \beta_{(2),\rm sym}^{i_1,i_2}(x+\varepsilon, x-\varepsilon, t) = \frac{1}{2}\Big[\beta_{(2)}^{i_1,i_2}(x+\varepsilon, x-\varepsilon, t+\epsilon) +\\+ \beta_{(2)}^{i_1,i_2}(x-\varepsilon, x+\varepsilon, t-\epsilon)\Big]\,,
\end{multline}
remains smooth even as $\epsilon \to 0$. Since charge and current density operators are $\mathcal{PT}$-symmetric \cite{de2019diffusion}, they couple exclusively to this smooth symmetric part in Eq.~\eqref{eq: def_2pointfunc_delta_1}. Consequently, the integral projects out the singular antisymmetric term, rendering the limit $\epsilon \to 0$ regular
\begin{multline}
    \langle j_{i_1}(x_1) \rangle = \langle j_{i_1}(x_1)\rangle _1+ \\
    + \beta_{(2),\rm sym}^{i_2,i_3}(x_1^+,x_1^-)  \frac{\delta^2\langle j_{i_1}(x_1)\rangle_1}{\delta\beta^{i_2}_{(1)}\delta\beta^{i_3}_{(1)}}
     +O(\xi^{-2})\,.
\end{multline}
This calculation follows analogously for the ensemble average of any $\mathcal{PT}$-symmetric observable.
Also, we used that in a local GGE correlation functions of local densities can be obtained by differentiation with respect to \(\beta_{(1)}\). Applying \eqref{eq:def_g2_beta2} and the definition of the inverse susceptibility matrix, we obtain
\begin{multline}
\label{eq: def_2pointfunc_delta_3}
    \langle j_{i_1}(x_1) \rangle = \langle j_{i_1}(x_1)\rangle_1+\\
    + \frac{1}{2} \frac{\delta^2\langle j_{i_1}(x_1)\rangle_1}{\delta\langle q_{i_2}\rangle_1\delta\langle q_{i_3}\rangle_1}g^{(2),\rm sym}_{i_2,i_3}(x_1^+,x_1^-)+O(\xi^{-2})\,.
\end{multline}
In analogy, we define
\begin{equation}
\label{eq: g2sym_definition}
    g^{(2),\rm sym}_{i_2,i_3}(x_1^+,x_1^-)=\frac{g^{(2)}_{i_2,i_3}(x_1^+,x_1^-)+g^{(2)}_{i_2,i_3}(x_1^-,x_1^+)}{2}\,,
\end{equation}
where we dropped the splitting in time, since the correlation functions are always smooth for any $t$.
Finally, introducing the notation
\begin{equation}
  H_{i_1}^{i_2,i_3}\equiv \frac{\delta^2\langle j_{i_1}\rangle_1}{\delta\langle q_{i_2}\rangle_1\delta\langle q_{i_3}\rangle_1} \,,
\end{equation}
we obtain the double-projection formula
\begin{multline}
\label{eq: def_2pointfunc_delta_3}
    \langle j_{i_1}(x_1) \rangle = \langle j_{i_1}(x_1)\rangle_1+\\+ \frac{1}{2} H_{i_1}^{i_2,i_3}g^{(2),\rm sym}_{i_2,i_3}(x_1^+,x_1^-)+O(\xi^{-2})\,.
\end{multline}
Similarly, for the current-charge correlator we find
\begin{multline}
\label{eq:def_symm_g2}
        \langle j_{i}(x_1) q_{j}(x_2)\rangle^{c}=A_i^k\langle q_{k}(x_1) q_{j}(x_2)\rangle^{c}+\\+\frac{1}{2} H_i^{k,l}g^{(3),\rm sym}_{k,l,j}(x^+_1,x^-_1,x_2)+O(\xi^{-3})\,,
\end{multline}
where we used the chain rule to introduce the flux Jacobian \(A_i^k=\delta \langle j_i\rangle_1/\delta \langle q_k\rangle_1\), and where \(g^{(3),\rm sym}_{k,l,j}(x^+_1,x^-_1,x_2)\) is defined in analogy with \eqref{eq: g2sym_definition}. This simplification is crucial, as it allows one to express the full hierarchy in terms of few-point charge correlators.
We also note that the generalized currents \(j_{i,0}\) appearing on the right-hand side of Eq.~\eqref{eq: evo_eq_implicit_deltas_1}, \eqref{eq: evo_eq_implicit_deltas_2} and \eqref{eq: evo_eq_implicit_deltas_3} are linear in the charge density, therefore the cumulant expansion truncates after the first term,
\begin{equation}
    \langle j_{i_1,0} q_{i_2}\ldots \rangle^{c}=A_{i_1,0}^k\langle q_{k} q_{i_2}\ldots\rangle^{c},
    \label{eq:integral_nonlocalGGE_j1}
\end{equation}
with $A_{i_1,0}^k={\delta \langle j_{i_1,0}\rangle_1}/{\delta \langle q_k\rangle_1}\,.$

\subsection{The gBBGKY hierarchy}\label{sec:finalhyear}
In this section we make use of the assumptions and results introduced in Sec \ref{sec: corr_fluid_cell_ens} in order to write Eqs.~\eqref{eq: evo_eq_implicit_deltas_1},\eqref{eq: evo_eq_implicit_deltas_2} and \eqref{eq: evo_eq_implicit_deltas_3} as a hierarchy of equations for the smooth few point functions $g^{(n)}_{i_1, \ldots, i_n}$.  The formulas \eqref{eq: def_2pointfunc_delta} and \eqref{eq: def_3pointfunc_delta}, a direct consequence of taking expectation values within the ensemble \eqref{eq: longrange_gge_state}, reveal that the correlators, and thus the hierarchy, feature singular terms. Thus, to understand the dynamics it is crucial to plug these expressions into the hierarchy alongside with the expression for the correlation functions containing the currents. To simplify the equations, we restrict to \textit{a homogeneous setting}, where the one-point functions are space independent, and two-point functions depend only on the relative distances. The equations~\eqref{eq: evo_eq_implicit_deltas_1},\eqref{eq: evo_eq_implicit_deltas_2} and \eqref{eq: evo_eq_implicit_deltas_3} reduce then to the following first layers of the gBBGKY hierarchy
\begin{widetext}
    \begin{equation}
    \label{eq: evo_eq_explicit_deltas_1}
        \partial_t \langle q_{i_1}\rangle = -  \int {\rm d}x_2 V'(x_1-x_2) A_{i_1,0}^k g^{(2)}_{k,0}(x_1,x_2),
    \end{equation}
    \begin{equation}
    \label{eq: evo_eq_explicit_deltas_2}
    \begin{aligned}
        &\partial_t g^{(2)}_{i_1,i_2}(x_1,x_2) + \Big[\partial_{x_1} A_{i_1}^jg^{(2)}_{j,i_2}(x_1,x_2)+\frac{1}{2}\partial_{x_1} H_{i_1}^{j,k}g^{(3),\rm sym}_{j,k,i_2}(x^+_1,x^-_1,x_2)\Big]_{(1,2)} =\\&-\Big[V'(x_1-x_2)A_{i_1,0}^j\left(\langle q_j\rangle C_{i_2,0}+ g^{(2)}_{j,k}(x_1,x_2)C^{k,m}C_{m,i_2,0}\right)+\int {\rm d}x_3 V'(x_1-x_3) A_{i_1,0}^j\left(\langle q_{j}\rangle g^{(2)}_{i_2,0}(x_2,x_3)  +  g^{(3)}_{j,i_2,0} \right)\Big]_{(1,2)}
        \\&
        -\delta(x_1-x_2) \int\dd{x_3}V'(x_1-x_3)
        \Big[-C_{i_1,i_2,j}C^{j,k}A_{k,0}^{n}+\left(A_{i_1,0}^j\delta_{i_2}^k+\delta_{i_1}^jA_{i_2,0}^k\right)C_{j,k,m}C^{m,n}\Big]g^{(2)}_{n,0}(x_1,x_3)\,,
    \end{aligned}
    \end{equation}
    \begin{equation}
    \label{eq: evo_eq_explicit_deltas_3}
    \begin{aligned}
        &\partial_t g^{(3)}_{i_1,i_2,i_3}(x_1,x_2,x_3) + \left[\partial_{x_1} A_{i_1}^{j}g^{(3)}_{j,i_2,i_3}(x_1,x_2,x_3)+\frac{1}{2}\partial_{x_1}H_{i_1}^{j,k}g^{(2)}_{j,i_2}(x_1,x_2)g^{(2)}_{k,i_3}(x_1,x_3)\right]_{(1,2,3)}=
        \\&
        =\bigg[ -\delta(x_1-x_2)\Big(  A_{i_1}^{k}
    C_{k,i_2,m} C^{m,n}- 
    C_{i_1,i_2,k} C^{k,m}A_m^{n}+ H_{i_1}^{m,n}C_{m,i_2}\Big)\partial_{x_1}g^{(2)}_{n,i_3}(x_1,x_3)\bigg]_{(1,2,3)}+O(V_0^2\xi^2,V_0\xi^{-3})\,,
    \end{aligned}
    \end{equation}
\end{widetext}
where, for brevity, in Eq.~\eqref{eq: evo_eq_explicit_deltas_3} we truncated higher-order terms in both parameters $V_0$ and $1/\xi$. Let us comment on the structure of this complex hierarchy of equations. Firstly, in homogeneous systems the evolution equation for the density, eq. ~\eqref{eq: evo_eq_explicit_deltas_1}, does not contain the Vlasov term and therefore all its dynamics is given by the evolution of the two-point functions, Eq.~\eqref{eq: evo_eq_explicit_deltas_2}. Here, on the LHS are the kinetic term describing the ballistic transport of two-point function and its diffusive corrections. On the RHS we observe that the singular contribution to the two-point function creates a contact forcing proportional to $V_0/\xi$. Finally, the dynamics of the three-point function is written in ~\eqref{eq: evo_eq_explicit_deltas_3} up to order $O(V_0/\xi)$. Again, its LHS include the kinetic terms for the ballistic transport of correlations, while the RHS behave as singular forcing, which is given the integrable dynamics, i.e. by the integrable contact interactions. In Appendix \ref{app:comment_eq_charges} more details are given in order to derive Eq.~\eqref{eq: evo_eq_explicit_deltas_1},\eqref{eq: evo_eq_explicit_deltas_2} and \eqref{eq: evo_eq_explicit_deltas_3}. 
Finally, we observe that the assumption of large hydrodynamic length scale $\xi$ is sufficient to have a well-defined truncation scheme for the hierarchy.

\subsection{The hierarchy for the (quasi)particle occupations and correlations}
\label{sec: gBBGKYfor__quasiparticles}
In this section we express the hierarchy in terms of quasiparticles of the Hamiltonian $\hat{H}_a^{\rm int}$. In particular, for integrable systems, the quasiparticles are characterized by Bethe rapidities $\theta$, whose distribution function in the fluid cell $(x,t)$ is denoted by $\rho_{\theta}(x,t)$ 
(in the rest of the text, we use Greek indices for rapidities, while Latin ones for charge labels).
The rapidities are related to the quasiparticles momentum through $k_\theta=k(\theta)$. The density $\rho_{\theta}(x,t)$ is defined through its relation with the hydrodynamic variables
\begin{equation}
    \langle q_i \rangle(x,t) = \int\dd{\theta}\rho_{\theta}(x,t) h_i(\theta)\,,
\end{equation}
where $h_i(\theta)$ is the single-particle eigenvalue of the charge $Q_i$. With a complete set of charges, it is possible to invert this expression, having a bijection between the quasiparticle density $\rho_{\theta}(x,t)$ and the expectation values of the charge density $\langle q_i \rangle$. Hence, it is possible to recast the infinite tower of hydrodynamic equations for the conservation of charges, labelled by the integer $i$, to the single evolution of $\rho_{\theta}(x,t)$.
We now define the dressing operator 
acting on a test function $h(\theta)$
\begin{equation}
    h^{\rm dr}=(1-Tn)^{-1} \cdot h\,,
\end{equation}
where the action of a linear operator is defined as $(A\cdot h)(\theta)=\int\dd{\gamma}A(\theta,\gamma)h(\gamma)=A_{\theta}^{\gamma}h_{\gamma}$, assuming the Einstein convention for the integration over repeated indexes.  In particular $T=T_{\theta,\gamma}$ is the scattering shift of the model (which vanishes in the limit of zero contract interactions $a\to 0$), being independent on the state. The filling function defines the occupation of rapidities respect to the total density of states $\rho^{\rm t} = (k')^{\rm dr}/2\pi $ and it is given by $n=2\pi\rho/(k')^{\rm dr}$.  Moreover, we shall make extensive use of compact matrix notation, for example by introducing the rotation matrix
\begin{equation}
    R_{\theta_1,\theta_2} = [1 - Tn]_{\theta_1,\theta_2} =  \delta_{\theta_1,\theta_2} -T_{\theta_1,\theta_2} n_{\theta_2},
\end{equation}
which implements the dressing operation as $h^{\rm dr}=R^{-1}\cdot h$. Crucially, it is also possible to express the density of currents, as
\begin{equation}
    \langle j_i(x)\rangle = \int\dd{\theta}\rho_{\theta}(x)v^{\rm eff}_{\theta}(x)h_i(\theta) + O(1/\xi)\,,
\end{equation}
with the effective velocity $v^{\rm eff}_{\theta}=(h_2')^{\rm dr}(\theta)/(h_1')^{\rm dr}(\theta)$, where for Galilean particles with unit mass $h_2(\theta)=\theta^2/2$ and $h_1(\theta)=\theta$.
Similarly, we can define quasiparticle expression for few point functions as 
\begin{equation}
    \langle q_{i_1}\ldots q_{i_n}\rangle = \int\prod_{k=1}^n\Big[\dd{\theta_k}h_{i_k}(\theta_k)\Big]\langle \rho_{\theta_1}\ldots \rho_{\theta_n}\rangle\,.
\end{equation}
Another important object is the flux Jacobian, that in quasiparticles reads
\begin{equation}
    A_{i}^{j}=\int\dd{\theta_1}\dd{\theta_2}h_{i}(\theta_1)R^{-t}_{\theta_1,\gamma}v^{\rm eff}_{\gamma}R^t_{\gamma,\theta_2}h^{j}(\theta_2)\,,
\end{equation}
where $h^j$ is defined as the inverse of $h_j$, such that $h_j(\theta)h^j(\theta')=\delta_{\theta,\theta'}$. For completeness we also recall $A_{\theta}^{\theta'}\equiv R^{-t}_{\theta, \gamma}v^{\rm eff}_{\gamma}R^t_{\gamma, \theta'}$, with $R^t=1-nT$.
In a similar manner we compute the second derivative of current as
\begin{equation}
    H_{i}^{j,k}=\int\dd{\theta}\dd{\alpha}\dd{\beta}h_{i}(\theta)H_{\theta}^{\alpha,\beta}h^{j}(\alpha)h^{k}(\beta)\,,
\end{equation}
where we defined (from below onwards we shall use the notation $v_{\mu, \gamma} \equiv v^{\rm eff}_{\mu}-v^{\rm eff}_{\gamma}$)
\begin{equation}
    H_{\theta}^{\alpha,\beta} = \big[R^{-t}_{\theta,\gamma}v_{\mu,\gamma}\frac{T^{\rm dr}_{\gamma,\mu}}{\rho^{\rm t}_{\gamma}}R^t_{\mu,\alpha}R^t_{\gamma,\beta}\big]_{(\alpha, \beta)}\,.
\end{equation}
Analogously to what was done in the previous section, we consider the connected few point correlators and separately recognize the GGE contribution from the regular long range part. For the two-point function we have
\begin{equation}
    \langle\rho_{\theta_1}(x_1)\rho_{\theta_2}(x_2)\rangle^c=\delta(x_1-x_2)C_{\theta_1,\theta_2}+g^{(2)}_{\theta_1,\theta_2}(x_1,x_2)\,,
\end{equation}
where $C_{\theta_1,\theta_2}=R^{-t}_{\theta_1,\gamma}\rho_{\gamma}f_{\gamma}R^{-1}_{\gamma,\theta_2}$ is the susceptibility matrix in quasiparticle representation, and $g^{(2)}_{\theta_1, \theta_2}(x_1,x_2)$ is the non-singular contribution. Also, the generic symbol $f_{\theta}=\{1,1-n_\theta,1+n_\theta, n_\theta^{-1}\}$ denotes the statistical factor (classical, fermionic, bosonic and wave statistics, respectively).
Analogously, for the three-point functions, we can write 
\begin{equation}
    \begin{split}
    \langle\rho_{\theta_1}(x_1)&\rho_{\theta_2}(x_2)\rho_{\theta_3}(x_3)\rangle^c=\\&
    =\delta(x_1-x_2)\delta(x_1-x_3)C_{\theta_1,\theta_2,\theta_3}\\&
    +\Big[\delta(x_1-x_2) C_{\theta_1,\theta_2}^{\gamma}g^{(2)}_{\gamma,\theta_3}(x_1,x_3)\Big]_{(1,2,3)}\\&+g^{(3)}_{\theta_1,\theta_2,\theta_3}(x_1,x_2,x_3)\,,
\end{split}
\end{equation}
where $C_{\theta_1,\theta_2,\theta_3}$ is the local three-point function in a GGE. We also introduce the tensor given by the functional derivative of the susceptibility respect to the local density
\begin{align}
    %\begin{split}
        \nonumber C_{\theta_1,\theta_2}^{\theta_3}&\equiv\frac{\delta C_{\theta_1,\theta_2}}{\delta \rho_{\theta_3}} = R^{-t}_{\theta_1,\alpha}\Big[\delta_{\alpha,\beta}\delta_{\alpha,\theta_3}f_{\alpha}-\delta_{\alpha,\beta}\varepsilon n_{\alpha} R^t_{\alpha,\theta_3} \\&+ T^{\rm dr}_{\alpha,\beta}\frac{R^t_{\alpha,\theta_3}}{\rho^{\rm t}_{\alpha}}f_{\beta}\rho_{\beta}
    +
    f_{\alpha}\rho_{\alpha} T^{\rm dr}_{\alpha,\beta}\frac{R^t_{\beta,\theta_3}}{\rho^{\rm t}_{\beta}}
    \Big]R^{-1}_{\beta,\theta_2}\,,
    %\end{split}
\end{align}
with $\varepsilon=\{0,-1,1\}$ respectively for classical, Fermi-Dirac or Bose-Einstein statistics and we also introduce the tensor giving the kinetic term of the three-point correlations
\begin{equation}
\label{eq:M_tensor}
M_{\theta_1,\theta_2}^{\theta_3}
      = \bigl[ R^{-t}_{\theta_1,\mu} R^{-t}_{\theta_2,\gamma}\,
    T^{\text{dr}}_{\mu,\gamma}\,
    \frac{\rho_\gamma f_\gamma}{\rho^{\rm t}_{\mu}}\,
   v_{\mu, \gamma}\,
    R^{t}_{\mu,\theta_3} \bigr]_{(\theta_1,\theta_2)}\,,
\end{equation}
which is proportional to the strength of the local interactions $a$ (see Appendix \ref{app:der_tensor_M} for a complete derivation).

With this, we can write down the first three layers in the quasiparticle representation:
\vspace{3cm}
\begin{widetext}
\begin{equation}\label{eq:BBGKY_quasi_rho}
    \partial_{t}\rho_{\theta_1}= \int\dd{x_2}\dd{\theta_2}V'(x_1-x_2)\partial_{\theta_1}g^{(2)}_{\theta_1,\theta_2}(x_1,x_2)\,,
\end{equation}
\begin{equation}
\label{eq:BBGKY_quasi_g2}
\begin{split}
    \partial_t g^{(2)}_{\theta_1, \theta_2}(x_1,x_2)&+ \Big[\partial_{x_1}\big(A_{\theta_1}^{\gamma} g^{(2)}_{\gamma, \theta_2}(x_1,x_2)\big) +\frac{1}{2}\partial_{x_1} \big(H_{\theta_1}^{\gamma,\eta} g^{(3),\rm sym}_{ \gamma,\eta,\theta_2}(x_1^+,x_1^-,x_2)\big)\Big]_{(1,2)}=
    \\&
    =\Big[V'(x_1-x_2) \int \dd{\theta_3} \partial_{\theta_1}\Big( \rho_{\theta_1}C_{\theta_2,\theta_3}+ C_{\theta_2,\theta_3}^{\gamma}g^{(2)}_{\theta_1,\gamma}(x_1,x_2)\Big)
    \\&+
    \int \dd x_3 \dd{\theta_3} V'(x_1-x_3) \partial_{\theta_1}\Big( \rho_{ \theta_1}g^{(2)}_{\theta_2,\theta_3}(x_2,x_3)+g^{(3)}_{\theta_1,\theta_2,\theta_3}(x_1,x_2,x_3)\Big)\Big]_{(1,2)}
    \\&+\delta(x_1-x_2)\int\dd{x_3}\dd{\theta_3}V'(x_1-x_3)\Big(-C_{\theta_1,\theta_2}^{\gamma}\partial_{\gamma}+(\partial_{\theta_1}+\partial_{\theta_2})C_{\theta_1,\theta_2}^{\gamma}\Big)g^{(2)}_{\gamma,\theta_3}(x_1,x_3)\,,
\end{split}
\end{equation}
% \begin{equation}\label{eq:BBGKY_quasi_g3}
% \begin{split}
%     &\partial_{t}g^{(3)}_{\theta,\theta',\theta''}(x,x',x'')+\partial_{x}(A_{\theta}^{\gamma}g^{(3)}_{\gamma,\theta',\theta''})+\partial_{x'}(A_{\theta'}^{\gamma}g^{(3)}_{\theta,\gamma,\theta''})+\partial_{x''}(A_{\theta''}^{\gamma}g^{(3)}_{\theta,\theta,\gamma})=
%     \\&
%     =\delta(x-x')M_{\theta,\theta'}^{\gamma}\partial_xg^{(2)}_{\gamma,\theta''}(x,x'')+\delta(x'-x'')M_{\theta',\theta''}^{\gamma}\partial_{x'}g^{(2)}_{\gamma,\theta}(x',x)+\delta(x-x'')M_{\theta,\theta''}^{\gamma}\partial_xg^{(2)}_{\gamma,\theta'}(x,x')+O(V_0,1/\xi)
% \end{split}
% \end{equation}
\begin{equation}\label{eq:BBGKY_quasi_g3}
\begin{split}
    \partial_{t}g^{(3)}_{\theta_1,\theta_2,\theta_3}(x_1,x_2,x_3)&+\Big[\partial_{x_1}A_{\theta_1}^{\gamma}g^{(3)}_{\gamma,\theta_2,\theta_3}(x_1,x_2,x_3)\Big]_{(1,2,3)}=\Big[\delta(x_1-x_2)M_{\theta_1,\theta_2}^{\gamma}\partial_{x_1}g^{(2)}_{\gamma,\theta_3}(x_1,x_3)\Big]_{(1,2,3)}+O(V_0^2\xi^2,V_0\xi^{-3})\,.
\end{split}
\end{equation}
\end{widetext}

{Let us comment on the kinetic terms for the two- and three-point functions in Eq.~\eqref{eq:BBGKY_quasi_g2} and \eqref{eq:BBGKY_quasi_g3}. The first terms with the matrix $A_{\theta}^{\gamma} =  [R^{-1} v^{\rm eff} R]_{\theta}^{\gamma}$ represent convective spreading with velocities associated to the different quasiparticles. While in an interacting model this matrix is non-diagonal (due to mode mixing), we shall see in the next section that the latter simply reduces to a diagonal matrix containing the velocities $v^{\rm br}(\theta)$ as eigenvalues in the limit $a \to 0$. The second term in eq. \eqref{eq:BBGKY_quasi_g2} is instead non-zero only in interacting systems. The latter can be seen as an effective diffusion induced by the motion of the three-point functions for the two-point functions. Indeed, its form becomes more familiar in the small times limit, i.e. for states close to the local GGE. In this limit, we can indeed show the relation (see \cite{SM} for a complete derivation) 
\begin{equation}
\label{eq: small_time_diffusion}    \lim_{t\to0^+}H_{\theta_1}^{\gamma,\eta} g^{(3),\rm sym}_{ \gamma,\eta,\theta_2}(x_1^+,x_1^-,x_2;t)=\mathfrak{D}_{\theta_1}^{\gamma}\partial_{x_1}g^{(2)}_{\gamma,\theta_2}(x_1,x_2)\,,
\end{equation}
where $\mathfrak{D}_{\theta}^{\gamma}$ is indeed the diffusion matrix \cite{de2019diffusion,medenjak2020diffusion,PhysRevB.98.220303} associated with the local equilibrium state around which we are linearising.  
As the system evolves, the validity of this relation gets worse, due to the build-up of nontrivial long-range three-point correlations, yet this contribution is fundamental to thermalization of the two-point functions and, consequently, of the 1-point functions as we shall see in the next sections.  In the limit $a \to 0$  this contribution vanishes, as there is indeed no internal (convective) diffusion $\mathfrak{D}\approx a^2$, as also confirmed by the analysis in the next section \ref{ssec:standrd_BBGKY}.

\subsection{Recovering the standard BBGKY hierarchy in the limit $a \to 0$} 

\label{ssec:standrd_BBGKY}
The standard BBGKY hierarchy is recovered in the limit of non-interacting unperturbed systems, i.e. $\hat{H}_a^{\rm int}\to\hat{H}^{\rm free}$. It is important here to stress that the gBBGKY formalism is valid both for classical and quantum systems and difference between the two scenarios is fully incorporated into the statistical factor $f=f(n)$. Here, in order to compare with standard textbook expressions \cite{Balescu_1975}, we specialize to the classical case and set $f=1$.\\
The limit $\hat{H}_a^{\rm int}\to\hat{H}^{\rm free}$ corresponds to zero scattering shift $T\to0$. We now carefully analyze the effect of this limit on the objects defined in Section \ref{sec: gBBGKYfor__quasiparticles}. Firstly, we observe that this limit trivialize the dressing operator as $R_{\theta, \theta'}\to\delta_{\theta, \theta'}$, and hence $h^{\rm dr}\to h$. As a consequence, the effective velocity reduces to the particle's velocity $v^{\rm eff}_\theta\to v^{\rm br}_\theta=\theta$. In a similar way the flux Jacobian becomes $A_{\theta}^{\gamma}\to v^{\rm br}_{\theta}\delta_{\theta,\gamma}$. Since non-interacting particles don't exhibit non-trivial correlations, we also have
\begin{equation}
\begin{aligned}
&C_{\theta_1,\theta_2}\to\delta_{\theta_1,\theta_2}\rho_{\theta_1}\,,\qquad
C^{\theta_3}_{\theta_1,\theta_2}\to\delta_{\theta_1,\theta_2}\delta_{\theta_1,\theta_3}\,.
\end{aligned}
\end{equation}
As a consequence we observe that the delta contribution appearing in Eq~\eqref{eq:BBGKY_quasi_g2} is vanishing, being proportional to 
\begin{equation}
    \int\dd{\gamma}\big[\delta_{\theta_1,\theta_2}\delta_{\theta_1,\gamma}\partial_\gamma-(\partial_{\theta_1}+\partial_{\theta_2})\delta_{\theta_1,\theta_2}\delta_{\theta_1,\gamma}\big]h_{\gamma}=0\,.
\end{equation}
We also observe that, in this limit, we have $H\to 0$ and $M\to0$, both being proportional to $T$. We also stress that any higher-order hydrodynamic correction in the cumulant expansion is vanishing 
since non-interacting particles simply spread ballistically.
Hence, in the non-interacting limit $T\to0$, Eq.~\eqref{eq:BBGKY_quasi_rho}, \eqref{eq:BBGKY_quasi_g2} and \eqref{eq:BBGKY_quasi_g3} reduce to 
\begin{equation}\label{eq:BBGKY_quasi_rho_free}
    \partial_{t}\rho_{\theta_1}=\int\dd{x}\dd{\theta_2}V'(x)\partial_{\theta_1}g^{(2)}_{\theta_1,\theta_2}(x),
\end{equation}
\begin{equation}\label{eq:BBGKY_quasi_g2_free}
\begin{split}
    &\partial_t g^{(2)}_{\theta_1, \theta_2}(x_1,x_2)+\Big[\partial_{x_1} \big(v^{\rm br}_{\theta_1} g^{(2)}_{\theta_1, \theta_2}(x_1,x_2) \big)\Big]_{(1,2)}=
    \\&
    = \Big[V'(x_1-x_2)\partial_{\theta_1}\Big(\rho_{ \theta_1}\rho_{ \theta_2}+ g^{(2)}_{\theta_1,\theta_2}\Big)+
    \\&+
    \int \dd{x_3} \dd{\theta_3} V'(x_1-x_3) \partial_{\theta_1} \rho_{\theta_1}g^{(2)}_{\theta_2,\theta_3}(x_2,x_3)+
    \\&+
    \int \dd{x_3} \dd{\theta_3} V'(x_1-x_3) \partial_{\theta_1} g^{(3)}_{\theta_1,\theta_2,\theta_3}(x_1,x_2,x_3)\Big]_{(1,2)}\,,
\end{split}
\end{equation}
\begin{equation}\label{eq:BBGKY_quasi_g3_free}
\begin{split}
    &\partial_{t}g^{(3)}_{\theta_1, \theta_2 ,\theta_3}(x_1,x_2,x_3) +\\&+\Big[\partial_{x_1} \big(v^{\rm br}_{\theta_1} g^{(3)}_{\theta_1, \theta_2 ,\theta_3}(x_1,x_2,x_3) \big)\Big]_{(1,2,3)} +\ldots =0\,,
\end{split}
\end{equation}
that perfectly coincide with the first equations of celebrated BBGKY hierarchy for a homogeneous system. It is also easy to prove that this equivalence is valid for any level of the hierarchy and that can be extended also in inhomogeneous settings. It is important to stress that, through the gBBGKY formalism, we provided a new alternative derivation of BBGKY in terms of correlated fluid cells that doesn't require to use the full system phase space distribution function.

 \subsection{The $V \to 0$ limit of the hierarchy}
It is natural to ask what happens to the dynamics of correlations when the $V \to 0$ limit is considered and the only possible source of interaction is due to integrable part of the Hamiltonian. Firstly, in the homogenous setting with no initial long-range correlations, i.e. $g^{(n)}(t=0)=0$ for $n>1$, there is no dynamics at all in the system, as expected. This is a trivial confirmation that a GGE is stationary state in our framework when there is no integrability breaking.

This picture changes when the initial state is inhomogenous. In such case our hierarchy extends the equations put forward in~\cite{Hubner2024} and recovers them upon truncation to two layers. These two equations provide the basis for the new understanding of diffusion in integrable models which occurs under the presence of long-range correlations generated by integrable dynamics~\cite{10.21468/SciPostPhys.15.4.136}. Let us mention that gBBGKY is also capable of tracking the dynamics of higher point functions, even if higher order hydrodynamic matrices are not explicitly computed in this manuscript. For instance, we expect to observe long-range structures in three-point correlations, generated due to the gradients in two-point function (similarly as two-point correlations are generated by gradients of one-point function in~\cite{Hubner2024}). This can be seen from solution~\eqref{eq: integrated_g3_10} of the Eq.~\eqref{eq:BBGKY_quasi_g3}. There, even in the homogenous case, we see a generation of three-point correlations solely by integrable dynamics, powered by gradients of $g^{(2)}$ (generated in homogenous setting by presence of $V$).

\section{The generalised Landau equation and late-time dynamics}
\label{s:Landau}
The gBBGKY equations presented above are clearly complicated to analyze and to fully numerically simulate. However, to extract the thermalisation dynamics, some approximations can be performed, leading to what is known as Landau equation, namely the Boltzmann equation in the limit of small momentum exchange, also known as the grazing collisions limit.

We shall perform a macroscopic rescaling of space $x\to \xi x$ and time $t\to \xi t$, and importantly, in what follows we will assume that the long-range coupling is weak by treating $V_0$ as a small parameter. Thus, scattering integral description have a narrower range of applicability than the hierarchy, which assumes only large $\xi$. However, as we show in Sec. \ref{s:hard-rods}, the scattering integral turns out to be quantitatively predictive also for quite large values of $V_0$, provided that $V_0/\xi\ll1$. 
Nevertheless, this approach is sufficient to understand the main mechanisms behind the thermalisation of the system.

We start the derivation by considering the gBBGKY hierarchy written in terms of quasiparticles occupations and correlations, as introduced in Sec.~\ref{sec: gBBGKYfor__quasiparticles}. More precisely, in this section, we explicitly make use of homogeneity by expressing correlations as uniquely functions of relative distance. Hence, for the two-point function we have $g^{(2)}_{\theta_1,\theta_2}(x_1,x_2)=g^{(2)}_{\theta_1,\theta_2}(z= x_1-x_2)$ and for the three-point function
$g^{(3)}_{\theta_1,\theta_2,\theta_3}(x_1,x_2,x_3)=g^{(3)}_{\theta_1,\theta_2,\theta_3}(z,z')$, with $z= x_1-x_2$ and $z'= x_2-x_3$.

Furthermore, we perform a decomposition of the two-point function $g^{(2)}$ into leading and subleading contributions in the $1/\xi$ expansion, namely
\begin{equation}
    g^{(2)}_{\theta_1,\theta_2}(z)=\frac{1}{\xi}g^{(2),1}_{\theta_1,\theta_2}(z)+\frac{1}{\xi^2}g^{(2),2}_{\theta_1,\theta_2}(z).
\end{equation}
%and they are the leading and subleading contributions to $g^{(2)}$ in the $1/\xi$ expansion.
We also explicitly factorize the scaling parameter of the three-point functions using $g^{(3)}\to g^{(3)}/\xi^2$. Now, using the definition of the interaction potential~\eqref{eq:potential}, the $1/\xi$ dependence is fully explicit.
Plugging the decomposition into equation~\eqref{eq:BBGKY_quasi_g2} for the two-point function and requiring equality both at the order $1/\xi$ and $1/\xi^2$ produces two equations:
\begin{equation}
\label{eq: evolution_for_g2(1)}
\begin{split}
    &\partial_t g^{(2),1}_{\theta_1, \theta_2}(z)+\partial_z \Big(A_{\theta_1}^{\gamma} g^{(2),1}_{\gamma, \theta_2}(z)-A_{\theta_2}^{\gamma} g^{(2),1}_{\theta_1, \gamma}(z) \Big)=\\&V_0\varphi'(z) \int \dd{\gamma} \Big(\partial_{\theta_1} \rho_{\theta_1}C_{\gamma,\theta_2}-\partial_{\theta_2} \rho_{\theta_2}C_{\theta_1,\gamma }\Big)\,,
\end{split}
\end{equation}
\begin{equation}
\label{eq: evolution_for_g2(2)}
\begin{split}
    &\partial_t g^{(2),2}_{\theta_1, \theta_2}(z)+\partial_z \Big(A_{\theta_1}^{\gamma} g^{(2),2}_{\gamma, \theta_2}(z)-A_{\theta_2}^{\gamma} g^{(2),2}_{\theta_1, \gamma}(z) \Big)=\\
    &-\frac{1}{2}\partial_z \Big(H_{\theta_1}^{\gamma_1,\gamma_2} g^{(3),\rm sym}_{\gamma_1,\gamma_2, \theta_2 }(0^+,z)-H_{\theta_2}^{\gamma_1, \gamma_2} g^{(3),\rm sym}_{\theta_1,\gamma_1, \gamma_2 }(z,0^+)\Big)\,.
\end{split}
\end{equation}
These equations are the effective evolution laws for $g^{(2),1}$ and $g^{(2),2}$. We can already observe that, meanwhile $g^{(2),1}$ is rapidly evolving due to the coupling with the density, the subleading correction $g^{(2),2}$ slowly builds up forced by the three-point function.
We start with considering four equations \eqref{eq:BBGKY_quasi_rho}, \eqref{eq: evolution_for_g2(1)}, \eqref{eq: evolution_for_g2(2)} and \eqref{eq:BBGKY_quasi_g3}. The goal is to reduce them to a single equation for quasiparticle distribution $\rho_\theta$. We will follow a procedure in spirit similar to classic derivations of Boltzmann scattering kernel from BBGKY hierarchy \cite{Fouvry2019,Fouvry2020}. The central idea here is the separation of timescales over which different correlations relax. For instance, one-point function is expected to relax on much longer time scale than two-point function. Schematically, the derivation will proceed as follows:
\begin{enumerate}
    \item Integrate \eqref{eq: evolution_for_g2(1)} in time expressing $g^{(2),1}_{\theta_1, \theta_2}(z,t)$ as explicit functional of $\rho_\theta(\tau)$ for all earlier times $\tau<t$.
    \item Plug the result of previous step into \eqref{eq:BBGKY_quasi_g3} and again integrate in time expressing $g^{(3)}_{\theta_1,\theta_2,\theta_3}(z,z',t)$ as  functional of $\rho_\theta(\tau)$.
    \item Use this result to compute $g^{(3)}_{\theta_1,\theta_2,\theta_3}(z,z',t)$, plug into equation \eqref{eq: evolution_for_g2(2)} and integrate in time expressing $g^{(2),2}_{\theta_1,\theta_2}(z,t)$ as  functional of $\rho_\theta(\tau)$.
    \item Plug this result into \eqref{eq:BBGKY_quasi_rho} and perform approximations related to separation of timescales which finally yield generalized Landau kinetic equation.
\end{enumerate}
We elaborate on the derivation in what follows. We assume that at $t=0$ the state is GGE state with $g^{(n)}=0$ for $n\geq 2$. This assumption is not necessary as the system will eventually forget about the initial correlations, but simplifies the calculation which anyway leads to universal, late-time physics. 

The first step consists of considering \eqref{eq: evolution_for_g2(1)} and viewing it as a linear equation (we assume here that $A_\theta^\gamma$ is constant operator) for $g^{(2),1}_{\theta_1 \theta_2}(z)$ with a $\rho$ - dependent source term on RHS given by
% \begin{equation}\label{eq:hierarchy_der1}
% \begin{aligned}
%     \partial_t g^{(2),1}_{\theta_1 \theta_2}(z)+\partial_z (A_{\theta_1}^{\gamma} g^{(2),1}_{\gamma, \theta_2}(z)-A_{\theta_2}^{\gamma} g^{(2),1}_{\theta_1, \gamma}(z))=F^{(2),1}_{\theta_1 \theta_2}[\rho](z),
% \end{aligned}
% \end{equation}
% with
\begin{equation}
    \mathcal{S}^{(2),1}_{\theta_1,\theta_2}[\rho](z)=V_0\varphi'(z) \int \dd{\gamma} \Big(\partial_{\theta_1} \rho_{\theta_1} C_{\gamma,\theta_2}-\partial_{\theta_2} \rho_{\theta_2}C_{\theta_1,\gamma }\Big).
\end{equation}
Thus, \eqref{eq: evolution_for_g2(1)} can formally be solved using Green's function method (for details see Appendix~\ref{app:Green}), which gives
\begin{equation}\label{eq:g21solution}
\begin{aligned}
    &g^{(2),1}_{\theta_1, \theta_2}[\rho](z,t) =R_{\theta_1,\xi_1}^{-t} R_{\theta_2,\xi_2}^{-t}  \int_0^t \dd \tau  \int \frac{\dd k}{2 \pi} \int \dd z' \times \\
    &e^{ik(z+z'-v_{\xi_1, \xi_2}\tau)}R_{\xi_1,\eta_1}^t R_{\xi_2,\eta_2}^t \mathcal{S}^{(2),1}_{\eta_1, \eta_2}[\rho](z',t-\tau).
\end{aligned}
\end{equation}
In this way we have expressed $g^{(2),1}_{\theta_1, \theta_2}(z,t)$ with $\rho_\theta$ at all earlier times. The structure of a linear equation with $\rho$-dependent source term turns out to be typical in this derivation and we encounter it also in the second step, which amounts to plugging \eqref{eq:g21solution} into the equation \eqref{eq:BBGKY_quasi_g3}. The linear equation features a source term on the RHS which reads
% \begin{equation}
% \begin{aligned}
%     &\partial_t g_{\theta_1,\theta_2\theta_3}^{(3)}(z,z')+\partial_z (A_{\theta_1}^\gamma g_{\gamma \theta_2 \theta_3}(z,z')-A_{\theta_2}^\gamma g_{\theta_1 \gamma \theta_3}(z,z'))+\\
%     &\partial_{z'}(A_{\theta_2}^\gamma
% g_{\theta_1 \gamma \theta_3}(z,z')-A_{\theta_3}^\gamma g_{\theta_1 \theta_2 \gamma}(z,z'))=F^{(3)}_{\theta_1,\theta_2,\theta_3}[\rho](z,z')
% \end{aligned}
% \end{equation}
\begin{equation}
\begin{split}
    &\mathcal{S}^{(3)}_{\theta_1, \theta_2, \theta_3}[\rho](z,z') = \delta(z)M_{\theta_1,\theta_2}^{\gamma}\partial_{z}g^{(2),1}_{\gamma,\theta_3}(z+z')+
    \\&
    -\delta(z')M_{\theta_2,\theta_3}^{\gamma}\partial_{z}g^{(2),1}_{\theta_1,\gamma}(z)+\delta(z')M_{\theta_2,\theta_3}^{\gamma}\partial_{z'}g^{(2),1}_{\theta_1,\gamma}(z+z')\,.
\end{split}
\end{equation}
Using a similar technique as used to get \eqref{eq:g21solution} we explicitly find the functional $g_{\theta_1,\theta_2,\theta_3}^{(3)}[\rho](z,z',t)$. In the third step we compute $g^{(3),\rm{sym}}_{\theta_1, \theta_2, \theta_3}[\rho]$, and plug it to equation \eqref{eq: evolution_for_g2(2)}. Once again, this equation is integrated in time yielding $g^{(2),1}_{\theta_1, \theta_2}[\rho](z,t)$. In the last step we insert the expressions for $g^{(2),1}_{\theta_1, \theta_2}[\rho](z,t)$ and $g^{(2),2}_{\theta_1, \theta_2}[\rho](z,t)$ into the equation for $\rho_\theta$, which now becomes closed as all the higher correlations have been eliminated through time integration. We find that the equation has two contributions
\begin{equation}
\label{eq:implicit_kinetic_equation}    \partial_t\rho_\theta=\frac{1}{\xi}\mathcal{I}^{(2),1}[\rho](\theta)+\frac{1}{\xi^2}\mathcal{I}^{(2),2}[\rho](\theta).
\end{equation}
Restoring the microscopic time variable, the two terms are, respectively, proportional to $V_0^2/\xi^2$ and $a^2V_0^2/\xi^3$. In sections \ref{sec:kineticblocking} and \ref{sec:landau_main_text} we derive them and show that $\mathcal{I}^{(2),1}$ is vanishing, proving that the leading thermalization rate is $a^2V_0^2/\xi^3$.

\begin{figure}[h!]
\includegraphics[width=1\linewidth]{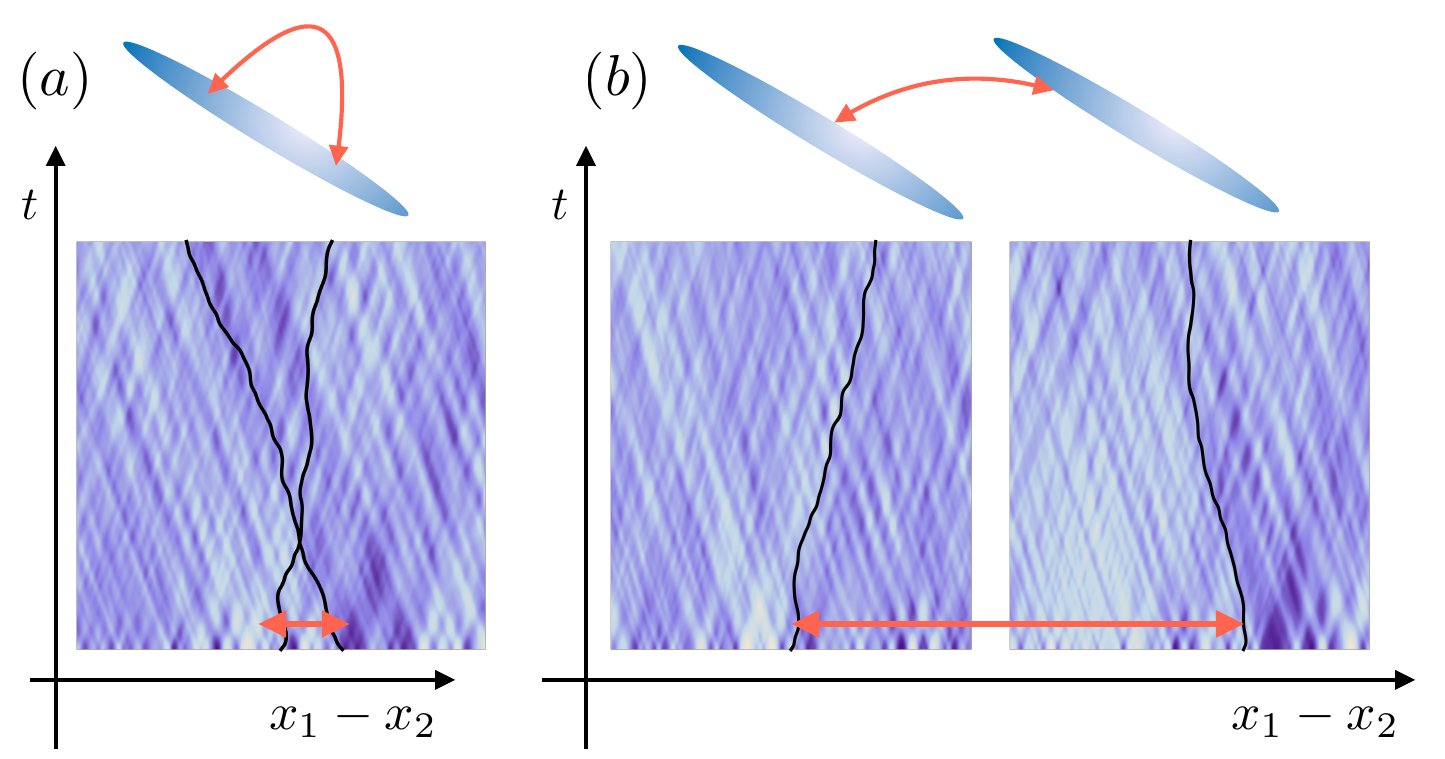}
\caption{
  Schematic representation of the two thermalisation mechanisms leading to the two terms in eq. \eqref{eq: Landau Equation}. On the \textit{left} is represented the dynamics giving the \textit{self} + \textit{cross} term: two correlated particles at a distance $\xi$ experience fluctuations due to convective waves carried by the other quasiparticles, giving cross diffusion, and they also scatter among each other, giving a self-diffusion to the two-point functions. On the \textit{right}, the dynamics giving the \textit{cross} term: two correlated particles never interact with contact interactions; therefore, they only experience the convective density waves through the system, giving standard diffusion to the two-point functions.    }
\label{fig: Correlations dynamicscartoon}
\end{figure} 

\subsection{Kinetic blocking at times $ \xi^2/V_0^2$}\label{sec:kineticblocking}
We start by considering the first term which reads
\begin{equation}
    \mathcal{I}^{(2),1}[\rho](\theta_1) = \int \dd z \dd \theta_2 V_0\varphi'(z) \partial_{\theta_1} g^{(2,1)}_{\theta_1, \theta_2}[\rho](z),
\end{equation}
with $g^{(2,1)}_{\theta_1, \theta_2}[\rho](z)$ given by \eqref{eq:g21solution}. We perform an approximation which amounts to replacing $\rho_\theta(z,t-\tau)$ with $\rho_\theta(z,t)$ under the integral and extending the integration range in time to infinity. Next, we use an identity $\int_0^\infty d\tau e^{i\tau x}=\pi \delta(x)+i\mathcal{P}(\frac{1}{x})$, where $\mathcal{P}$ denotes principal value. The term with principal value vanishes due to the symmetry of the integrand, the term with delta function yields (we also use $\int \dd \theta R^{-t}_{\theta,\gamma}=2 \pi \rho^{\rm t}_\gamma$)
\begin{equation}\label{eq:I21}
\begin{aligned}
    &\mathcal{I}^{(2),1}[\rho](\theta_1)=  \int \frac{\dd k}{2 \pi} |k|V_0^2\hat{\varphi}^2(k)\, \partial_{\theta_1} R_{\theta_1,\gamma_1}^{-t} \rho^{\rm t}_{\gamma_2}  \delta(v_{\gamma_1, \gamma_2}) \times \\
   & R_{\gamma_1, \eta_1}^{t}R^t_{\gamma_2 ,\eta_2} \int \dd \gamma_3 (\partial_{\eta_1} \rho_{\eta_1}C_{\gamma_3, \eta_2}-\partial_{\eta_2} \rho_{\eta_2}C_{\eta_1, \gamma_3}),
\end{aligned}
\end{equation}
where $\hat{\varphi}(k) = \int \dd x \, e^{ikx}\varphi(x)$. Crucially, the Dirac delta condition $v_{\gamma_1 \gamma_2}=0$ implies $\gamma_1=\gamma_2$ and hence, the integral \textit{vanishes} identically
\begin{equation}
    \mathcal{I}^{(2),1}[\rho](\theta_1)=0.
\end{equation}
This is a feature of 1D systems known under the name of \textit{kinetic blocking}. It is strictly related to the nature of two-particle scattering in 1D, which can only induce momentum exchanges between particles. The kinetic blocking can be avoided in systems with non-monotonous effective velocity, for instance in lattice. Then, so called Umklapp scatterings allow the system to thermalize due to two-body collisions~\cite{PhysRevLett.115.180601,Bertini2016IB}. For the present discussion, however, we focus on systems with monotonous effective velocity. {Another scenario involves systems of coupled 1D gases, as described in Section~\ref{s:dipolar}. If 1D gases have different densities or different strengths of integrable interactions, this opens a channel for scattering that redistribute the momenta, albeit not very effectively, leading only to prethermal stationary states~\cite{Lebek2024a,lisiak2025burgers}.} Hence, here, in order to capture the thermalization phenomena we have to include higher-order terms, which describe three-particle scatterings. These processes are encoded in $\mathcal{I}^{(2),2}[\rho](\theta)$, which we now analyze.

\subsection{(generalised) Thermalization at times $ \xi^3/(V_0 a)^2$}
\label{sec:landau_main_text}
The derivation of $\mathcal{I}^{(2),2}$ proceeds as sketched above and the approximations made along the way are of the same form as in derivation of $\mathcal{I}^{(2),1}$. Namely, every time we time-integrate the equation for correlation function, we replace $\rho_{\theta_1}(z,t-\tau)$ with $\rho_{\theta_1}(z,t)$ under the integral and extend integration range in time to infinity. In order to obtain the collision term, three such integrations have to performed leading to much more complex collision term as compared to \eqref{eq:I21}. The details of this quite long calculation are exposed in~\cite{SM}. Here we present the final result. We simplify the notation $\mathcal{I} \equiv \mathcal{I}^{(2),2}$ and find two type of terms
\begin{equation}
\label{eq: Landau Equation}
    \mathcal{I}=\mathcal{I{^{\rm cross}}}+ \mathcal{I{^{\rm self}}},
\end{equation}
Both terms are given by  $\mathcal{I}^\alpha = \partial_{\theta_1}R^{-t}_{\theta_1,\gamma}\mathcal{I}^{\alpha{\rm}}_0(\gamma)$, where $\alpha\in \{\rm cross, self\}$ and using the notation $v_{\theta_2, \theta_1} = v^{\rm eff}_{\theta_2} - v^{\rm eff}_{\theta_1}$, we have 
\begin{widetext}
\begin{equation}\label{eq:Idiagonal}
\begin{aligned}
    &\mathcal{I}^{\rm cross}_0[\rho](\theta_1)= 2 \pi V_0^2 \int  \dd k k^2 \hat{\varphi}^2(k)  \int \dd \theta_2 \dd \theta_3   \left[\frac{(\rho^{\rm t}_{\theta_3})^2}{\rho^{\rm t}_{\theta_1}} (T^{\rm dr}_{\theta_2, \theta_1})^2 \frac{|v_{\theta_2, \theta_1}|}{v_{\theta_3, \theta_2} v_{ \theta_1, \theta_3}^2} -\rho^{\rm t}_{\theta_1}(T^{\rm dr}_{\theta_3, \theta_2})^2 \frac{|v_{\theta_3, \theta_2}|}{v_{\theta_1, \theta_3} v_{\theta_2, \theta_1}^2}\right]\mathcal{B}_{\theta_1,\theta_2,\theta_3},
\end{aligned}
\end{equation}

\begin{equation}\label{eq:Ioffdiagonal}
\begin{aligned}
    \mathcal{I}^{\rm self}_0[\rho](\theta_1)= 2 \pi  V_0^2 \int  \dd \theta_2 \dd \theta_3 \int \dd k k^2  \hat{\varphi}(k)\rho^{\rm t}_{\theta_3}T^{\rm dr}_{\theta_1, \theta_2} \frac{v_{\theta_2, \theta_1}}{v_{\theta_1,\theta_3}} \bigg[&\hat{\varphi}\left(\frac{v_{\theta_2, \theta_1 }}{v_{ \theta_1, \theta_3}}k \right)\frac{\rho^{\rm t}_{\theta_2} }{\rho^{\rm t}_{\theta_1}}  \frac{T^{\rm dr}_{\theta_3, \theta_1}}{|v_{\theta_1, \theta_3}|v_{\theta_3, \theta_2 }}-\\&-2\hat{\varphi}\left(\frac{v_{\theta_2, \theta_1 }}{v_{  \theta_3, \theta_2 }}k \right) \frac{T^{\rm dr}_{\theta_2, \theta_3}}{v_{\theta_1, \theta_3} |v_{\theta_2, \theta_3}|}\bigg]  \mathcal{B}_{\theta_1,\theta_2,\theta_3},
\end{aligned}
\end{equation}
\end{widetext}
with the $\mathcal{B}$ tensor given by
\begin{equation}
\begin{aligned}
    &\mathcal{B}_{\theta_1,\theta_2,\theta_3}=\Big[\frac{v_{\theta_3, \theta_2}}{\rho^{\rm t}_{\theta_1}} R^t_{\theta_1, \eta} \partial_\eta \rho_\eta \rho_{\theta_3} f_{\theta_3} \rho_{\theta_2} f_{\theta_2}\Big]_{(\theta_1,\theta_2,\theta_3)}.
\end{aligned}
\end{equation}
It can be shown that the derived collision integral has all the expected features: thermal states are stationary states, it conserves particle number, momentum and energy and satisfies H-theorem with positive entropy increase.

Restoring the microscopic times in the kinetic equation \eqref{eq:implicit_kinetic_equation} and using that $T^{\rm dr} \sim a$, we can finally observe that the scattering integral given by Eq~\eqref{eq:Idiagonal} and Eq~\eqref{eq:Ioffdiagonal} predict a thermalization rate proportional to $(V_0/\xi)^2 \, a^2 /\xi$. The scaling can be understood as following: two factors of $V_0/\xi$ are necessary to form correlated two-point functions on large scales, then, two local scatterings with other particles carry $a^2$, and the spatial inhomogeneity of the two-point functions, inducing its convective diffusion in space, carry another factor $1/\xi$. See also in Fig. \ref{fig:1}: particles correlate due to the long-range potential and then diffuse in the convective flow carried by the sea of ballistic particles. This picture may look counter-intuitive, as, due to integrability breaking, one would naively expect the ballistic channels to decay at large times. However, this is not the case, as higher-order correlations preserve the ballistic structure to much longer times than $\xi^3$, giving an effective bath of convective particles inducing density waves which is the main and only mechanism for particle diffusion in interacting integrable models, see for example \cite{medenjak2020diffusion,Krajnik2022,Gopalakrishnan2024,Hubner2024,mcculloch2024,yoshimura_anomalous_2024,Doyon2025-2}.

The reason for distinguishing the two terms is  motivated by two different dynamical mechanisms, see Fig. \ref{fig: Correlations dynamicscartoon} related to convective diffusion. The first term, the cross term, originates from the diffusion of the two-point correlations where the two correlated particles never interact via contact interaction. This can be the case when for example two quasi-one dimensional systems interact with a long-range potential that only acts between the two different tubes. The second term, the self term, originates from the self-induced diffusion of the two-particles that can also interact via contact interaction: namely one of the two particles acts as a convective density wave for the other and vice-versa. As shown in \cite{SM}, this scattering term is absent when the long-range interaction acts only between different tubes, but is present when it acts also on the same tube. %Clearly, the situation where the long-range potential acts across different tubes and on the same one leads to a scattering integral given by the sum of the two term as in eq. \eqref{eq: Landau Equation}. 
{In the situation where the long-range potential acts across different tubes and on the same one, the scattering integral from eq.~\eqref{eq: Landau Equation} should include also the cross terms due to other tubes. We will discuss such scenarios in Section~\ref{s:dipolar} when we apply this formalism to experimental setups with dipolar quantum gases.}

It is important to stress that, even if one-point functions thermalize on time scales of order \(\xi^{3}/(V_{0} a)^{2}\), \textit{higher-point functions remain far from thermal}, which is why we have dubbed this process \emph{generalized thermalization}, in contrast with standard thermalization. In particular, three-point functions exhibit long-range correlations that do not decay on these time scales and arise entirely from local integrable interactions, mirroring the behavior of two-point functions in fully integrable dynamics \cite{Hubner2024}. In order to derive the collision integral, in \cite{SM} we derive an explicit expression for the dynamics of the three-point functions given by eq. \eqref{eq:BBGKY_quasi_g3}; here we report its large-time limit, namely in the regime where time is of the order $ \xi^3(V_0 a)^{-2}$:
\begin{widetext}
\begin{equation}\label{eq: integrated_g3_10}
\begin{aligned}
 &   g^{(3)}_{\theta_1, \theta_2, \theta_3}(z,z',t \gg \xi^3(V_0 a)^{-2}) =\\& = \Big[\frac{1}{2}R^{-t}_{{\theta_1},{\gamma_1}}R^{-t}_{{\theta_2},{\gamma_2}}R^{-t}_{{\theta_3},\gamma_3} \frac{1}{v_{\gamma_1, \gamma_2}}   \left[\text{sgn}(z)+ \text{sgn}(v_{\gamma_1, \gamma_2})\right]   R^t_{{\gamma_1},{\mu_1}}R^t_{{\gamma_2},{\mu_2}}R^t_{{\gamma_3},{\mu_3}}M_{\mu_1, \mu_2}^\beta \partial_1g^{(2),1}_{\beta,\mu_3} \left(
    \frac{v_{\gamma_2, \gamma_3}}{v_{\gamma_1, \gamma_2}}z-z'\right) \Big] + \\& +  \Big[\frac{1}{2}R^{-t}_{{\theta_1},{\gamma_1}} R^{-t}_{{\theta_2},{\gamma_2}}R^{-t}_{{\theta_3},\gamma_3}\frac{1}{v_{\gamma_1, \gamma_3}}\left[\text{sgn}(z+z')+ \text{sgn}(v_{\gamma_1,\gamma_3})\right]   R^t_{{\gamma_1},{\mu_1}}R^t_{{\gamma_2},{\mu_2}}R^t_{{\gamma_3},{\mu_3}}M_{\mu_1, \mu_3}^\beta \partial_1 g^{(2),1}_{\beta,\mu_2} \left( \frac{v_{\gamma_3, \gamma_2}}{v_{\gamma_1, \gamma_3}}z+ \frac{v_{\gamma_1, \gamma_2}}{v_{\gamma_1, \gamma_3}}z'\right) \Big] + \\&  + \Big[ \frac{1}{2}R^{-t}_{{\theta_1},{\gamma_1}} R^{-t}_{{\theta_2},{\gamma_2}}R^{-t}_{{\theta_3},\gamma_3} \frac{1}{v_{\gamma_2, \gamma_3}}\left[\text{sgn}(z')+ \text{sgn}(v_{\gamma_2,\gamma_3})\right]  R^t_{{\gamma_1},{\mu_1}}R^t_{{\gamma_2},{\mu_2}}R^t_{{\gamma_3},{\mu_3}} M_{\mu_2, \mu_3}^\beta \partial_1 g^{(2),1}_{\beta,\mu_1} \left( z+ \frac{v_{\gamma_2, \gamma_1}}{v_{\gamma_2, \gamma_3}}z'\right) \Big]\,,
\end{aligned}
\end{equation}
\end{widetext}
where the symbol $\partial_1$ means that the derivative is performed on the spatial argument of the function, the symbol $M_{\theta_1,\theta_2}^{\theta_3} \sim a $ is defined in eq. \eqref{eq:M_tensor} and the two-point functions $g^{(2),1}_{\theta_1,\theta_2}(z)$ are given by their asymptotic value in time. Given that the latter decays rapidly in \(z\), the expression above shows that there exist distinct directions in the \((z,z')\) plane, e.g., along lines \(z' \simeq \frac{v_{\gamma_{2},\gamma_{3}}}{v_{\gamma_{1},\gamma_{2}}}\, z\) from the first term, along which long-range three-point correlations persist well beyond the thermalization time scales. Therefore, at times of order \(\xi^{3}/(V_{0} a)^{2}\), thermalization is incomplete (generalized), and the deviation from a standard thermal state can be diagnosed by higher-point functions.

\subsection{Thermalization dynamics in the limit $a\to0$.}
\label{sec: Landau_a=0}
In the limit \(a \to 0\), the generalized Landau equation, i.e., the contributions \eqref{eq:Idiagonal} and \eqref{eq:Ioffdiagonal}, vanishes, as it is proportional to \(a^{2}\). As noted above, the thermalization mechanism encoded by the generalized Landau equation is present only for perturbed interacting integrable models. In the case \(a=0\), thermalization is driven solely by the long-range interaction. A kinetic equation for weak long-range potentials in this regime was derived in \cite{Fouvry2019,Fouvry2020}. Using the same approximations introduced in Sec.~\ref{s:Landau}, the standard BBGKY equations governing thermalization, analogous to~\eqref{eq: evolution_for_g2(1)} and~\eqref{eq: evolution_for_g2(2)}, read:
\begin{multline}
\label{eq: evolution_for_g2(1)_a=0}
    \partial_t g^{(2),1}_{\theta_1, \theta_2}(x_1,x_2)+[v^{\rm br}(\theta_1)\partial_{x_1}  g^{(2),1}_{\theta_1, \theta_2}(x_1,x_2)]_{(1,2)}=\\=[V_0\varphi'(x_1-x_2) \partial_{\theta_1} \rho_{\theta_1}\rho_{\theta_2}]_{(1,2)}\,,
\end{multline}
\begin{equation}
\label{eq: evolution_for_g2(2)_a=0}
\begin{split}
    &\partial_t g^{(2),2}_{\theta_1, \theta_2}(x_1,x_2)+[v^{\rm br}(\theta_1)\partial_{x_1}  g^{(2),2}_{\theta_1, \theta_2}(x_1,x_2)]_{(1,2)}=\\
    &=\Big[\int\dd{\theta_3}\dd{x_3}V_0\varphi'(x_1-x_3)\partial_{\theta_1}g^{(3)}_{\theta_1,\theta_2,\theta_3}(x_1,x_2,x_3)\Big]_{(1,2)}\,.
\end{split}
\end{equation}
Similarly, the leading equation for the evolution of \(g^{(3)}\) contains a forcing term proportional to the potential. As a consequence, the four nested equations for \(\rho\), \(g^{(2),1}\), \(g^{(2),2}\), and \(g^{(3)}\), which must be integrated to derive the scattering integral, have the interaction potential on the right hand side, yielding a thermalization rate proportional to \((V_{0})^{4}/\xi^{4}\), as also demonstrated numerically in Fig.~\ref{fig:scaling_plot}.
In this case the picture is simpler, see Fig.~\ref{fig: Correlations dynamicscartoon}, the thermalization rate is set first by the formation of two-point functions, namely \((V_{0}/\xi)^{2}\), and then by the formation of three-point functions, necessary for thermalisation, giving this way another factor of \((V_{0}/\xi)^{2}\) and contributing to the final thermalization rate \((V_{0}/\xi)^{4}\).

\section{Hard spheres with long-range interactions} \label{s:hard-rods}

In order to numerically verify the gBBGKY hierarchy of equations, we apply it to a gas of classical impenetrable hard spheres (Tonks gas) with diameter $a$ \cite{Spohn1991,Flicker68,Robledo1986,Doyon2017HR}, interacting with the long-range potential of Eq. \eqref{eq:potential}. In particular, we consider the potential 
\begin{equation}\label{eq:potentialHR}
    V(z) = \frac{V_0}{\xi} \frac{1}{1+|z/\xi|^3}.
\end{equation}
which mimics the dipolar interactions emerging in one-dimensional cold atoms, as we shall see in the next section, as well as standard Lennard-Jones interacting potentials \cite{Lepri2005,Mareschal1988,Eltohfa2024,Parameshwaran2025,Benettin2023} (especially in the case of $V_0 < 0$). 
Hard spheres (or rods) gas in one dimension is integrable in the absence of other interactions, namely in the limit $V_0=0$, with a well-known hydrodynamic description ~\cite{SpohnBook,Boldrighini1983,Boldrighini1997,Doyon2017HR} which share the same structure as quantum integrable systems, modulus changing the statistical factor~\cite{Doyon2020}, related to quasiparticles' statistics $f=1$, and constant scattering shift $T_{\theta,\theta'}=-a/(2 \pi)$. We focus on a homogeneous system with constant density of particles, and we compare gBBGKY prediction with exact microscopic simulations.  

%We initialize the time-evolution from two boosted finite temperature states $ \rho_{\pm; \beta; \theta_0} \equiv \sum_{\sigma = \pm } e^{-\beta (\theta- \sigma \theta_0)^2} /Z$, and with $g^{(2)}_{\theta_1,\theta_2}(t=0)=0$. 
{We initialize the time-evolution from a finite temperature states in which half of the particles is boosted with velocity $\theta_0$ and half with velocity $-\theta_0$, $\rho_{\beta; \theta_0} \equiv \sum_{\sigma = \pm } e^{-\beta (\theta- \sigma \theta_0)^2} /Z$, and with $g^{(2)}_{\theta_1,\theta_2}(t=0)=0$. }
The state is stationary under the unperturbed Hamiltonian $\hat{H}^{\rm int}_a$ but it evolves non-trivially as long as $V_0$ is finite. As a probe of thermalisation dynamics, we compute the kurtosis of the (normalised) quasiparticle distribution and the two-point correlations. The kurtosis is expected to be zero in the thermal state, where quasiparticle distribution is Gaussian.\\
Solving the set of gBBGKY equations \eqref{eq:BBGKY_quasi_rho}, \eqref{eq:BBGKY_quasi_g2} and \eqref{eq:BBGKY_quasi_g3} is numerically too expensive in terms of computational time and memory usage. In fact, it requires to save and perform operations on the three-point function $g^{(3)}_{\theta_1,\theta_2,\theta_3}(x_1-x_2,x_2-x_3)$, being, even in a homogeneous system, a five order tensor. In order to reduce the complexity of the numerical computation we separately solve the dynamics at small and large times. For small times $t<t_{L}$, we solve the gBBGKY hierarchy \eqref{eq:BBGKY_quasi_rho} and \eqref{eq:BBGKY_quasi_g2} by replacing the contribution proportional to $H$ in~\eqref{eq:BBGKY_quasi_g2} with the known Kubo diffusion term (explicitly known for integrable systems), as given by Eq.~\eqref{eq: small_time_diffusion}, and neglecting the remaining contribution from $g^{(3)}$.
For large times $t\geq t_{L}$, we evolve the system by solving the Landau equation~\eqref{eq: Landau Equation}, using as initial state the gBBGKY solution obtained at time $t=t_{L}$.
See Appendix \ref{sec: additional_details_numerical_solution} for additional informations about the numerical solution of gBBGKY hierarchy.\\
In Fig. \ref{fig: Diffusive correction} we show how gBBGKY reproduces the full molecular dynamics, predicting the evolution of Kurtosis as well as the evolution of two-point function. We can observe two regimes in the quasiparticle evolution: a fast evolution at short times $t<t_L$, driven by the change in the interaction energy due to the build-up of the density-density correlations in the system and the \textit{kinetic regime} $t>t_L$, where the system slowly evolves toward the thermal state. In this regime, the kinetic energy is mostly constant and the dynamics is well described by the Landau equation, which well describes the (generalized) thermalisation with rate $ (aV_0 )^2/\xi^2$ , see Fig. \ref{fig:scaling_plot}, as predicted in the previous section. 

We can observe the striking difference between the dynamics of point particles $a=0$ and the ones with finite $a$ already from a single trajectory snapshot, see Fig. \ref{fig:trajHR}: the dynamics with $a=0$ and $V_0$ finite is given by smoothly  long-range interacting particles, while the one with finite $a$ is given both by long-range interactions and local scatterings, even at long times, where the thermalisation mechanisms start to kick in. 
{The difference between these two cases is also highlighted in Fig. \ref{fig: Diffusive correction}(a), where we show how the solution gBBGKY with $a=0$ (dashed lines), i.e. BBGKY, is far from the hard spheres molecular dynamics from any arbitrary macroscopic small time. Interestingly, the short-time dynamics with $a=0$ appears to be faster, compared to the interacting case with $a=1$, namely in the presence of contact interaction the system appears to be jammed. However the scaling of the thermalisation, long-times, rates in the two cases shows the opposite behavior: $V_0^4/\xi^4$ in the free case and} $(aV_0)^2/\xi^3$, resulting in a much faster (generalized) thermalisation in the interacting case $a >0$.} 

\begin{figure}
    \centering
    \includegraphics[width=1.0\linewidth]{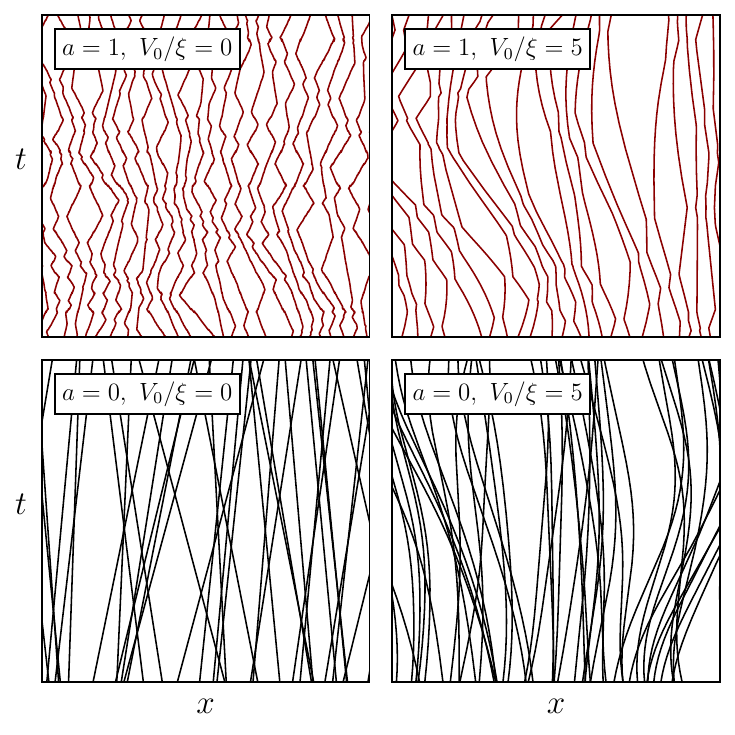}
    \caption{Examples of microscopic trajectories of hard spheres  with long-range interaction given by the potential \eqref{eq:potentialHR}, both with $a=0$ (free particles) and finite $a=1$. In the latter case we can notice in the trajectories with finite $V_0$ the interplay of local scatterings and long-range interaction.  }
    \label{fig:trajHR}
\end{figure}

\begin{figure*}[t]
\includegraphics[width=\linewidth]{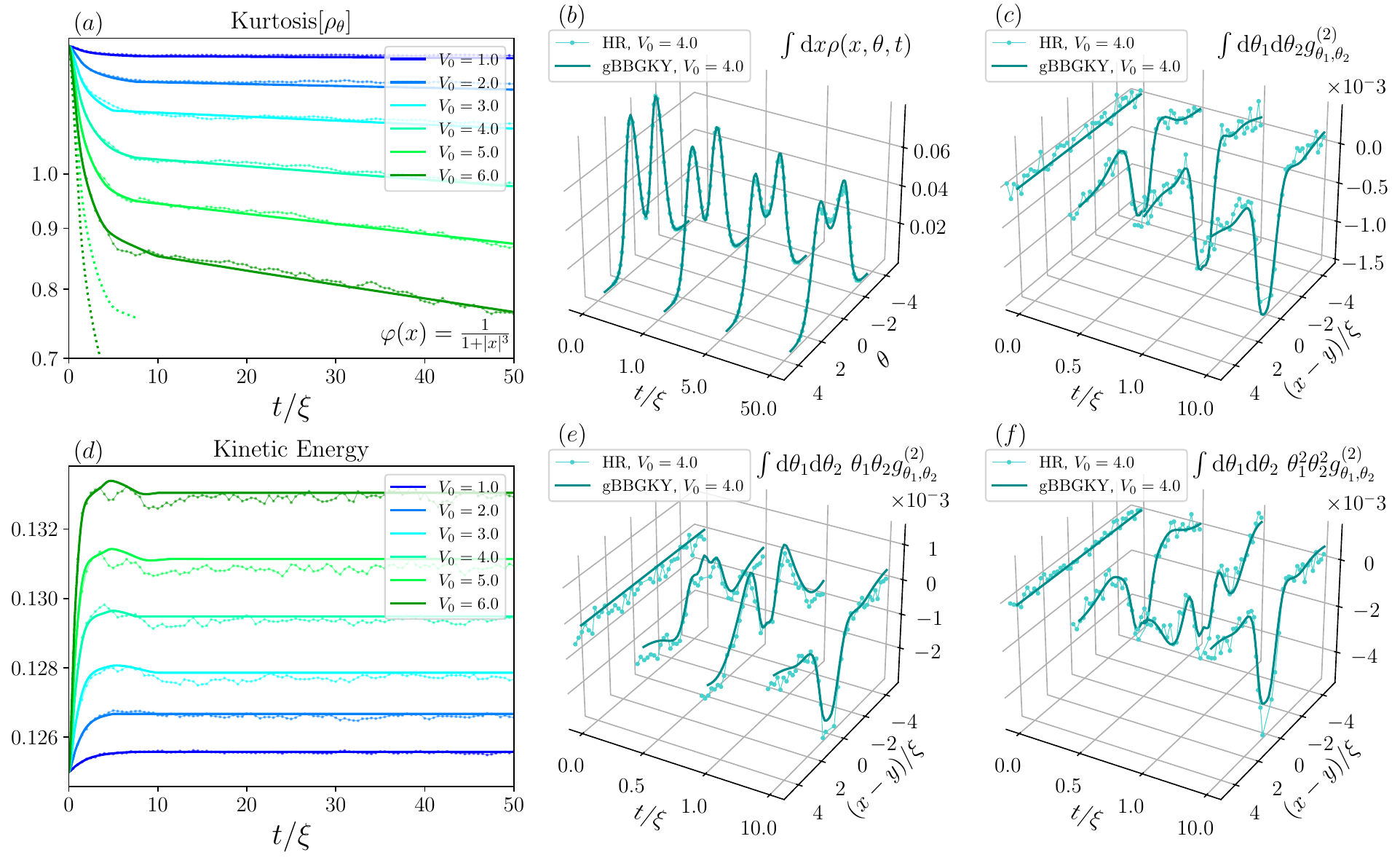}
\caption{
Numerical results for long-range interacting hard spheres  (HR) with $a=1$, $\xi=20$ and $\varphi=1/(1+|x|^3)$, starting from the two boosted thermal states $\rho_{\beta; \theta_0} $ with $\beta = 2.5$, $\theta_0=1$, $\bar{\rho}=0.2$. The system is homogeneous on a ring. In all the figures, the dots represent the molecular dynamics of the system, while the solid line represent the gBBGKY prediction. 
Figure $(a)$ shows the dynamics of kurtosis of rapidity distribution, for $\xi=20$ and with $V_0=\{1,2,\ldots,6\}$, as function of the rescaled time $t/\xi$. The numerical solution of gBBGKY hierarchy is detailed in Appendix \ref{sec: additional_details_numerical_solution}. In particular, for $t<t_L$ the solid line represents the solution of the two coupled equations ~\eqref{eq:BBGKY_quasi_rho_numerical} and~\eqref{eq:BBGKY_quasi_g2_numerical}, while for $t>t_L$ it shows the solution of generalised Landau equation~\eqref{eq: Landau Equation}. We observe a convincing agreement between molecular dynamics and theoretical predictions for all the values of integrability breaking potential $V_0$.
The additional dashed lines represent solutions of the standard BBGKY hierarchy, i.e. gBBGKY with $a=0$, truncated to two first levels.  As expected, this purely perturbative theory doesn't predict the correct system's evolution. 
Similarly, Fig. $(d)$ shows the dynamics of Kinetic Energy $\int\dd{\theta} \rho_{\theta} \theta^2/2$.
Fig. $(b)$ shows the evolutions of distribution of rapidity $\rho_{\theta}$ for different times. Fig. $(c,e,f)$ represent the evolution two-point correlation function $\int\dd{\theta_1}\dd{\theta_2}\theta_1^{\alpha}\theta_2^{\alpha}g^{(2)}_{\theta_1,\theta_2}$ respectively with $\alpha=\{0,1,2\}$. As expected, the gBBGKY is not only able to predict the evolution of both one and two-point functions.  
}
\label{fig: Diffusive correction}
\end{figure*}

\begin{figure}
    \centering
    \includegraphics[width=1.0\linewidth]{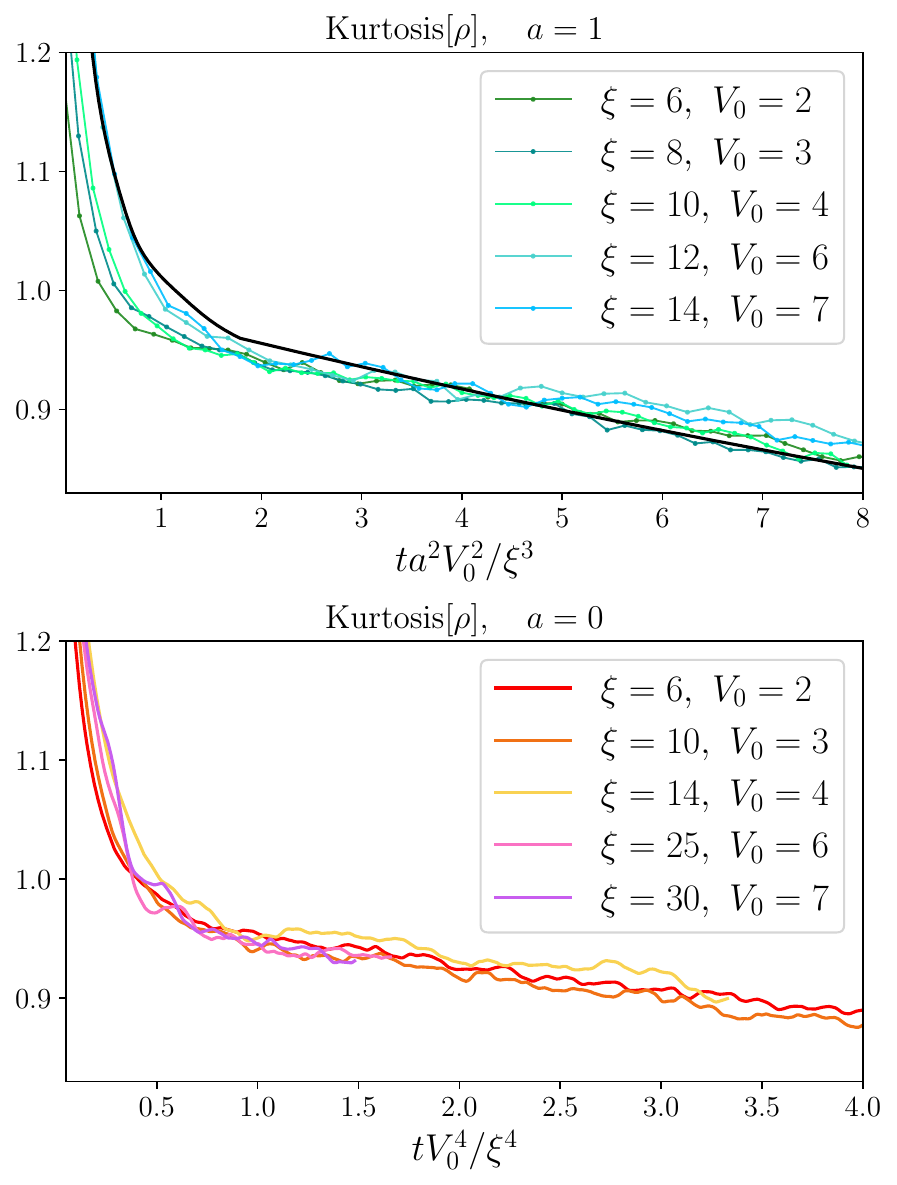}
    \caption{{Numerical confirmation of the predicted thermalization rate $\propto a^2V_0^2/\xi^3$, when $a \neq 0$ and $\propto V_0^4/\xi^4$, when $a=0$. 
    The two plots show the kurtosis of $\rho_{\theta}$ for different values of $V_0$, $\xi$, in the case $a=1$ (top) and $a=0$ (bottom), respectively as
    function of the rescaled times $\bar{t}=ta^2V_0^2/\xi^3$ and $\bar{t}=tV_0^4/\xi^4$. The dynamics firstly undergoes a fast small time evolution and then, for $\bar{t}>1$, the late time thermalization dynamics. We show that, under this appropriate time scaling, at late time the dynamics induced by different potentials parameters perfectly collapse, confirming that the thermalization rate is $\propto a^2V_0^2/\xi^3$ for $a \neq 0$ and $\propto V_0^4/\xi^4$ for $a=0$. More precisely, in this figure we introduce a constant offset parameter that takes into account the differences in the small time behavior, in order to renormalize the starting point for the late time dynamics. The black curve represents the solution of gBBGKY for $\xi=14$, $V_0=7$, being in good agreement with the numerical simulation.
    }}
    \label{fig:scaling_plot}
\end{figure}

\section{Application to dipolar one-dimensional quantum gases} \label{s:dipolar}

Integrability breaking due to additional long-range interactions arises naturally in quantum gases of atoms or molecules with large magnetic moments~\cite{Chomaz2023}. In these systems, the basic Hamiltonian is given by Eq.~\eqref{eq:lll0}. When confined to one dimension, the van der Waals interaction can be modeled effectively by a contact potential~\cite{Olshanii1998}, and the gas is described by the integrable Lieb--Liniger (LL) Hamiltonian~\cite{Lieb1963,Lieb1963a}:
\begin{equation}
  H_{\rm LL}
  = -\frac{\hbar^2}{2m}\sum_i \partial_{x_i}^2
    + c \sum_{i<j}\delta(x_i - x_j)\,,
\end{equation}
where \(m\) is the mass of the bosons and \(c\) the interaction strength.

If the atoms carry a nonzero magnetic moment, they also interact via a dipolar potential, which decays more slowly in space and must be treated separately. In the geometry considered by us (see Fig.~\ref{fig:thermalizationrate}), dipolar interactions couple both the atoms inside a tube (intratube interactions) as well as atoms in neighbouring tubes (intertube interactions). We discuss first the intratube potential. Upon integrating out the two “frozen” transverse degrees of freedom, the effective one-dimensional dipolar coupling becomes~\cite{Deuretzbacher2010,Deuretzbacher2013}
\begin{equation}
  V_{\rm DDI}^{\rm 1D}(z)
  = V(\theta_{\rm DDI}) \Bigl[v_{\rm DDI}^{\rm 1D}(u) - \tfrac{8}{3}\,\delta(u)\Bigr]\,,
  \quad u = \frac{z}{l_\perp}\,,
\end{equation}
where \(l_\perp = \sqrt{\hbar/(m\omega_\perp)}\) is the transverse oscillator length. The prefactor is
\begin{equation}
  V(\theta_{\rm DDI})
  = \frac{\mu_0 \mu^2}{4\pi}
    \frac{1 - 3\cos^2\!\theta_{\rm DDI}}{4\,l_\perp^3}\,,
\end{equation}
and the dimensionless shape function is
\begin{equation}
  v_{\rm DDI}^{\rm 1D}(u)
  = -2|u| + \sqrt{2\pi}\,(1+u^2)\,e^{u^2/2}\,\mathrm{erfc}\bigl(|u|/\sqrt{2}\bigr)\,.
\end{equation}
Here \(\mu\) is the dipole moment, and \(\theta_{\rm DDI}\) is the angle between the dipoles (aligned by an external field) and the tube axis; by tuning \(\theta_{\rm DDI}\), one effectively controls the strength of the dipolar interaction.

Following Ref.~\cite{PhysRevX.8.021030}, we consider $^{162}\mathrm{Dy}$ atoms ($\mu=9.93\mu_B$) loaded into a two-dimensional lattice of one-dimensional tubes (lattice spacing $a_l = 370\,\mathrm{nm}$), each with average density $\bar\rho = 0.8\,\mu\mathrm{m}^{-1}$ (see sketch in Fig.~\ref{fig:thermalizationrate}). Transversal confinement is provided by harmonic trap with oscillator length $l_\perp=57 \, \rm{nm}$. In addition to intratube dipolar interactions, atoms in different tubes $(i,j)$ interact with ones in the tube (0,0) via
\begin{equation}
  V_{\rm inter}^{i,j}(x)
  = \frac{\mu_0 \mu^2}{4\pi}
    \frac{1 - 3(\hat r\cdot\hat B)^2}{r^3}\,,
\end{equation}
where $\vec r = (x,\,i\,a_l,\,j\,a_l)$ and $\hat B = (\cos\theta_{\rm DDI},\,0,\,\sin\theta_{\rm DDI})$ denotes the magnetic field direction.

At $t=0$, a Bragg pulse splits the atoms into two momentum components $\pm 2\hbar k_D$, with $k_D = 2\pi/\lambda$ and $\lambda=741\,\mathrm{nm}$. The gas then prethermalizes to a state described by a reference density $\rho_{\rm Init}(\theta)$ (Fig.~\ref{fig:thermalizationrate}), which closely matches the experimental observations at the end of the first stage of the dynamics.

Although the dipolar potential decays as 
%$V_{\rm DDI}^{\rm 1D}(x)\sim(x/\xi)^{-3}$
$V_{\rm DDI}^{\rm 1D}(x) \sim x^{-3}$ at large distances, it also contains a short-range component. To separate the integrable hard-core part from the genuine long-range contribution, we note that the LL coupling $c$ has two parts: the van der Waals term
\begin{equation}
  c_{\rm vdW} = -\frac{2\hbar^2}{m\,a_{\rm 1D}}\,,\quad
  a_{\rm 1D} = -435\,\mathrm{nm}\,,
\end{equation}
and a dipolar term
\begin{equation}
  c_{\rm DDI}(\theta_{\rm DDI})
  = \frac{4\,V(\theta_{\rm DDI})}{3}\,.
\end{equation}
Hence $c(\theta_{\rm DDI})=c_{\rm vdW}+c_{\rm DDI}(\theta_{\rm DDI})$. %\ml{The dipolar contribution to $c$ comes from treating the entire intratube interaction as a contact coupling $c_{\rm DDI}(\theta_{\rm DDI} )=V(\theta_{\rm DDI}) \int {\rm d}u v^{1D}_{\rm DDI}(u)$, we comment on that later in the text.} 

To estimate the effect of the long-range potential we invoke the characteristic distance $\xi$. In the LL model the scattering shift defines an effective hard-sphere diameter for the quasiparticle interaction \cite{schemmer2019generalized,Urilyon2025}
\begin{equation}
  \mathfrak{a}(\theta-\theta')
  = \frac{2(c/m)}{(c/\hbar)^2+(\theta-\theta')^2}
\end{equation}
Estimating $\theta-\theta'\approx\hbar k_D/m$ (reflecting the initial momentum spread) and comparing with the dipolar length scales yields
\begin{equation}
  \xi_{\rm intra} = \frac{l_\perp}{\mathfrak{a}} = 0.81\,,\quad
  \xi_{\rm inter} = \frac{a_l}{\mathfrak{a}} = 5.28\,.
\end{equation}
Thus, intratube dipolar interactions can be absorbed into the contact coupling, while intertube interactions act as a truly long-range perturbation. This is consistent with the modeling of recent experimental results in \cite{Li2023,Yang2024} where the authors show that intratube dipolar effects mostly renormalize the coupling $c$ at equilibrium \cite{Li2023} and during dynamics, slightly in contrast with the original interpretation in~\cite{PhysRevX.8.021030}. Moreover, we observe that summing the contributions to the Landau collision integral from tubes with indices $-2\leq i,j\leq2$ (we assume that each tube is in the same state) suffices to capture the relaxation rate, since more distant tubes have negligible effect.

To model the two-stage dynamics observed under the Bragg-split protocol (cf. the hard-rod gas), we adapt our theory to the experimental parameters. The key observable in~\cite{PhysRevX.8.021030} is the thermalization rate as a function of $\theta_{\rm DDI}$. 
{We consider a measure of distance to (generalized) thermalization  $\text{DT}(t)$ very similar to the one studied in~\cite{PhysRevX.8.021030}. We fit a Gaussian distribution $\tilde{\rho}_\theta(t)$ to the state $\rho_\theta(t)$ at each time $t$, then $\text{DT}(t)$ is computed as 
\begin{equation}
    \text{DT}(t) = \int \dd \theta(\rho_\theta(t)-\tilde{\rho}_\theta(t))^2,
\end{equation}
measuring the distance of the state at time $t$ from a Gaussian distribution.  After the first phase of dynamics, an exponential decay is observed, with 
\begin{equation}\label{eq:thermalization rate}
\log \text{DT}(t) \approx -t/t_{\rm therm}  +\ldots\,.
\end{equation}
We compute $t_{\rm therm}$ from the expression of the Landau collision integral, giving a very good agreement with the experimental rates, see Fig.~\ref{fig:thermalizationrate}, providing this way a first direct application of gBBGKY to an experimental setting. We should stress that in our approach, we track the Gaussianity of the rapidity distribution $\rho_\theta$, whereas in~\cite{PhysRevX.8.021030} the momentum distribution is measured. However, these two are expected to be close due to high temperature of the state. This is especially clear in the final state $\rho_{\rm th}$, see inset of Fig.~\ref{fig:thermalizationrate} which is very well approximated by a Gaussian, just like final momentum distributions measured in~\cite{PhysRevX.8.021030}.

What is more, our result can also be used to deduct important aspects of the thermalization rates. 
Indeed, a more conventional theoretical approach would linearize the collision integral around the thermal distribution $\rho_{\rm th}$ and study spectrum of the linearized collision operator. Equivalently, one can track perturbations in the conserved charges $\delta q_n$:
\begin{equation}
  \partial_t \langle\delta q_i\rangle = -\Gamma_i^{\,j}\,\langle\delta q_j\rangle\,,
\end{equation}
where matrix $\Gamma$ is symmetric, positive semidefinite, and has three zero modes (particle number, momentum, energy). See~\cite{SM} for the explicit expression and the derivation of matrix $\Gamma$. The first nonzero eigenvalue $\gamma_{\rm gap}$ sets $t_{\rm therm}=\gamma_{\rm gap}^{-1}$. However, this approach predicts thermalization times of order $10^3\,\mathrm{s}$—far longer than experimentally accessible—as it captures the slowest decaying mode, whereas the current-state of the art experiments can only probe the fastest relaxation channels.

\begin{figure}[h]
  \centering
  \includegraphics[width=0.9\linewidth]{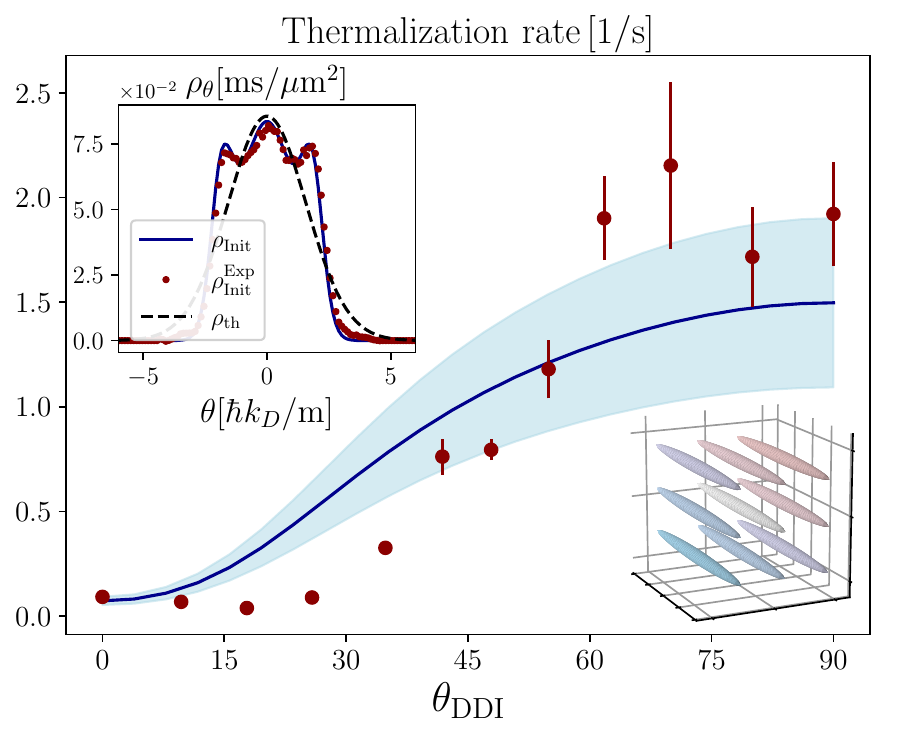}
  \caption{Thermalization rate (blue solid line) as defined in~\eqref{eq:thermalization rate} for the experimental parameters of~\cite{PhysRevX.8.021030}, obtained from gBBGKY, together with shaded area indicating uncertainty resulting from $13\%$ density variation in the experiment. The bottom right inset shows the geometry of the quasi-one-dimensional tubes loaded with dipolar atoms. The top left inset shows the initial state $\rho_{\rm Init}$ (solid line) compared with the experimentally measured initial state (dots)  (determined from the data presented in Figs.2(c), 2(d) and 2(e) of~\cite{PhysRevX.8.021030}) together with the final, thermal state $\rho_{\rm th}$ (dashed line). }
  \label{fig:thermalizationrate}
\end{figure}

\subsection{Fermi's Golden rule approach and check of the generalised Landau equation }
The advantage of restricting to this case of perturbed Lieb-Liniger gas also offer us the possibility of giving a non-trivial check of the generalised Landau equation \eqref{eq: Landau Equation} and therefore or our gBBGKY via a different approach. Indeed, the dynamics of a quantum nearly integrable systems can be also addressed with the Fermi's Golden Rule (FGR), which can be used to write the Boltzmann equation~\cite{Mallayya2021,Durnin2021,Panfil2023}. In this section, we give a short review of these results and compare them to generalized Landau equation obtained from our hierarchy.

The FGR approach relies heavily on the knowledge of form factors, which are the matrix elements of the perturbing operator computed for integrable theory. In the case of integrability breaking perturbation in the form of additional long-range interactions, the relevant form factors are one- and two-point functions of the density operator. In general, these objects are not known, however in the limit of small excitations between the states, there are exact results in the LL model
~\cite{FFDeNardis2015,FFDeNardis2016,FFDeNardis2018,Cubero2019,Panfil2021}. As we will see, the small excitation limit will be automatically guaranteed by the long-range limit of the integrability-breaking potential, rendering FGR Boltzmann kernel computable in the regime of interest of the present work.

Once again we assume that the Hamiltonian is $\hat{H}=\hat{H}_{\rm LL}+\hat{V}$ where $\hat{V}=\int {\rm d}x \hat{v}(x)$ is the perturbation:
%While $\hat{H}_0$ conserves an infinite number of charges, $\hat{H}$ commutes only with particle number $\hat{N}$, momentum $\hat{P}$ and energy $\hat{H}$.
\begin{equation}
    \hat{v}(x) = \frac{1}{2}\int {\rm d}yV(x-y) \hat{q}_0(x) \hat{q}_0(y).
\end{equation}
The LL model has fermionic statistics and we define density of holes $\rho^{\rm h} = \rho^{\rm t} - \rho$. In homogeneous system, the dynamics is given by Boltzmann-like equation~\cite{Durnin2021,Panfil2023}
\begin{equation}\label{eq:Istructure1}
\begin{aligned}
    &\partial_t \rho_\theta = \mathcal{I}[\rho](\theta)=\\
    &\int {\rm d} \mathbf{p} {\rm d} \mathbf{h} \, \mathbb{F}_{\theta, \mathbf{p}} B( \mathbf{p} \to \mathbf{h})[\rho^{\rm h}_{\mathbf{p}} \rho_{ \mathbf{h}}-\rho^{\rm h}_{ \mathbf{h}}\rho_{ \mathbf{p})}]\,,
\end{aligned}
\end{equation}
% where the collision term has the following structure
% \begin{equation}\label{eq:Istructure}
%     \mathcal{I}[\rho_{\rm p}](\theta)= \int {\rm d} \mathbf{p} {\rm d} \mathbf{h} \, \mathbb{F}(\theta, \mathbf{p}) B( \mathbf{p} \to \mathbf{h})[\rho_{\rm h}(\mathbf{p}) \rho_{\rm p} (\mathbf{h})-\rho_{\rm h}( \mathbf{h})\rho_{\rm p}( \mathbf{p})],
% \end{equation}
with $\rho^{\rm h}(\mathbf{p}) = \prod_{p \in \mathbf{p}} \rho^{\rm h}_p, \, \rho(\mathbf{h}) = \prod_{h \in \mathbf{h}} \rho_h $ and the measure involving different particle-hole (particle-hole) scattering processes reads
\begin{equation}
    {\rm d} \mathbf{p} {\rm d} \mathbf{h}= \sum_{N=0}^\infty \frac{1}{(N!)^2} \prod_{i=1}^{N} {\rm d} p_i {\rm d} h_i.
\end{equation}
% \begin{equation}
%     {\rm d} \mathbf{p} {\rm d} \mathbf{h}= \sum_{N=0}^\infty \frac{1}{(N!)^2} \prod_{i=1}^{N} {\rm d} p_i {\rm d} h_i.
%    %{\rm d} \mathbf{p} {\rm d} \mathbf{h}= \sum_{N=0}^\infty \frac{1}{N!} \sum_{N_p=0}^N \prod_{i=1}^{N_p} {\rm d} p_i \prod_{j=1}^{N-N_p}{\rm d} h_j.
% \end{equation}
Apart from the density factors there are two ingredients in \eqref{eq:Istructure1}. Firstly, function $B(\mathbf{p} \to \mathbf{h})$ is related to particle-hole form-factors of the perturbing operator $\langle \rho | v | \rho; \mathbf{p}, \mathbf{h} \rangle$ \footnote{Note the additional missing factor of $2 \pi$ with respect to~\cite{Durnin2021}.}
\begin{equation}
    B( \mathbf{p} \to \mathbf{h})=(2 \pi)^2 \delta(k) \delta (\varepsilon) | \langle \rho | v | \rho; \mathbf{p}, \mathbf{h} \rangle |^2.
\end{equation}
Assuming particle-hole symmetry we have $B( \mathbf{p} \to \mathbf{h}) = B( \mathbf{h} \to \mathbf{p})$. The Dressed energy and momentum are 
\begin{equation}
\begin{aligned}
    &k_\theta = \theta - \int {\rm d} \mu\, n_\mu F_{\mu,\theta}, \\
    &\varepsilon_\theta = \frac{\theta^2}{2}  - \int {\rm d} \mu\,
    \mu \ n_\mu F_{\mu,\theta},
\end{aligned}
\end{equation}
where $F_{\mu,\theta}$ is the backflow function~\cite{Korepin1993} of the underlying integrable model. Explicitly, the terms inside Dirac delta functions are then
    $k = \sum_{p \in \mathbf{p}} k_p - \sum_{h \in \mathbf{h}} k_h$, $ \varepsilon = \sum_{p\in \mathbf{p}} \varepsilon_p - \sum_{h \in \mathbf{h}} \varepsilon_h$.
For later use, we note that $k'_{\lambda}=2 \pi \rho^{\rm t}_\lambda$ and $v_\lambda=\varepsilon'_\lambda/k'_\lambda$.
The second ingredient in \eqref{eq:Istructure1} is the backflow effect implemented by the term $\mathbb{F}_{\theta, \mathbf{p}}$ which reads
\begin{equation}
    \mathbb{F}_{\theta, \mathbf{p}} = \sum_{p \in \mathbf{p}} \mathbb{F}_{\theta, p}, \, \, \mathbb{F}_{\theta, p}= \delta_{\theta,p} + \partial_{\theta} \left( n_\theta F_{\theta, p} \right).
\end{equation}
The main practical difficulty for evaluating the collision term is the resulting infinite sum over different numbers of particle-hole excitations. Indeed, writing the sum in integration measure explicitly, we have
\begin{equation}
    \mathcal{I}[\rho](\theta) =\mathbb{F}_{\theta, \nu}\sum_{m=1}^\infty \mathcal{I}^{(m)}_0[\rho](\nu),
\end{equation}
where we have factored out the dressing operator $\mathbb{F}$ defining bare collision integrals $\mathcal{I}_0^{(m)}$. Due to this structure, an additional small parameter is needed to effectively truncate the sum. Such situation will happen in our case of long-range interactions, where the potential allows only for small momentum transfer, which is equivalent to considering $\xi$ as a large parameter. This allows us to truncate the sum to the leading term, which corresponds to 3-particle-hole ($m=3$) excitations. The $m=2$ term is zero, due to the kinetic blocking effect described in Sec.~\ref{sec:kineticblocking}.

From now on, we will assume that all transferred momenta $p_i-h_i$  are small (this approximation is discussed in the Supplementary Materials~\cite{SM}). We therefore introduce new variables $\lambda_i=(p_i+h_i)/2$ and $q_i=k(p_i)-k(h_i)$ and obtain $m=3$ collision integral in the form
\begin{equation}\label{eq:Istructure_smallmom}
\begin{aligned}
    \mathcal{I}^{(3)}_0[\rho](\theta) &= \frac{1}{3!} \frac{1}{(2\pi)^3} \partial_\theta\int {\rm d} \boldsymbol{\lambda} {\rm d}\mathbf{q}\,  B(\boldsymbol{\lambda}, \mathbf{q}) \times \\
    &\left(\sum_{i=1}^3 \frac{q_i}{2k'_{\lambda_i}} \delta(\theta - \lambda_i)\right) \left(\sum_{i=1}^3 q_i \frac{\varepsilon'_{\lambda_i}}{k'_{\lambda_i}} \right),
\end{aligned}
\end{equation}
with  integration measure ${\rm d} \boldsymbol{\lambda} {\rm d}\mathbf{q} = \prod_{i=1}^3 {\rm d}\lambda_i {\rm d}q_i\, \rho_{\lambda_i} f_{\lambda_i}$ where for fermionic statistics $f=1-n$. Moreover, in that limit the back-flow operator becomes $ \mathbb{F}_{\theta, p}  = \frac{q}{2k'_{\lambda}} \partial_\theta R_{\theta, \lambda}^{-1}$ and implements the usual dressing procedure. To move forward, we have to analyze now $m=3$ form factors of the density-density operator. Consider form-factor of bi-local operator $\langle \rho| \hat{q}_0(x) \hat{q}_0(0)| \rho ; \mathbf{p}, \mathbf{h}\rangle$. 
We plug the resolution of the identity between the two operators to obtain
\begin{equation}
\begin{aligned}
    \langle \rho| \hat{q}_0(x)& \hat{q}_0(0)| \rho_; \mathbf{p}, \mathbf{h}\rangle =\\
    &\sum_{\boldsymbol{\alpha}} \langle \rho| \hat{q}_0(x) | \rho; \boldsymbol{\alpha}\rangle\langle \rho; \boldsymbol{\alpha} | \hat{q}_0(0)| \rho ; \mathbf{p}, \mathbf{h} \rangle.
\end{aligned}
\end{equation}
The leading contribution to the composite form-factor comes then from bipartite partitions of the 3-particle-hole excitation $(\mathbf{p}, \mathbf{h})$. More explicitly, for $(\mathbf{p}, \mathbf{h}) = (p_1, p_2, p_3, h_1, h_2, h_3)$ there are $6$ relevant partitions. Three of them are 1-particle-hole excitations: $\boldsymbol{\alpha} = (p_1, h_1)$, $\boldsymbol{\alpha} = (p_2, h_2)$ and $\boldsymbol{\alpha} = (p_3, h_3)$.
% \begin{equation}
%     \boldsymbol{\alpha}_+ = (p_1, h_1), \qquad \boldsymbol{\alpha}_+ = (p_2, h_2), \qquad \boldsymbol{\alpha}_+ = (p_3, h_3),
% \end{equation}
The other three are 2-particle-hole excitations $\boldsymbol{\alpha} = (p_1, p_2, h_1, h_2)$, $\boldsymbol{\alpha} = (p_2, p_3, h_2, h_3)$ and $\boldsymbol{\alpha} = (p_1, p_3, h_1, h_3)$.
% \begin{equation}
% \begin{aligned}
%     &\boldsymbol{\alpha}_+ = (p_1, p_2, h_1, h_2), \qquad \boldsymbol{\alpha}_+ = (p_2, p_3, h_2, h_3), \\
%     &\boldsymbol{\alpha}_+ = (p_1, p_3, h_1, h_3).
% \end{aligned}
% \end{equation}
Using reparametrization invariance of the thermodynamic form-factors~\cite{crossing_symmetry_JHEP}, a property that an excitation common to the bra and ket can be absorbed into the thermodynamic state $\rho$, the contributions from 1-particle-hole and 2-particle-hole excitations are pairwise identical up to the $x$-dependent term.
% \begin{equation}
%         \langle \mathbf{p}, \mathbf{h}; \rho| \hat{V}|\rho \rangle = L \left(\tilde{V}(k(p_1, p_2, h_1, h_2)) + \tilde{V}(k(p_3, h_3)) \right)\langle p_1, p_2, h_1, h_2; \rho| \hat{\rho}(0) |\rho\rangle \langle p_3, h_3, \rho |\hat{\rho}(0)|\rho \rangle + {\rm cycl}(1,2,3).
% \end{equation}
% and abbreviating the momenta by $q_i = k(p_i, h_i)$, we get
% \begin{align}
%         \langle \mathbf{p}, \mathbf{h}; \rho| \hat{V}|\rho \rangle = \left(\tilde{V}(q_1 + q_2) + \tilde{V}(q_3)\right) \langle p_1, p_2, h_1, h_2; \rho| \hat{\rho}(0) |\rho\rangle \langle p_3, h_3, \rho |\hat{\rho}(0)|\rho \rangle + {\rm cycl}(1,2,3),
% \end{align}
The expressions for the form-factors (assuming small momentum transfer) are~\cite{FFDeNardis2015,FFDeNardis2018}
\begin{align}
    \langle \rho| \hat{q}_0(0) |\rho; p_1,h_1\rangle &= k'_{\lambda_1}, \\
    \langle \rho| \hat{q}_0(0) |\rho; p_1, p_2, h_1, h_2\rangle &= 2\pi T^{\rm dr}_{\lambda_1, \lambda_2} \frac{(q_1 + q_2)^2}{q_1 q_2}\,,
\end{align}
and lead to six relevant contributions that can be written as follows
% \begin{equation}
% \begin{aligned}
%     &\langle \mathbf{p}, \mathbf{h}; \rho| \hat{V}|\rho \rangle = 2\pi \left( \tilde{V}(q_1 + q_2) + \tilde{V}(q_3) \right) \times\\
%     &T_{12}^{\rm dr} k_3' \frac{(q_1 + q_2)^2}{q_1 q_2}+ {\rm cycl}(1,2,3).
% \end{aligned}
% \end{equation}
\begin{equation}
\begin{aligned}
    &\langle \rho| \hat{v}|\rho; \mathbf{p}, \mathbf{h}
     \rangle = (2\pi)^2  \left(\hat{V}(q_3) T_{\lambda_1 \lambda_2}^{\rm dr} k_{\lambda_3}' \frac{(q_1 + q_2)^2}{q_1 q_2}\right)_{(1,2,3)}\,.
\end{aligned}
\end{equation}
The factor $B$ appearing in the collision integral is then
% \begin{equation}
% \begin{aligned}
%     &B(\boldsymbol{\lambda}, \mathbf{q}) = 4 (2\pi)^3 \delta(k) \delta(\varepsilon) \times\\
%     &\left( T_{12}^{\rm dr} k_3' \tilde{V}(q_3) \frac{q_3^2}{q_1 q_2}+ {\rm cycl}(1,2,3) \right)^2,
% \end{aligned}
% \end{equation}
\begin{equation}
\begin{aligned}
    B(\boldsymbol{\lambda}, \mathbf{q}) =  (2\pi)^4 \delta(q) \delta(vq)\left[\left( T_{\lambda_1 \lambda_2}^{\rm dr} k_{\lambda_3}' \hat{V}(q_3) \frac{q_3^2}{q_1 q_2}\right)_{(1,2,3)}\right]^2,
\end{aligned}
\end{equation}
where we defined $\delta(q)=\delta(q_1+q_2+q_3)$, $\delta(vq)=\delta(v_1q_1+v_2q_2+v_3q_3)$ and  used that $\hat{V}(k) = \hat{V}(-k)$. 
Crucially, expanding the square we find two types of contributions, which correspond one-to-one to the two contributions in eq. \eqref{eq: Landau Equation}. The cross ones are proportional to $(\hat{V}(q_i))^2$, whereas the self terms involve $\hat{V}(q_i)\hat{V}(q_{j \neq i})$ factors. The collision integral has consequently two contributions, which turn out to be exactly equal to the two contributions of eq. \eqref{eq:Idiagonal} and \eqref{eq:Ioffdiagonal}, therefore giving an independent, and mathematically highly non-trivial, verification of the gBBGKY approach.

 \section{Conclusions }\label{s:conclusions}

In this work, we have extended the celebrated BBGKY hierarchy to systems with integrable contact interactions. We derived the hierarchy from a \emph{correlated fluid‐cell ensemble}, which enables accurate characterization of multi‐point correlations both in the standard BBGKY framework (where the unperturbed Hamiltonian is free) and in more general settings.

For classical hard spheres with long‐range interactions, we tested our theory against molecular‐dynamics simulations and observed excellent agreement, even at relatively strong coupling. For quantum gases described by the Lieb–Liniger model with long‐range interactions, we showed that the generalized Landau equation aligns with the Fermi golden rule approach, providing a rigorous validation of both our theory and the thermalization hypothesis. Finally, we established a theoretical foundation for the experimental results of Ref.~\cite{PhysRevX.8.021030}, demonstrating that dynamics in these quasi‐one‐dimensional systems are encompassed by gBBGKY hierarchy. The latter, in the presence of strong contact interactions, accurately captures short‐time dynamics during the \emph{prethermal} regime, driven by the rapid build-up of correlations that our formalism tracks precisely. Conventional kinetic equations fail to describe this transient phase, which is often the only accessible regime in experiments and simulations. Thus, the gBBGKY provides both a conceptual framework and quantitative predictions for prethermal behavior.

At late times, gBBGKY overcomes the well-known kinetic blocking in one dimension. For perturbed free systems, thermalization is driven by three-body scatterings generated by the external potential \(V\), yielding a rate that scales as \(V_0^{4}\). By contrast, with contact interactions, the combined effect of long-range interactions and convective diffusion among particles produces a thermalization rate of order \((V a)^2\). Importantly, such thermalization is not complete (\textit{generalized thermalisation}) as only one- and two-point functions reach their canonical thermal values. Higher-order correlations remain far from thermal over the same window of times and retain their integrable structure, leading to finite long-range correlations that would be absent in a thermal Gibbs ensemble but which are instead correctly captured by the correlated fluid-cell ensemble.

Our method is broadly applicable: its central ingredient is a thermodynamic description of the unperturbed integrable model, yielding a universal framework for diverse classical and quantum systems. Unlike approaches based on Fermi’s golden rule and explicit matrix elements, the gBBGKY hierarchy enables direct access to the Boltzmann scattering integrals and the time evolution of multi‐point correlations in a wide range of perturbed integrable models. It can be applied to lattice systems—such as long‐range spin chains and fermionic models~\cite{RevModPhys.93.025001,PhysRevB.95.245111,Jurcevic2014,PhysRevA.99.032114,PhysRevB.106.014306}—inspired by recent experiments. Additionally, it opens a path to connect kinetic theories of interacting waves (via wave turbulence)\cite{nazarenko2011wave,2412.14153} with general non‐linear wave systems \cite{PhysRevLett.134.147201,PhysRevLett.134.193804,2503.13030}. 
%We anticipate that the gBBGKY hierarchy will become a cornerstone for future studies of thermalization and prethermalization across a wide range of physical settings, from low‐dimensional cold‐atom experiments to complex condensed‐matter and optical systems with tunable interactions.
We expect that the gBBGKY hierarchy will be an important tool in further studies of thermalization and prethermalization across a wide range of physical settings, from low‐dimensional cold‐atom experiments to complex condensed‐matter and optical systems with tunable interactions.

\begin{acknowledgments}
We acknowledge inspiring discussions and feedbacks from B. Doyon, F. Essler, B. Bertini,  J.B. Fouvry, R. Konik and S. Gopalakrishnan and the anonymous Referee C. 
MŁ and MP acknowledge support by the National Science Centre (NCN), Poland via project 2022/47/B/ST2/03334, J.D.N. and L.B. are funded by the ERC Starting Grant 101042293 (HEPIQ) and the ANR-22-CPJ1-0021-01. 
\end{acknowledgments}

\section*{Appendix}

% ---------------------------
% End matter
% \clearpage
% \setcounter{section}{0}
% \setcounter{secnumdepth}{2}
% \section{Additional numerical data}
% \label{s:additional numerical data}

% \begin{figure}[ht]
% \includegraphics[width=\linewidth]{plot_prl_Corr.pdf}
% \caption{
% Numerical results for long-range interacting hard spheres  with $a=1$ and $\xi=20$ starting from the two boosted thermal states $\rho_{\pm; \beta; \theta_0} $ with $\beta = 2.5$, $\theta_0=1$, $\bar{\rho}=0.2$: Plot of the density-density correlations $\int\dd{\theta_1}\dd{\theta_2}g_{\theta_1,\theta_2}^{(2)}(z;\bar t)$ as a function of $z/\xi$ at different times for the interaction potential  $\varphi(x)=1/(1+x^6)$. This figures shows how the Eqs. \eqref{eq:fullg1}, \eqref{eq:fullg2} correctly capture the dynamics of $2$-point correlation function of the system. }
% \label{fig: Correlations dynamics}
% \end{figure} 
\makeatletter

\renewcommand{\appendix}{\par
  \setcounter{section}{0}
  \setcounter{subsection}{0}
  \gdef\thesection{\Alph{section}}
}
\appendix
\renewcommand{\thesubsection}{\thesection.\arabic{subsection}}
\section{Correlation functions in the correlated fluid cell ensemble}\label{app:corrfuns}
In this appendix we will derive the formulas for correlation functions in the ensemble \eqref{eq: longrange_gge_state}. We will also connect the various ingredients of these correlators to Lagrange multipliers of the ensemble.
The main idea behind this calculation consists on expanding the measure \eqref{eq: longrange_gge_state} in $1/\xi$ (according to~\eqref{eq:scaling_xi_corr}) in order to compute the corrections to the local GGE prediction. Firstly, we consider the equal time two-point function
\begin{equation}
\label{eq: app_2point_func_cal}
\begin{split}
    &\langle q_i(x)q_j(x')\rangle^c=\delta(x-x')C_{i,j}+\\&+\frac{1}{2}\int\dd{y}\dd{y'}\beta^{k,l}_{(2)}(y,y')\langle \delta q_k(y)\delta q_l(y')q_i(x)q_j(x')\rangle_1\,.
\end{split}
\end{equation}
In order to write the correlator in a cumulant expansion, we now use the relation
\begin{multline}
    \langle \delta a\delta b\,c d\rangle = \langle a b c d\rangle^c
    + \langle a c\rangle^c\langle b d\rangle^c+\\+
    \langle a d\rangle^c\langle b c\rangle^c
    +\langle a c d\rangle^c\langle \delta b\rangle
    +\langle b c d\rangle^c\langle \delta a\rangle\,,
\end{multline}
where we observe that $\langle \delta q_i\rangle=0$ and that $\langle a b c d\rangle^c$ is higher order in $\xi^{-1}$.
Hence, Eq.~\eqref{eq: app_2point_func_cal} reduces to 
\begin{align}
\label{eq: app_2point_func_cal2}
%\begin{split}
    &\nonumber
    =\delta(x-x')C_{i,j}+\\&\nonumber+\int\dd{y}\dd{y'}\beta_{(2)}^{k,l}(y,y')\langle q_k(y)q_i(x)\rangle^c_1\langle q_l(y')q_j(x')\rangle^c_1
    \\&\nonumber
    =\delta(x-x')C_{i,j}+\beta_{(2)}^{k,l}(x,x')C_{i,k}(x)C_{j,l}(x') \\&=\delta(x-x')C_{i,j}+g^{(2)}_{i,j}(x,x')\,,
%\end{split}
\end{align}
where in the last line we identified $g^{(2)}_{i,j}(x,x')\equiv\beta_{(2)}^{k,l}(x,x')C_{i,k}(x)C_{j,l}(x')+O(\xi^{-2})$. Also, in the last two lines, we explicitly computed the correlation functions on the local GGE.\\
\\
We now compute the three-point function using the same technique
\begin{widetext}
\begin{equation}
\begin{split}
    &\langle q_i(x)q_j(x')q_k(x'')\rangle^c=\delta(x-x')\delta(x-x'')C_{i,j,k}(x)+\frac{1}{2}\int\dd{y}\dd{y'}\beta_{(2)}^{m,n}(y,y')\langle \delta q_m(y)\delta q_n(y') \,q_i(x) q_j(x') q_k(x'')\rangle_1
    \\&
    \qquad\qquad+\frac{1}{3}\int\dd{y}\dd{y'}\dd{y''}\beta_{(3)}^{m,n,p}(y,y',y'')\langle \delta q_m(y)\delta q_n(y')\delta q_p(y'')\,q_i(x)q_j(x')q_k(x'')\rangle_1
    \\
    &=\delta(x-x')\delta(x-x'')C_{i,j,k}(x)+\Big[\delta(x'-x'')\beta_{(2)}^{m,n}(x,x')
    C_{m,i}(x) C_{n,j,k}(x')\Big]_{(i,j,k)}
    \\&\qquad\qquad+\beta_{(3)}^{m,n,p}(y,y')C_{i,m}(x)C_{j,n}(x')C_{k,p}(x'')
    \\
    &=\delta(x-x')\delta(x-x'')C_{i,j,k}(x)+\Big[\delta(x'-x'')g^{(2)}_{i,q}(x,x')C^{q,n}(x')
    C_{n,j,k}(x')\Big]_{(i,j,k)}+g^{(3)}_{i,j,k}(x,x',x'')\,.
\end{split}
\end{equation}
\end{widetext}
In particular, in the last line we defined $g^{(3)}_{i,j,k}(x,x',x'')=\beta_{(3)}^{m,n,p}(y,y')C_{i,m}(x)C_{j,n}(x')C_{k,p}(x'')$, and where we used the formula for $g^{(2)}$ derived in Eq.~\eqref{eq: app_2point_func_cal2}.
Finally, we observe that the last line coincides with Eq.~\eqref{eq: def_3pointfunc_delta}.

\section{\lowercase{g}BBGKY hierarchy and generalised Landau equation for long-range interacting hard spheres }
\label{sec: Explicit_Hard_Rods_equations}
In this section we explicitly write the main hydrodynamic functions for the hard spheres  model that permits to explicitly write the gBBGKY hierarchy and the generalised Landau equation in this simpler case.
The hard spheres gas is a classical system of billiard balls with radius $a$ and unitary mass. The rapidities of the particles in the system are defined as the set of particles velocities. The scattering shift is simply $T_{\theta_1,\theta_2}=-a/2\pi$. Hence, the rotation matrix is given by $R_{\theta_1,\theta_2}=\delta_{\theta_1,\theta_2}+a {/2\pi}1_{\theta_1}n_{\theta_2}$, giving $\rho^{\rm t}=(1+a\bar\rho)/2\pi$ and $n_{\theta}=2\pi\rho_{\theta}/(1-a\bar\rho)$, being $\bar\rho=\int\dd{\theta}\rho_{\theta}$. Since in this model the rotation matrix can be explicitly inverted, the dressing operator can be computed explicitly $R^{-1}_{\theta_1,\theta_2}=\delta_{\theta_1,\theta_2}-a 1_{\theta_1}\rho_{\theta_2}$, as well as the effective velocity 
\begin{equation}
    v^{\rm eff}_{\theta_1}=\frac{\theta_1-a\int\dd{\theta_2}\theta_2\rho_{\theta_2}}{1-a\bar\rho}.
\end{equation}
The susceptibility matrix for this system is given by 
\begin{equation}
    C_{\theta_1,\theta_2}=\delta_{\theta_1,\theta_2}\rho_{\theta_1}+a\bar\rho(a\bar\rho-2)\rho_{\theta_1}\rho_{\theta_2}.
\end{equation}
Hence, we can also compute its functional derivative with respect to the density
\begin{multline}
 C_{\theta_1,\theta_2}^{\theta_3}= \delta_{\theta_1,\theta_3}\delta_{\theta_2,\theta_3}+a^2\rho_{\theta_1}\rho_{\theta_2}1_{\theta_3}+\\+a(a\bar\rho-2)\rho_{\theta_2}\delta_{\theta_1,\theta_3}+a(a\bar\rho-2)\rho_{\theta_1}\delta_{\theta_2,\theta_3}\,.
\end{multline}
This tensor is entering in the gBBGKY hierarchy as 
\begin{multline}
    \int\dd{\theta_3}C_{\theta_2,\theta_3}^{\gamma}g^{(2)}_{\theta_1,\gamma}(x_1,x_2)=(1-a\bar\rho)^2g^{(2)}_{\theta_1,\theta_2}(x_1,x_2)-\\-2a(1-a\bar\rho)\rho_{\theta_1}\int\dd{\gamma}g^{(2)}_{\theta_3,\theta_2}(x_1,x_2)\,.
\end{multline}
The flux Jacobian $A_{\theta_1}^{\theta_2}$ can be easily computed using the explicit relations for the dressing operator, the rotation matrix and the effective velocity.
We also define $\mathcal{D}$ as the linear response diffusion matrix of the system, that for hard spheres  reads
\begin{equation}
    \mathcal{D}_{\theta_1}^{\theta_2}=a^2(1-a\bar\rho)\Big[\delta_{\theta_1,\theta_2}\int\dd{\theta_3}\rho_{\theta_3}|v_{\theta_1,\theta_3}|-\rho_{\theta_1}|v_{\theta_1,\theta_2}|\Big]\,.
\end{equation}

Finally, we can explicitly write the generalised Landau equation for the hard spheres model. For simplicity, we write it considering systems with vanishing total momentum $\int\dd{\theta_2}\theta_2\rho_{\theta_2}=0$
\begin{equation}
\begin{aligned}
    \mathcal{I}^{\rm cross}_{\theta_1}&= \frac{1}{2\pi}a^2(1-a\bar\rho)^3\int {\dd}k k^2 (\hat{V}(k))^2  \\&\times\int \dd{\theta_2} \dd{\theta_3} \Big(\frac{|v^{\rm br}_{\theta_3 \theta_1}|}{v^{\rm br}_{\theta_2 \theta_3} (v^{\rm br}_{ \theta_1 \theta_2})^2}-\frac{|v^{\rm br}_{\theta_2 \theta_3}|}{v^{\rm br}_{\theta_1 \theta_2} (v^{\rm br}_{\theta_3 \theta_1})^2}\Big)\\&\times\mathcal{B}_{\theta_1,\theta_2,\theta_3}^{\rm br}\,,
\end{aligned}
\end{equation}
\begin{equation}
\begin{aligned}
    \mathcal{I}^{\rm self}_{\theta_1}&=  \frac{1}{2\pi}a^2(1-a\bar\rho)^3\int  \dd{\theta_3} \dd{\theta_2} \int {\dd}k k^2   \hat{V}(v^{\rm br}_{\theta_1\theta_3}k) \\
    &\times\Big(\hat{V}(v^{\rm br}_{ \theta_1 \theta_2}k)\frac{v^{\rm br}_{\theta_3 \theta_1}v^{\rm br}_{ \theta_1 \theta_2}}{ v^{\rm br}_{\theta_3 \theta_2}} -2 \hat{V}(v^{\rm br}_{  \theta_2 \theta_3 } k )\frac{(v^{\rm br}_{  \theta_2 \theta_3 })^2v_{\theta_3 \theta_1}}{(v^{\rm br}_{\theta_1 \theta_2})^2 }\Big) 
    \\&\times\mathcal{B}_{\theta_1,\theta_2,\theta_3}^{\rm br}\,,
\end{aligned}
\end{equation}
where we defined the bare velocity difference as $v^{\rm br}_{\theta_1,\theta_2}=\theta_1-\theta_2$ and where 
\begin{multline}
    \mathcal{B}_{\theta_1,\theta_2,\theta_3}^{\rm br}\equiv v^{\rm br}_{\theta_3 \theta_2} \partial_{\theta_1} \rho_{\theta_1} \rho_{\theta_3}  \rho_{\theta_2}  +\\+ v^{\rm br}_{\theta_2 \theta_1} \rho_{\theta_1} \partial_{\theta_3} \rho_{\theta_3} \rho_{\theta_2} + v^{\rm br}_{\theta_1 \theta_3}\rho_{\theta_1} \rho_{\theta_3}\partial_{\theta_2} \rho_{\theta_2}.
\end{multline}
\section{Additional details on the numerical solution of \lowercase{g}BBGKY hierarchy}
\label{sec: additional_details_numerical_solution}
In this section, we detail the numerical methods used to produce the results shown in Fig. \ref{fig: Diffusive correction}. Let us start from the microscopic simulation of hard spheres  system with long-range interactions. The molecular dynamics is solved using the algorithm introduced in \cite{Biagetti2024}, combining symplectic integration of equation of motion and the explicit solution of hard core collisions. More precisely, we proceeded setting a time interval $\Delta t_S$ for the symplectic integrator and computing the minimal time before the first hard core interaction in the system $\Delta t_{\rm Int}\approx \min_i\{(x_{i+1}-x_i)/(\theta_{i+1}-\theta_i)\}$. At any time, if $\Delta t_S>\Delta t_{\rm Int}$, we perform the symplectic integration of equation of motion using an explicit Runge-Kutta method of order 8 implemented by the \textit{DOP853} method in \texttt{scipy.integrate.solve\_ivp} function with time step $\Delta t_S$. If, instead, $\Delta t_S<\Delta t_{\rm Int}$, we perform the symplectic integration with time step $\Delta t_{\rm Int}$, and then we explicitly compute the hard core collision between the two colliding rods, i.e. exchanging the colliding rods' velocities. In particular, we choose $\Delta t_S=5\times 10^{-3}$. This value guarantees the correct convergence of the numerical method for the set of parameters considered in the simulations. The results are averaged over $N_{\rm sym}=1000$ initial configurations. The initial state is a GGE state, see \cite{Urilyon2025} for a detailed explanation. \\
\\
We now discuss the solution of gBBGKY equations. In the results shown in Fig. \ref{fig: Diffusive correction}, we split the dynamics in two stages, solving the system separately for $t<t_L$ and $t>t_L$. For the small time dynamics $t<t_L$, we solved the following two gBBGKY equations for a hard spheres  system (see Appendix \ref{sec: Explicit_Hard_Rods_equations} for the explicit functions)
\begin{equation}
\label{eq:BBGKY_quasi_rho_numerical}
    \partial_{t}\rho_{\theta_1}=\frac{1}{\xi}\int\dd{x_2}\dd{\theta_2}V'(x_1-x_2)\partial_{\theta_1}g^{(2)}_{\theta_1,\theta_2}(x_1,x_2)\,,
\end{equation}

\begin{align}
\label{eq:BBGKY_quasi_g2_numerical}
%\begin{split}
    &\partial_t g^{(2)}_{\theta_1, \theta_2}(x_1,x_2)+\\&\nonumber+ \Big[\partial_{x_1}A_{\theta_1}^{\gamma} g^{(2)}_{\gamma, \theta_2}(x_1,x_2) -\frac{1}{2}\partial_{x_1} \mathfrak{D}_{\theta_1}^{\gamma} g^{(2)}_{ \gamma,\theta_2}\Big]_{(1,2)}=
    \\&\nonumber=
    \int \dd{\theta_3}\partial_{\theta_1}\Big[\int \dd x_3 V'(x_1-x_3)\rho_{ \theta_1}g^{(2)}_{\theta_2,\theta_3}(x_2,x_3)
    \\&\nonumber
    +V'(x_1-x_2)  \Big( \rho_{\theta_1}C_{\theta_2,\theta_3}+ C_{\theta_2,\theta_3}^{\gamma}g^{(2)}_{\theta_1,\gamma}(x_1,x_2)\Big)\Big]_{(1,2)}
    \\&\nonumber+\delta(x_1-x_2)\int\dd{x_3}\dd{\theta_3}V'(x_1-x_3)\Big(-C_{\theta_1,\theta_2}^{\gamma}\partial_{\gamma}\\&\nonumber\qquad\quad+(\partial_{\theta_1}+\partial_{\theta_2})C_{\theta_1,\theta_2}^{\gamma}\Big)g^{(2)}_{\gamma,\theta_3}(x_1,x_3)\,,
%\end{split}
\end{align}
using a $4-$th order Runge-Kutta method with time interval $=0.02$. In particular, we used a grid of $100$ points in space and rapidity in the intervals, with $-10\xi<x<10\xi$ and $-7<\theta<7$.
The space interval has been chosen large enough to have vanishing correlations at the boundaries at any time, in order to avoid spurious reflective effects in the dynamics of $g^{(2)}$. For $t>t_L$, we instead solved the generalised Landau equation. The precise value of $t_L$ has been considered as a "fitting" parameter, hence fixed in the interval $\xi<t_L<\xi^2$, in order to accurately reproduce the molecular dynamics results.

For the data presented in Fig.~\ref{fig:thermalizationrate} we use grid of 150 points in rapidity space with $-4.4\hbar k_d/m \leq \theta \leq 4.4\hbar k_d/m$. The initial state presented in inset of Fig.~\ref{fig:thermalizationrate} is then evolved for a short time evolved using expression for inter-tube collision integral provided in SM. From that dynamics, we extract the rate of decay of $\log DT$.

\section{\lowercase{g}BBGKY hierarchy for inhomogeneous systems}
\label{sec:Inhomogeneous_systems}
In this section we write the first equations of gBBGKY hierarchy for inhomogeneous systems. The derivation follows exactly the same procedure introduced in Sec. \ref{s:gBBGKY}. For brevity reasons, we show here only the first two equations of the hierarchy. higher-order equations can be derived using the same technique.
We first consider the set of generic equations~\eqref{eq: evo_eq_implicit_deltas_1} and~\eqref{eq: evo_eq_implicit_deltas_2}. Hence, the next steps are to use the cumulant expansion for charge-currents correlators (see Sec. \ref{sec:currents}) and to factorize the singular component of few point functions (see Sec. \ref{sec: corr_fluid_cell_ens}). After these steps, the equations read

\begin{widetext}
    \begin{equation}
    \label{eq: evo_eq_explicit_deltas_1_inhom}
        \partial_t \langle q_{i_1}(x_1)\rangle +\partial_{x_1} \Big(A_{i_1}^k \langle q_k(x_1)\rangle+\frac{1}{2} H_{i_1}^{j,k}g^{(2),\rm sym}_{j,k}(x^+_1,x^-_1)\Big)= - \int {\rm d}x_2 V'(x_1-x_2) A_{i_1,0}^k \big(\langle q_k(x_1)\rangle\langle q_0(x_2)\rangle+g^{(2)}_{k,0}(x_1,x_2)\big),
    \end{equation}
    \begin{equation}
    \label{eq: evo_eq_explicit_deltas_2_inhom}
    \begin{aligned}
        &\partial_t g^{(2)}_{i_1,i_2}(x_1,x_2) + \Big[\partial_{x_1} A_{i_1}^jg^{(2)}_{j,i_2}(x_1,x_2)+\frac{1}{2}\partial_{x_1} H_{i_1}^{j,k}g^{(3),\rm sym}_{j,k,i_2}(x^+_1,x^-_1,x_2)\Big]_{(1,2)} =\\&=-\delta(x_1-x_2)\langle q_{i_1}q_{i_2}j^-_{k}\rangle^c_1 C^{kl}\partial_{x_1}\langle q_{l}(x_1)\rangle-\Big[V'(x_1-x_2)A_{i_1,0}^j\left(\langle q_j(x_1) \rangle C_{i_2,0}+g^{(2)}_{j,k}(x_1,x_2)C^{k,m}C_{m,i_2,0}\right)+\\&+\int {\rm d}x_3 V'(x_1-x_3) A_{i_1,0}^j\Big(\langle q_{j}(x_1) \rangle g^{(2)}_{i_2,0}(x_2,x_3)  +\langle q_{0}(x_3) \rangle g^{(2)}_{j,i_2}(x_1,x_2)  + g^{(3)}_{j,i_2,0} \Big)\Big]_{(1,2)}
        \\&
        -\delta(x_1-x_2)\int\dd{x_3}V'(x_1-x_3)
        \Big[-C_{i_1,i_2,j}C^{j,k}A_{k,0}^{n}+\left(A_{i_1,0}^j\delta_{i_2}^k+\delta_{i_1}^jA_{i_2,0}^k\right)C_{j,k,m}C^{m,n}\Big]g^{(2)}_{n,0}(x_1,x_3)\,.
    \end{aligned}
    \end{equation}
\end{widetext}
with $j^-_i=j_i-A_i^jq_j$. In the last equation, we can observe few important differences with respect to the homogeneous case. Firstly, the dynamical equation for density~\eqref{eq: evo_eq_explicit_deltas_1_inhom} gets more involved, since now it non-trivial flow of particles are also allowed. It is important to observe that the first term on RHS is a generalization to the celebrated Vlasov forcing term.
We also observe that Eq.~\eqref{eq: evo_eq_explicit_deltas_1_inhom} is coupled to two-point function also through the second term in currents cumulant expansion, in the exact way as for purely integrable evolution, see the recent results in \cite{Hubner2024}.
The equation for two-point function~\eqref{eq: evo_eq_explicit_deltas_2_inhom}, has an additional non-trivial contribution proportional to a Dirac delta function, originating from the complex propagation of initial local fluctuations, also already introduced in \cite{Hubner2024}.

\section{Extra details on the derivation of the gBBGKY hierarchy}
% Eq.~\eqref{eq: evo_eq_explicit_deltas_1},~\eqref{eq: evo_eq_explicit_deltas_2} and~\eqref{eq: evo_eq_explicit_deltas_3}}
\label{app:comment_eq_charges}
In this section we detail the main steps needed in order to derive Eq.~\eqref{eq: evo_eq_explicit_deltas_1},~\eqref{eq: evo_eq_explicit_deltas_2} and~\eqref{eq: evo_eq_explicit_deltas_3} from the hierarchy  Eq.~\eqref{eq: evo_eq_implicit_deltas_1},~\eqref{eq: evo_eq_implicit_deltas_2} and~\eqref{eq: evo_eq_implicit_deltas_3} written in terms of complete correlation functions.
Firstly, we specialize these equations to homogeneous systems, hence Eq.~\eqref{eq: evo_eq_explicit_deltas_1} gets particularly simple for $q_i$ being independent on the space coordinate. In addition, the disconnected forcing term is vanishing because of the symmetry of the potential. Similarly, the singular contribution to two-point function  
is not entering in the equation since $\varphi'(0)=0$.
Eq.~\eqref{eq: evo_eq_explicit_deltas_2} describe the evolution of regular connected two-point function. This equation is derived from~\eqref{eq: evo_eq_implicit_deltas_2} by performing three main steps: expanding currents through hydrodynamic projections, subtract disconnected component and finally separate singular and regular part of correlators. The most involved part of the calculation is related to the calculation of singular terms in RHS, i.e. proportional to the Dirac delta. Hence in this section we describe the main steps of their derivation. Firsly, we compute the contributions coming from the delta part in LHS of ~\eqref{eq: evo_eq_implicit_deltas_2}
\begin{multline}
    \delta(x_1-x_2)\partial_t C_{i_1,i_2}+\Big[\partial_{x_1}\delta(x_1-x_2)A_{i_1}^{j}C_{j,i_2}]=
    \\
    =\delta(x_1-x_2)\int\dd{x_3}V'(x_1-x_3)\times\\\times C_{i_1,i_2,j}C^{j,k}A_{k,0}^{n}g^{(2)}_{n,0}(x_1,x_3)\,,
\end{multline}
where we used $\delta C_{i_1,i_2}/\langle\delta q_k \rangle_1= C_{i_1,i_2,j}C^{j,k}$ and where we used Eq.~\eqref{eq: evo_eq_explicit_deltas_1}. The additional singular contribution to the RHS of~\eqref{eq: evo_eq_implicit_deltas_2} comes from the singular part of the three-point function $\langle j_{i,0}(x)q_{j}(y)q_0\rangle^c$ for coinciding $x$, $y$. All the remaining delta contributions from RHS are integrated, i.e. appear as regular terms.
We now focus on Eq.~\eqref{eq: evo_eq_explicit_deltas_3}. The procedure to derive this equation follows the same logic as Eq.~\eqref{eq: evo_eq_explicit_deltas_2}, but here we simply stop at the leading order terms in $O(V_0,1/\xi)$. Firstly, we observe that in the case of three-point function the hydrodynamic projection formula is more involved already for Euler scales as, for any local observable $\hat{o}$, we have \cite{10.21468/SciPostPhys.15.4.136}
\begin{multline}
\label{eq: hydro proj 3}
    \langle {o}(x){q}_j(y){q}_k(z) \rangle^c
    =\frac{\delta\langle o(x)\rangle_1}{\delta\langle q_i(x)\rangle_1}\langle{q}_i(x){q}_j(y){q}_k(z)\rangle^c+\\+\frac{\delta^2\langle o(x)\rangle_1}{\delta \langle q_i(x)\rangle_1\delta \langle q_m(z)\rangle_1}\langle{q}_m(x){q}_k(z)\rangle^c\langle{q}_i(x){q}_j(y)\rangle^c
    \,.
\end{multline}
More precisely, in this equation, the regular part of second term on RHS appears as higher-order in $V_0$ with respect to the regular part of the first one, hence we did not explicitly include in Eq.~\eqref{eq: evo_eq_explicit_deltas_3}.
Then, we proceed separating the singular part from it. 
In doing that we used the following simplifications coming from homogeneity:
$\partial_tC_{i,j,k}\sim O(V_0^2)$, $\partial_{x}C_{i,j,k}=0$.
\subsection{Derivation of the equation for the three-point functions}
\label{app:der_tensor_M}
In this section we derive the RHS of Eq~\eqref{eq:BBGKY_quasi_g3}. 
As a starting point, we consider the RHS of Eq.~\eqref{eq: evo_eq_explicit_deltas_3}. 
In particular, it turns out to be writable in a more convenient way as 
\begin{align}
%\begin{split}
    \nonumber&A_i^{m}
    C_{m,j,p} C^{p,q}- 
    C_{i,j,m} C^{m,p}A_p^{q}+ H_{i}^{m,q}C_{m,j}=
    \\&\nonumber=
    \frac{\delta \langle j_i\rangle_1}{\delta \langle q_m\rangle_1} \frac{\delta}{\delta \langle q_q\rangle_1}\frac{\delta \langle q_m\rangle_1}{\delta \beta^j}-\frac{\delta}{\delta \langle q_p\rangle_1}\Big[\frac{\delta \langle q_i\rangle_1}{\delta \beta^j}\Big]\frac{\delta \langle j_p\rangle_1}{\delta \langle q_q\rangle_1}+\\&\nonumber+\frac{\delta^2 \langle j_i\rangle_1}{\delta \langle q_m\rangle_1\delta \langle q_q\rangle_1}\frac{\delta \langle q_m\rangle_1}{\delta \beta^j} =
    \\&\nonumber
    =\frac{\delta^2 \langle j_i\rangle_1}{\delta \beta^j\delta \beta^p}\frac{\delta \beta^p}{\delta \langle q_q\rangle_1}
    -\frac{\delta}{\delta \langle q_p\rangle_1}\Big[\frac{\delta \langle q_i\rangle_1}{\delta \beta^j}\Big]\frac{\delta \langle j_p\rangle_1}{\delta \langle q_q\rangle_1}\\&=\langle q_i q_j j_p \rangle^c_1 C^{p,q}-\langle q_i q_j q_m \rangle^c_1 C^{m,p}A_p^q
    =\langle q_i q_j j^-_p \rangle^c_1 C^{p,q}\,,
%\end{split}
\end{align}
where we used that $AC=CA^t$. Here  we defined $j^-_i=j_i-A_i^jq_j$.
Hence, as shown in \cite{Urilyon2025}, in quasiparticle picture it turns out to be equal to
\begin{equation}
\label{eq:M_tensor_implicit}
    \langle q_i q_j j^-_p \rangle_1^c C^{p,q} = \int\dd{\theta_1}\dd{\theta_2}h_{i}(\theta_1)h_j(\theta_2)  M_{\theta_1, \theta_2}^{\theta_3} h^q(\theta_3)\,,
\end{equation}
with the $M$ tensor defined in~\eqref{eq:M_tensor}.

\subsection{ Solutions with Green's function method}\label{app:Green}
The aim of this appendix is to justify the solution ~\eqref{eq:g21solution}, which is crucial for derivation of generalized Landau kinetic equation. Our starting point is the following equation (we suppress upper indices for lighter notation)
\begin{equation}\label{eq:hierarchy_der1}
\begin{aligned}
    \partial_t g_{\theta_1, \theta_2}(z)+\partial_z (A_{\theta_1}^{\gamma} g_{\gamma, \theta_2}(z)-A_{\theta_2}^{\gamma} g_{\theta_1, \gamma}(z))=\mathcal{S}_{\theta_1, \theta_2}[\rho](z).
\end{aligned}
\end{equation}
We start with moving to normal modes in order to diagonalize $A$ operators $g_{\theta_1, \theta_2}(z) =R^{-t}_{\theta_1,\xi_1}R^{-t}_{\theta_2,\xi_2}\bar{g}_{\xi_1, \xi_2}(z)$ and similarly for $\mathcal{S}_{\theta_1, \theta_2}[\rho](z)$. We get a equation
\begin{equation}\label{eq:hierarchy_der2}
\begin{aligned}
    \partial_t\bar{g}_{\theta_1, \theta_2}(z)+ v_{\theta_1,\theta_2}\partial_z\bar{g}_{\theta_1, \theta_2}(z)=\bar{\mathcal{S}}_{\theta_1, \theta_2}[\rho](z),
\end{aligned}
\end{equation}
for which the Green's function can be found with Fourier transform and reads
\begin{equation}
    G_{\theta_1, \theta_2}(z,z';t,t')=\delta(z+z'-v_{\theta_1 ,\theta_2}t)\theta_H(t-t'),
\end{equation}
where $\theta_H$ is Heaviside step function. Hence, we find
\begin{equation}
\begin{aligned}
      \bar{g}_{\theta_1 ,\theta_2}(z) =& \int \dd z' \int_0^t \dd \tau  \int \frac{\dd k}{2 \pi}  e^{ik(z+z'-v_{\theta_1,\theta_2}\tau)} \times \\
      &\bar{\mathcal{S}}_{\theta_1,\theta_2}[\rho](z',t-\tau).
\end{aligned}
\end{equation}
Lastly, we rotate back to $g_{\theta_1,\theta_2}(z)$ to obtain the final result \eqref{eq:g21solution}.

\section{Computational Complexity of gBBGKY and Generalization to Multi-Species Models}
In practical applications, evaluating the gBBGKY hierarchy requires numerically solving a set of coupled integro-differential equations. To assess the computational complexity of this framework, let us consider the numerical cost of solving the hierarchy truncated at the second layer, meaning we retain the coupled evolution of the one-point root densities and the two-point correlation structures ~\eqref{eq:BBGKY_quasi_rho} and~\eqref{eq:BBGKY_quasi_g2}, while higher-order connected correlations are set to zero. For an integrable model with a single quasiparticle species, such as the hard rods gas or the repulsive Lieb-Liniger gas, the state is parameterized by a single rapidity variable $\theta$. Numerically, the rapidity space is discretized into $N_\theta$ grid points. The computational bottleneck in solving the hierarchy lies in the standard Generalized Hydrodynamics (GHD) dressing operations, which require matrix inversions that scale polynomially with $N_\theta$. For more generic integrable models that feature bound states, such as the Heisenberg XXX spin chain or the attractive Lieb-Liniger model, the excitation spectrum is significantly richer. In the thermodynamic limit, these bound states are classified as "strings" of length $n$, where $n \in \mathbb{N}$. Consequently, the thermodynamic state is no longer described by a single root density, but rather by an infinite set of root densities $\rho_n(\theta)$. The algebraic structure of the gBBGKY hierarchy remains identical, but the integration over the rapidity space is formally upgraded to include a sum over all string species: $\int \text{d}\theta \to \sum_n \int \text{d}\theta$. In any numerical implementation, this sum must be truncated to a maximum string length $n_{\max}$. This truncation dictates the scaling of the computational cost. For the first layer~\eqref{eq:BBGKY_quasi_rho}, instead of a single density $\rho(\theta)$, the solver must track and update $n_{\max}$ coupled root densities $\rho_n(\theta)$. For the second layer~\eqref{eq:BBGKY_quasi_g2}, the two-point correlation structures now involve cross-correlations between different string species $n$ and $m$. Therefore, the solver must track an $n_{\max} \times n_{\max}$ matrix of coupled two-point functions. Effectively, the dimension of the local rapidity grid expands from $N_\theta$ to $n_{\max} N_\theta$. Hence, with this truncation, the total number of dynamic variables at each spatial grid point scales as $O(n_{\max}^2)$. Correspondingly, the matrix operations required for the dressing transformations and the convective transport will scale polynomially with this expanded grid size. While this $O(n_{\max}^2)$ scaling makes the numerical integration for generic multi-species models computationally heavier than for single-species models, it is crucial to emphasize that the scaling remains strictly polynomial, i.e. the framework does not introduce any exponential bottlenecks.

\let\oldaddcontentsline\addcontentsline
\renewcommand{\addcontentsline}[3]{}
\bibliography{bib}
\let\addcontentsline\oldaddcontentsline

\clearpage % Forces a page break
% \onecolumn % Switches the layout to single column

% Optional: If this is Supplemental Material, you usually want to reset 
% the counters and add an "S" prefix to equations and figures (e.g., Eq. S1)
\setcounter{equation}{0}
\setcounter{figure}{0}
\setcounter{table}{0}
\setcounter{page}{1}
\renewcommand{\theequation}{S\arabic{equation}}
\renewcommand{\thefigure}{S\arabic{figure}}

% Include your second file here (do not add the .tex extension)
\onecolumngrid

\begin{center}
\textbf{\large Supplemental Material:\\
Generalised BBGKY hierarchy for near-integrable dynamics }\\[5pt]
Leonardo Biagett${\rm i}^1$, Maciej Łebe${\rm k}^2$, Miłosz Panfi${\rm l}^2$ and Jacopo De Nardi${\rm s}^1$ \\

{\footnotesize \sl $^1\mathit{L}$\textit{aboratoire de Physique Th\'eorique et Mod\'elisation, CNRS UMR 8089, CY Cergy Paris Universit\'e, 95302 Cergy-Pontoise Cedex, France}
\newline
 $^2\mathit{F}$\textit{aculty of Physics, University of Warsaw, Pasteura 5, 02-093 Warsaw, Poland}}\\
\end{center}
In the Supplemental Material we provide additional computations supporting the results presented in the main text. The material is presented as follows

\section{Derivation of the generalised Landau equation}
\label{sec:Landau_Eq_SM}
In this section we derive in full details the generalised Landau equation from the gBBGKY hierarchy of equations introduced in the main text. As already introduced in the main text, we start considering the first three layers of the hierarchy and decompose the two-point function into leading $g^{(2),1}$ and subleading $g^{(2),2}$ contribution in $1/\xi$. Hence, according to this notation, we consider the following set of equations
\begin{equation}\label{eq:BBGKY_quasi_rho_SM}
	\partial_{t}\rho_{\theta_1}=\frac{1}{\xi^2}\int\dd{z}\dd{\theta_2}V_0\varphi'(z)\partial_{\theta_1}\big(g^{(2),1}_{\theta_1,\theta_2}(z)+\frac{1}{\xi}g^{(2),2}_{\theta_1,\theta_2}(z)\big)\,,
\end{equation}
\begin{equation}
	\label{eq: evolution_for_g2(1)_SM}
	\begin{split}
		&\partial_t g^{(2),1}_{\theta_1, \theta_2}(z)+\partial_z \Big(A_{\theta_1}^{\gamma} g^{(2),1}_{\gamma, \theta_2}(z)-A_{\theta_2}^{\gamma} g^{(2),1}_{\theta_1, \gamma}(z) \Big)=V_0\varphi'(z) \int \dd{\gamma} \Big(\partial_{\theta_1} \rho_{\theta_1}C_{\gamma,\theta_2}-\partial_{\theta_2} \rho_{\theta_2}C_{\theta_1,\gamma }\Big)\,,
	\end{split}
\end{equation}
\begin{equation}
	\label{eq: evolution_for_g2(2)_SM}
	\begin{split}
		&\partial_t g^{(2),2}_{\theta_1, \theta_2}(z)+\partial_z \Big(A_{\theta_1}^{\gamma} g^{(2),2}_{\gamma, \theta_2}(z)-A_{\theta_2}^{\gamma} g^{(2),2}_{\theta_1, \gamma}(z) \Big)=-\frac{1}{2}\partial_z \Big(H_{\theta_1}^{\gamma_1\gamma_2} g^{(3),\rm reg}_{\gamma_1, \gamma_2, \theta_2 }(0^+,z)-H_{\theta_2}^{\gamma_1, \gamma_2} g^{(3),\rm reg}_{\theta_1,\gamma_1, \gamma_2 }(z,0^+)\Big)\,,
	\end{split}
\end{equation}
\begin{equation}\label{eq:BBGKY_quasi_g3_SM}
	\begin{split}
		\partial_{t}g^{(3)}_{\theta_1,\theta_2,\theta_3}(z,z')&+\partial_{z}A_{\theta_1}^{\gamma}g^{(3)}_{\gamma,\theta_2,\theta_3}(z,z')+(\partial_{z'}-\partial_{z})A_{\theta_2}^{\gamma}g^{(3)}_{\theta_1,\gamma,\theta_3}(z,z')-\partial_{z'}A_{\theta_3}^{\gamma}g^{(3)}_{\theta_1,\theta_2,\gamma}(z,z')=\\&=\delta(z)M_{\theta_1,\theta_2}^{\gamma}\partial_{z}g^{(2),1}_{\gamma,\theta_3}(z+z')
		-\delta(z')M_{\theta_2,\theta_3}^{\gamma}\partial_{z}g^{(2),1}_{\theta_1,\gamma}(z)+\delta(z')M_{\theta_2,\theta_3}^{\gamma}\partial_{z'}g^{(2),1}_{\theta_1,\gamma}(z+z')\,.
	\end{split}
\end{equation}
The goal of this section is to integrate this set of equations in order to derive a unique closed kinetic equation for $\partial_t\rho$.
In order to do that, we crucially observe that all these equations are now written in an analogous form: as a convection equation with an external driving force. Hence, the main idea behind the derivation is to impose separation of time scales for the evolution of $\rho$, $g^{(2),1}$, $g^{(2),2}$ and $g^{(2),3}$. More precisely, we impose that the particle density evolve much slower then the few point functions. In this way we can decouple the set of equations and solve them separately. Hence, using a standard Green's function technique, we find the following time evolution relation
\begin{equation}\label{eq:g21solution_SM}
	\begin{aligned}
		&g^{(2),1}_{\theta_1, \theta_2}(z,t) =R_{\theta_1,\sigma_1}^{-t} R_{\theta_2,\sigma_2}^{-t}  \int_0^t \dd \tau  \int \frac{\dd k}{2 \pi} \int \dd{\gamma}\int {\dd}z'e^{ik(z+z'-v_{\sigma_1, \sigma_2}\tau)}R_{\sigma_1,\eta_1}^t R_{\sigma_2,\eta_2}^t V_0\varphi'(z')  \Big[\partial_{\eta_1} \rho_{\eta_1} C_{\gamma,\eta_2}-\partial_{\eta_2} \rho_{\eta_2}C_{\eta_1,\gamma }\Big]_{t-\tau},
	\end{aligned}
\end{equation}
where we defined $v_{\xi_1,\xi_2}\equiv v^{\rm eff}_{\xi_1}-v^{\rm eff}_{\xi_2}$ and where the subscript $t-\tau$ indicates the time at which must be computed the function inside the brackets.
We now perform an approximation which amounts to replacing $\rho_\theta(z,t-\tau)$ with $\rho_\theta(z,t)$ under the integral and extending the integration range in time to infinity. This simplification follows from the assumption of separation of scales. 
Next, we use the identity $\int_0^\infty \dd \tau e^{i\tau x}=\pi \delta(x)+i\mathcal{P}(\frac{1}{x})$, where $\mathcal{P}$ denotes principal value. The contribution coming from the Dirac delta function vanishes since the argument of the square bracket is antisymmetric with respect to $\eta_1$, $\eta_2$. Hence the result reads
\begin{equation}\label{eq:g21solution_2SM}
	\lim_{t\to\infty}g^{(2),1}_{\theta_1, \theta_2}(z,t) =V_0\varphi'(z')R_{\theta_1,\sigma_1}^{-t} R_{\theta_2,\sigma_2}^{-t}\frac{1}{v_{\sigma_1,\sigma_2}}R_{\sigma_1,\eta_1}^t R_{\sigma_2,\eta_2}^t   \Big(\partial_{\eta_1} \rho_{\eta_1} C_{\gamma,\eta_2}-\partial_{\eta_2} \rho_{\eta_2}C_{\eta_1,\gamma }\Big),
\end{equation}
where the principal value has been removed, for the function being analytic. Also, here we stress that the RHS is evaluated at time $t$.
In an analogous way, we can now proceed integrating Eq.~\eqref{eq:BBGKY_quasi_g3_SM}. More precisely, we use the same exact set of approximations in order to time integrate the evolution equation for $g^{(3)}$. In this case the result is slightly more complicated, since this equation is characterized by three forcing terms being proportional to space Dirac delta functions
\begin{equation}
	\label{eq: integrated_g3_1}
	\begin{aligned}
		g^{(3)}_{\theta_1, \theta_2, \theta_3}(z,z';t)&= \frac{1}{2}R^{-t}_{{\theta_1},{\gamma_1}}R^{-t}_{{\theta_2},{\gamma_2}}R^{-t}_{{\theta_3},\gamma_3} \frac{1}{v_{\gamma_1, \gamma_2}} \left[\text{sgn}(z)- \text{sgn}(z-v_{\gamma_1, \gamma_2}t)\right] \times \\
		& \times R^t_{{\gamma_1},{\mu_1}}R^t_{{\gamma_2},{\mu_2}}R^t_{{\gamma_3},{\mu_3}}M_{\mu_1, \mu_2}^\beta \partial_1g^{(2),1}_{\beta,\mu_3} \left(
		\frac{v_{\gamma_2, \gamma_3}}{v_{\gamma_1, \gamma_2}}z-z';t- \frac{z}{v_{\gamma_1, \gamma_2}}\right)+ \\
		&+\frac{1}{2}R^{-t}_{{\theta_1},{\gamma_1}} R^{-t}_{{\theta_2},{\gamma_2}}R^{-t}_{{\theta_3},\gamma_3}\frac{1}{v_{\gamma_1, \gamma_3}}\left[\text{sgn}(z+z')- \text{sgn}(z+z'-v_{\gamma_1,\gamma_3}t)\right] \times \\
		& \times R^t_{{\gamma_1},{\mu_1}}R^t_{{\gamma_2},{\mu_2}}R^t_{{\gamma_3},{\mu_3}}M_{\mu_1, \mu_3}^\beta \partial_1 g^{(2),1}_{\beta,\mu_2} \left( \frac{v_{\gamma_3, \gamma_2}}{v_{\gamma_1, \gamma_3}}z+ \frac{v_{\gamma_1, \gamma_2}}{v_{\gamma_1, \gamma_3}}z';t- \frac{z+z'}{v_{\gamma_1 ,\gamma_3}}\right)+ \\
		&+\frac{1}{2}R^{-t}_{{\theta_1},{\gamma_1}} R^{-t}_{{\theta_2},{\gamma_2}}R^{-t}_{{\theta_3},\gamma_3} \frac{1}{v_{\gamma_2, \gamma_3}}\left[\text{sgn}(z')- \text{sgn}(z'-v_{\gamma_2,\gamma_3}t)\right] \times \\
		& \times R^t_{{\gamma_1},{\mu_1}}R^t_{{\gamma_2},{\mu_2}}R^t_{{\gamma_3},{\mu_3}} M_{\mu_2, \mu_3}^\beta \partial_1 g^{(2),1}_{\beta,\mu_1} \left( z+ \frac{v_{\gamma_2, \gamma_1}}{v_{\gamma_2, \gamma_3}}z';t- \frac{z'}{v_{\gamma_2, \gamma_3}}\right)\,,
	\end{aligned}
\end{equation}
where the symbol $\partial_1$ means that the derivative is performed on the spacial argument of the function. Now, we make use of the assumption of initial state null correlations, for which $g^{(2),1}_{\theta_1,\theta_2}(z,t\leq0)=0$. Hence, the time argument in the correlation functions imposes
\begin{equation}
	t- \frac{z'}{v_{\alpha, \beta}}>0\,\longrightarrow\, \text{sgn}(z'-v_{\alpha,\beta}t)=- \text{sgn}(v_{\alpha,\beta})\,.
\end{equation}
In order to use this integrated correlator into evolution equation~\eqref{eq: evolution_for_g2(2)_SM}, we need to evaluate the result~\eqref{eq: integrated_g3_1} and compute the point split symmetrization, reading
\begin{equation}
	\label{eq: t_evolution_g_reg}
	\begin{aligned}
		&g^{(3), \text{reg}}_{\theta_1, \theta_2, \theta_3}(0^+,z';t) = \frac{1}{2}R^{-t}_{{\theta_1},{\gamma_1}} R^{-t}_{{\theta_2},{\gamma_2}}\frac{1}{|v_{\gamma_1, \gamma_2}|} R^t_{{\gamma_1},{\mu_1}}R^t_{{\gamma_2},{\mu_2}}M_{\mu_1, \mu_2}^\beta \partial_1g^{(2),1}_{\beta,\theta_3} \left(
		-z';t\right) \\
		&+\frac{1}{2}R^{-t}_{{\theta_1},{\gamma_1}} R^{-t}_{{\theta_2},{\gamma_2}}R^{-t}_{{\theta_3},\gamma_3}\frac{1}{v_{\gamma_1 ,\gamma_3}}\left[\text{sgn}(z')+\text{sgn}(v_{\gamma_1,\gamma_3})\right]R^t_{{\gamma_1},{\mu_1}}R^t_{{\gamma_2},{\mu_2}}R^t_{{\gamma_3},{\mu_3}}M_{\mu_1 ,\mu_3}^\beta \partial_1 g^{(2),1}_{\beta,\mu_2} \left( \frac{v_{\gamma_1 ,\gamma_2}}{v_{\gamma_1, \gamma_3}}z';t- \frac{z'}{v_{\gamma_1, \gamma_3}}\right) \\
		&+\frac{1}{2}R^{-t}_{{\theta_1},{\gamma_1}} R^{-t}_{{\theta_2},{\gamma_2}}R^{-t}_{{\theta_3},\gamma_3} \frac{1}{v_{\gamma_2, \gamma_3}}\left[\text{sgn}(z')+ \text{sgn}(v_{\gamma_2,\gamma_3})\right] R^t_{{\gamma_1},{\mu_1}}R^t_{{\gamma_2},{\mu_2}}R^t_{{\gamma_3},{\mu_3}} M_{\mu_2, \mu_3}^\beta \partial_1 g^{(2),1}_{\beta,\mu_1} \left( \frac{v_{\gamma_2 ,\gamma_1}}{v_{\gamma_2, \gamma_3}}z';t- \frac{z'}{v_{\gamma_2, \gamma_3}}\right)\,.
	\end{aligned}
\end{equation}
Here, we can already observe that the expression has two different kind of terms: the first one on RHS is simpler, and will generate the {\em cross} part of Landau equation, while the second and third lines will be part of the {\em self} contributions. 
Using the expression for the $M$ tensor introduced in the main text, the effective forcing term appearing in RHS of Eq.~\eqref{eq: evolution_for_g2(2)_SM} reads
\begin{equation}
	\label{eq: forcing_term_g2_1}
	\begin{aligned}
		\frac{1}{2}\partial_z \Big(&H_{\theta_1}^{\gamma_1,\gamma_2} g^{(3),\rm reg}_{\gamma_1, \gamma_2 ,\theta_2 }(0^+,z)-H_{\theta_2}^{\gamma_1 ,\gamma_2} g^{(3),\rm reg}_{\theta_1,\gamma_1 ,\gamma_2 }(z,0^+)\Big) =\\
		\frac{1}{4}\partial_z \Bigg(&-2\mathfrak{D}_{\theta_1,\beta} \partial_zg^{(2),1}_{\beta,\theta_2}(z)- 2\mathfrak{D}_{\theta_2,\beta} \partial_zg^{(2),1}_{\theta_1,\beta}(z)+\\
		&+ R^{-t}_{\theta_1,\eta}R_{\theta_2,\nu_1}  ^{-t}(\delta_{\sigma, \nu_2}\delta_{\eta, \nu_3}+\delta_{\eta, \nu_2}\delta_{\sigma, \nu_3})(\delta_{\nu_1, \zeta_1}\delta_{\nu_2, \zeta_2}+\delta_{\nu_2, \zeta_1}\delta_{\nu_1, \zeta_2})(\text{sgn}(z)+\text{sgn}(v_{\nu_1, \nu_2}))\times\\
		& \times\frac{T^{\rm dr}_{\sigma, \eta} T^{\rm dr}_{\zeta_1, \zeta_2}}{\rho^{\rm t}_\eta \rho^{\rm t}_{\zeta_1}} \frac{v_{\sigma, \eta} v_{{\zeta_1, \zeta_2}}}{v_{\nu_1, \nu_2}} \rho_{\zeta_2} f_{\zeta_2}R^t_{\nu_3, \mu_3} R^t_{\zeta_1, \beta} \frac{v_{\nu_1,\nu_2}}{v_{\nu_2, \nu_3}}\partial_z g^{(2),1}_{\beta,\mu_3} \Big( \frac{v_{\nu_2, \nu_3}}{v_{\nu_1,\nu_2}} z; t- \frac{z}{v_{\nu_1,\nu_2}} \Big)+\\
		&+ R^{-t}_{\theta_1,\eta}R_{\theta_2,\nu_1}  ^{-t}(\delta_{\sigma, \nu_2}\delta_{\eta, \nu_3}+\delta_{\eta, \nu_2}\delta_{\sigma, \nu_3})(\delta_{\nu_1, \zeta_1}\delta_{\nu_3, \zeta_2}+\delta_{\nu_3, \zeta_1}\delta_{\nu_1, \zeta_2})(\text{sgn}(z)+\text{sgn}(v_{\nu_1, \nu_3}))\times\\
		& \times\frac{T^{\rm dr}_{\sigma, \eta} T^{\rm dr}_{\zeta_1, \zeta_2}}{\rho^{\rm t}_\eta \rho^{\rm t}_{\zeta_1}} \frac{v_{\sigma, \eta} v_{{\zeta_1 ,\zeta_2}}}{v_{\nu_1, \nu_3}} \rho_{\zeta_2} f_{\zeta_2}R^t_{\nu_2, \mu_2} R^t_{\zeta_1, \beta} \frac{v_{\nu_1,\nu_3}}{v_{\nu_3, \nu_2}}\partial_z g^{(2),1}_{\beta,\mu_2} \Big( \frac{v_{\nu_3, \nu_2}}{v_{\nu_1,\nu_3}} z; t- \frac{z}{v_{\nu_1,\nu_3}} \Big)+  \\
		&- R^{-t}_{\theta_2, \eta}R_{\theta_1,\nu_3}^{-t}(\delta_{\sigma, \nu_1}\delta_{\eta, ,\nu_2}+\delta_{\eta, \nu_1}\delta_{\sigma, \nu_2})(\delta_{\nu_1, \zeta_1}\delta_{\nu_3, \zeta_2}+\delta_{\nu_3, \zeta_1}\delta_{\nu_1, \zeta_2})(\text{sgn}(z)+\text{sgn}(v_{\nu_1, \nu_3}))\times\\
		& \times\frac{T^{\rm dr}_{\sigma, \eta} T^{\rm dr}_{\zeta_1, \zeta_2}}{\rho^{\rm t}_\eta \rho^{\rm t}_{\zeta_1}} \frac{v_{\sigma, \eta} v_{{\zeta_1, \zeta_2}}}{v_{\nu_1, \nu_3}} \rho_{\zeta_2} f_{\zeta_2}R^t_{\nu_2, \mu_2} R^t_{\zeta_1, \beta}\frac{v_{\nu_1,\nu_3}}{v_{\nu_1, \nu_2}}\partial_z g^{(2),1}_{\beta,\mu_2} \Big( \frac{v_{\nu_1, \nu_2}}{v_{\nu_1,\nu_3}} z; t- \frac{z}{v_{\nu_1,\nu_3}} \Big)+ \\ 
		&- R^{-t}_{\theta_2,\eta}R_{\theta_1, \nu_3}  ^{-t}(\delta_{\sigma, \nu_1}\delta_{\eta, \nu_2}+\delta_{\eta, \nu_1}\delta_{\sigma, \nu_2})(\delta_{\nu_2, \zeta_1}\delta_{\nu_3 ,\zeta_2}+\delta_{\nu_3, \zeta_1}\delta_{\nu_2, \zeta_2})(\text{sgn}(z)+\text{sgn}(v_{\nu_2, \nu_3}))\times\\
		& \times\frac{T^{\rm dr}_{\sigma, \eta} T^{\rm dr}_{\zeta_1, \zeta_2}}{\rho^{\rm t}_\eta \rho^{\rm t}_{\zeta_1}} \frac{v_{\sigma, \eta} v_{{\zeta_1, \zeta_2}}}{v_{\nu_2, \nu_3}} \rho_{\zeta_2} f_{\zeta_2}R^t_{\nu_1, \mu_1} R^t_{\zeta_1, \beta} \frac{v_{\nu_2,\nu_3}}{v_{\nu_2, \nu_1}}\partial_z g^{(2),1}_{\beta,\mu_1} \Big( \frac{v_{\nu_2, \nu_1}}{v_{\nu_2,\nu_3}} z; t- \frac{z}{v_{\nu_2,\nu_3}} \Big) \bigg)\,.
	\end{aligned}
\end{equation}
In particular, we used the crucial relation
\begin{equation}
	\label{eq: kubo_diff_matrix_def}    H_{\theta_1}^{\eta_1,\eta_2}R^{-t}_{{\eta_1},{\gamma_1}} R^{-t}_{{\eta_2},{\gamma_2}}\frac{1}{|v_{\gamma_1, \gamma_2}|} R^t_{{\gamma_1},{\mu_1}}R^t_{{\gamma_2},{\mu_2}}M_{\mu_1, \mu_2}^\beta =2(R^{-t}_{\theta_1,\mu_1}-R^{-t}_{\theta_1,\mu_2}) (T^{\rm dr}_{\mu_2,\mu_1})^2\frac{\rho_{\mu_2} f_{\mu_2}}{(\rho^{\rm t}_{\mu_1})^2}|v_{\mu_2, \mu_1}| R^t_{\mu_1,\beta} = \mathfrak{D}_{\theta}^{\beta},
\end{equation}
with $\mathfrak{D}$ being the linear response diffusion matrix in integrable systems \cite{de2018hydrodynamic,de2019diffusion}, see \cite{Urilyon2025} for an explicit derivation.
The next step is to perform the large time limit $t\to\infty$ and use the solution \eqref{eq:g21solution_2SM} into~\eqref{eq: forcing_term_g2_1}
\begin{equation}
	\label{eq: forcing_term_g2_2}
	\begin{aligned}
		-\mathcal{S}^{(2),2}_{\theta_1,\theta_2}(z)&\equiv\lim_{t\to\infty}\frac{1}{2}\partial_z \Big(H_{\theta_1}^{\gamma_1,\gamma_2} g^{(3),\rm reg}_{\gamma_1, \gamma_2, \theta_2 }(0^+,z)-H_{\theta_2}^{\gamma_1, \gamma_2} g^{(3),\rm reg}_{\theta_1,\gamma_1, \gamma_2 }(z,0^+)\Big)=\\
		& -\frac{1}{2}V_0\partial^2_z\varphi(z)(\mathfrak{D}_{\theta_1,\beta} R_{\beta \alpha}^{-t}R_{\theta_2 ,\gamma}^{-t}+\mathfrak{D}_{\theta_2,\beta}R_{\theta_1, \alpha}^{-t}R_{\beta ,\gamma}^{-t})\frac{1}{v_{\alpha, \gamma}} (R_{\alpha, \eta}^t \partial_\eta \rho_\eta\rho^{\rm t}_\gamma \rho_\gamma f_\gamma -\rho^{\rm t}_\alpha \rho_\alpha f_\alpha R^t_{\gamma, \eta} \partial_\eta \rho_\eta)+\\ 
		&+\frac{1}{4} R^{-t}_{\theta_1,\eta}R_{\theta_2,\nu_1}  ^{-t}(\delta_{\sigma, \nu_2}\delta_{\eta, \nu_3}+\delta_{\eta ,\nu_2}\delta_{\sigma ,\nu_3})(\delta_{\nu_1 ,\zeta_1}\delta_{\nu_2, \xi_2}+\delta_{\nu_2, \zeta_1}\delta_{\nu_1, \xi_2})(\text{sgn}(z)+\text{sgn}(v_{\nu_1, \nu_2})) \times\\
		& \times\frac{T^{\rm dr}_{\sigma, \eta} T^{\rm dr}_{\zeta_1, \xi_2}}{\rho^{\rm t}_\eta \rho^{\rm t}_{\zeta_1}} \frac{v_{\sigma, \eta} v_{{\zeta_1 ,\xi_2}}}{v_{\nu_1, \nu_2}}\frac{v_{\nu_1,\nu_2}}{v_{\nu_2, \nu_3}} \rho_{\xi_2} f_{\xi_2}V_0\partial^2_z\varphi \left( \frac{v_{\nu_2 ,\nu_3}}{v_{\nu_1,\nu_2}}z \right) \frac{1}{v_{\zeta_1, \nu_3}}(R_{\zeta_1, \gamma}^t \partial_\gamma \rho_\gamma \rho^{\rm t}_{\nu_3} \rho_{\nu_3} f_{\nu_3}-\rho^{\rm t}_{\zeta_1} \rho_{\zeta_1} f_{\zeta_1} R^t_{\nu_3, \gamma} \partial_\gamma \rho_\gamma)+\\
		&+ \frac{1}{4} R^{-t}_{\theta_1,\eta}R_{\theta_2,\nu_1}  ^{-t}(\delta_{\sigma ,\nu_2}\delta_{\eta ,\nu_3}+\delta_{\eta, \nu_2}\delta_{\sigma, \nu_3})(\delta_{\nu_1 ,\zeta_1}\delta_{\nu_3, \xi_2}+\delta_{\nu_3, \zeta_1}\delta_{\nu_1, \xi_2})(\text{sgn}(z)+\text{sgn}(v_{\nu_1, \nu_3})) \times\\
		& \times\frac{T^{\rm dr}_{\sigma, \eta} T^{\rm dr}_{\zeta_1, \xi_2}}{\rho^{\rm t}_\eta \rho^{\rm t}_{\zeta_1}} \frac{v_{\sigma ,\eta} v_{{\zeta_1 ,\xi_2}}}{v_{\nu_1, \nu_3}}\frac{v_{\nu_1,\nu_3}}{v_{\nu_3 ,\nu_2}} \rho_{\xi_2} f_{\xi_2} V_0\partial^2_z\varphi \left( \frac{v_{\nu_3, \nu_2}}{v_{\nu_1,\nu_3}}z \right)\frac{1}{v_{\zeta_1, \nu_2}}(R_{\zeta_1, \gamma}^t \partial_\gamma \rho_\gamma \rho^{\rm t}_{\nu_2} \rho_{\nu_2} f_{\nu_2}-\rho^{\rm t}_{\zeta_1} \rho_{\zeta_1} f_{\zeta_1} R^t_{\nu_2, \gamma} \partial_\gamma \rho_\gamma)+  \\
		&- \frac{1}{4} R^{-t}_{\theta_2, \eta}R_{\theta_1,\nu_3}^{-t}(\delta_{\sigma, \nu_1}\delta_{\eta, \nu_2}+\delta_{\eta, \nu_1}\delta_{\sigma, \nu_2})(\delta_{\nu_1, \zeta_1}\delta_{\nu_3, \xi_2}+\delta_{\nu_3, \zeta_1}\delta_{\nu_1, \xi_2})(\text{sgn}(z)+\text{sgn}(v_{\nu_1 ,\nu_3})) \times\\
		& \times\frac{T^{\rm dr}_{\sigma, \eta} T^{\rm dr}_{\zeta_1, \xi_2}}{\rho^{\rm t}_\eta \rho^{\rm t}_{\zeta_1}} \frac{v_{\sigma, \eta} v_{{\zeta_1, \xi_2}}}{v_{\nu_1 ,\nu_3}}\frac{v_{\nu_1,\nu_3}}{v_{\nu_1, \nu_2}} \rho_{\xi_2} f_{\xi_2}V_0\partial^2_z\varphi \left( \frac{v_{\nu_1, \nu_2}}{v_{\nu_1,\nu_3}} z \right) \frac{1}{v_{\zeta_1, \nu_2}}(R_{\zeta_1, \gamma}^t \partial_\gamma \rho_\gamma \rho^{\rm t}_{\nu_2} \rho_{\nu_2} f_{\nu_2}-\rho^{\rm t}_{\zeta_1} \rho_{\zeta_1} f_{\zeta_1} R^t_{\nu_2, \gamma} \partial_\gamma \rho_\gamma) + \\ 
		&- \frac{1}{4} R^{-t}_{\theta_2,\eta}R_{\theta_1, \nu_3}  ^{-t}(\delta_{\sigma ,\nu_1}\delta_{\eta, \nu_2}+\delta_{\eta, \nu_1}\delta_{\sigma, \nu_2})(\delta_{\nu_2, \zeta_1}\delta_{\nu_3, \xi_2}+\delta_{\nu_3, \zeta_1}\delta_{\nu_2, \xi_2})(\text{sgn}(z)+\text{sgn}(v_{\nu_2, \nu_3})) \times\\
		& \times\frac{T^{\rm dr}_{\sigma, \eta} T^{\rm dr}_{\zeta_1, \xi_2}}{\rho^{\rm t}_\eta \rho^{\rm t}_{\zeta_1}} \frac{v_{\sigma, \eta} v_{{\zeta_1, \xi_2}}}{v_{\nu_2, \nu_3}} \frac{v_{\nu_2,\nu_3}}{v_{\nu_2, \nu_1}}\rho_{\xi_2} f_{\xi_2}V_0\partial^2_z\varphi\left( \frac{v_{\nu_2, \nu_1}}{v_{\nu_2,\nu_3}} z \right)  \frac{1}{v_{\zeta_1, \nu_1}}(R_{\zeta_1, \gamma}^t \partial_\gamma \rho_\gamma \rho^{\rm t}_{\nu_1} \rho_{\nu_1} f_{\nu_1}-\rho^{\rm t}_{\zeta_1} \rho_{\zeta_1} f_{\zeta_1} R^t_{\nu_1, \gamma} \partial_\gamma \rho_\gamma)\,.
	\end{aligned}
\end{equation}
In the last step we used the relation for susceptibility matrix
\begin{equation}
	\int\dd{\theta_2}C_{\theta_1,\theta_2}=2\pi R^{-t}_{\theta_1,\gamma}\rho_{\gamma}f_{\gamma}\rho^{\rm t}_{\gamma}\,.
\end{equation}
We now proceed by integrating Eq.~\eqref{eq:BBGKY_quasi_g3_SM} in time using the same Green's function technique as used before 
\begin{equation}\label{eq:g21solution_2}
	\begin{aligned}
		&\lim_{t\to\infty}g^{(2),2}_{\theta_1, \theta_2}[\rho](z,t) =R_{\theta_1,\sigma_1}^{-t} R_{\theta_2,\sigma_2}^{-t}  \int_0^\infty d \tau  \int \frac{dk}{2 \pi} \int {\dd}z' e^{ik(z+z'-v_{\sigma_1, \sigma_2}\tau)}R_{\sigma_1,\eta_1}^t R_{\sigma_2,\eta_2}^t \mathcal{S}^{(2),2}_{\eta_1 \eta_2}[\rho](z',t)\,,
	\end{aligned}
\end{equation}
where $\mathcal{S}^{(2),2}_{\eta_1, \eta_2}[\rho]$ is defined in~\eqref{eq: forcing_term_g2_2}, and where, we again used the time scale separation assumption as $\mathcal{S}^{(2),2}_{\eta_1, \eta_2}[\rho](z,t-\tau)\approx\mathcal{S}^{(2),2}_{\eta_1, \eta_2}[\rho](z,t)$.
Under this assumption, we can perform the time integral, finding 
\begin{equation}\label{eq:g21solution_2_2}
	\begin{aligned}
		&\lim_{t\to\infty}g^{(2),2}_{\theta_1, \theta_2}[\rho](z,t) =R_{\theta_1,\sigma_1}^{-t} R_{\theta_2,\sigma_2}^{-t}\frac{1}{v_{\sigma_1,\sigma_2}}R_{\sigma_1,\eta_1}^t R_{\sigma_2,\eta_2}^t \mathcal{S}^{(2),2}_{\eta_1, \eta_2}[\rho](z,t)\,.
	\end{aligned}
\end{equation}
The final step of the derivation consists in plugging the solution~\eqref{eq:g21solution_2_2} into the quasiparticle density evolution equation~\eqref{eq:BBGKY_quasi_rho_SM}, together with the forcing tensor definition~\eqref{eq: forcing_term_g2_2}
\begin{equation}\label{eq:BBGKY_quasi_rho_SM_2}
	\partial_{t}\rho_{\theta_1}=-\frac{1}{\xi^2}\int\dd{k}\dd{z}\dd{\theta_2}V_0\hat{\varphi}(k)e^{-ikz}\partial_{\theta_1}R_{\theta_1,\sigma_1}^{-t} R_{\theta_2,\sigma_2}^{-t}\frac{1}{v_{\sigma_1,\sigma_2}}R_{\sigma_1,\eta_1}^t R_{\sigma_2,\eta_2}^t \mathcal{S}^{(2),2}_{\eta_1, \eta_2}[\rho](z)=\frac{1}{\xi^2}\partial_{\theta_1}R_{\theta_1,\sigma_1}^{-t}(\mathcal{I}^{\rm cross}_{\sigma_1}+ \mathcal{I}^{\rm self}_{\sigma_1})\,.
\end{equation}
In the last step we also used that, as shown in the main text, the contribution from $g^{(2),1}$ in Eq.~\eqref{eq:BBGKY_quasi_rho_SM} is vanishing. In particular, Eq.~\eqref{eq:BBGKY_quasi_rho_SM_2} already represent a kinetic equation, that is correctly depending only on quasiparticle occupation function. In the next steps, we simply leverage algebraic manipulations in order to recast it in a more explicit form.
Firstly, we observe that Eq.~\eqref{eq:BBGKY_quasi_rho_SM_2} has two different type of contributions, that we call cross and self. Cross terms involve the diffusion matrix $\mathfrak{D}$ and are proportional to the square product $T^{\rm dr}_{\alpha,\beta}T^{\rm dr}_{\alpha,\beta}$. Meanwhile, the self terms only show product between dressed scattering shift $T^{\rm dr}_{\alpha,\beta}T^{\rm dr}_{\alpha,\gamma}$. Firstly we compute the cross part the scattering integral
\begin{multline}\label{eq:BBGKY_quasi_rho_SM_2}
	\mathcal{I}^{\rm cross}_{\theta_1}=\int\dd{k}k^2V_0^2\hat{\varphi}^2(z)\int\dd{\theta_2} \frac{\rho^{\rm t}_{\theta_2}}{v_{\theta_1,\theta_2}}R_{\theta_1,\eta_1}^tR_{\theta_2,\eta_2}^t 
	(\mathfrak{D}_{\eta_1,\beta} R_{\beta, \alpha}^{-t}R_{\eta_2, \gamma}^{-t}+\mathfrak{D}_{\eta_2,\beta}R_{\eta_1, \alpha}^{-t}R_{\beta, \gamma}^{-t})\times\\\times\frac{1}{v_{\alpha \gamma}} (R_{\alpha, \eta}^t \partial_\eta \rho_\eta\rho^{\rm t}_\gamma \rho_\gamma f_\gamma -\rho^{\rm t}_\alpha \rho_\alpha f_\alpha R^t_{\gamma, \eta} \partial_\eta \rho_\eta)\,.
\end{multline}
We now use the explicit formula for diffusion matrix expressed in~\eqref{eq: kubo_diff_matrix_def} to rewrite the term
\begin{equation}
	\label{eq:BBGKY_quasi_rho_SM_2}
	\begin{split}
		\mathcal{I}^{\rm cross}_{\theta_1}&=2\pi\int\dd{k}k^2\hat{V}^2(z)\int
		(\delta_{\theta_1,\mu_1}-\delta_{\theta_1,\mu_2}) (T^{\rm dr}_{\mu_2,\mu_1})^2\frac{\rho_{\mu_2} f_{\mu_2}\rho^{\rm t}_{\theta_2} }{(\rho^{\rm t}_{\mu_1})^2}\frac{|v_{\mu_2, \mu_1}|}{v_{\mu_1, \theta_2}v_{\theta_1,\theta_2}} (R_{\mu_1, \eta}^t \partial_\eta \rho_\eta\rho^{\rm t}_{\theta_2} \rho_{\theta_2} f_{\theta_2} -\rho^{\rm t}_{\mu_1} \rho_{\mu_1} f_{\mu_1} R^t_{\theta_2, \eta} \partial_\eta \rho_\eta)
		\\&+
		2\pi\int\dd{k}k^2\hat{V}^2(z)\int 
		(\delta_{\theta_2,\mu_1}-\delta_{\theta_2,\mu_2}) (T^{\rm dr}_{\mu_2,\mu_1})^2
		\frac{\rho_{\mu_2} f_{\mu_2}\rho^{\rm t}_{\theta_2}}{(\rho^{\rm t}_{\mu_1})^2} \frac{|v_{\mu_2, \mu_1}|}{v_{\theta_1, \mu_1}v_{\theta_1,\theta_2}} (R_{\theta_1, \eta}^t \partial_\eta \rho_\eta\rho^{\rm t}_{\mu_1} \rho_{\mu_1} f_{\mu_1} -\rho^{\rm t}_{\theta_1} \rho_{\theta_1} f_{\theta_1} R^t_{\mu_1 ,\eta} \partial_\eta \rho_\eta)\,,
	\end{split}
\end{equation}
where we used the definition of inverse rotation matrix as $R^tR^{-t}=1$. The last expression can be finally expressed as 
\begin{equation}\label{eq:Idiagonal_SM}
	\mathcal{I}^{\rm cross}_{\theta_1}= 2\pi\int {\dd}k k^2 \hat{V}^2(k)  \int {\dd} \sigma d \zeta   \left[\rho^{\rm t}_{\theta_1}(T^{\rm dr}_{\sigma, \zeta})^2 \frac{|v_{\sigma, \zeta}|}{v_{\theta_1, \sigma} v_{\zeta, \theta_1}^2}-\frac{(\rho^{\rm t}_\sigma)^2}{\rho^{\rm t}_{\theta_1}} (T^{\rm dr}_{\zeta, \theta})^2 \frac{|v_{\zeta, \theta_1}|}{v_{\sigma, \zeta} v_{ \theta_1, \sigma}^2} \right]\mathcal{B}(\theta_1,\zeta,\sigma),
\end{equation}
with the definition
\begin{equation}
	\mathcal{B}(\theta,\zeta,\sigma)\equiv\frac{v_{\sigma, \zeta}}{\rho^{\rm t}_{\theta}} R^t_{\theta, \eta} \partial_\eta \rho_\eta \rho_{\sigma} f_{\sigma} \rho_{\zeta} f_{\zeta} +\frac{v_{\zeta, \theta}}{\rho^{\rm t}_{\sigma}} \rho_{\theta} f_{\theta} R^t_{\sigma, \eta} \partial_\eta \rho_\eta \rho_{\zeta} f_{\zeta} + \frac{v_{\theta, \sigma}}{\rho^{\rm t}_{\zeta}} \rho_{\theta} f_{\theta} \rho_{\sigma} f_{\sigma}R^t_{\zeta, \eta} \partial_\eta \rho_\eta.
\end{equation}
We now compute the self contributions to the scattering integral from Eq~\eqref{eq:BBGKY_quasi_rho_SM_2} and~\eqref{eq: forcing_term_g2_2}
\begin{equation}
	\begin{aligned}
		\mathcal{I}^{\rm self}_{\theta_1}&= -2\pi\int (\delta_{\sigma, \nu_2}\delta_{\theta_1, \nu_3}+\delta_{\theta_1, \nu_2}\delta_{\sigma, \nu_3})(\delta_{\nu_1, \zeta_1}\delta_{\nu_2, \xi_2}+\delta_{\nu_2, \zeta_1}\delta_{\nu_1, \xi_2})\frac{T^{\rm dr}_{\sigma, \theta_1} T^{\rm dr}_{\zeta_1, \xi_2}}{\rho^{\rm t}_{\theta_1} \rho^{\rm t}_{\zeta_1}} \frac{1}{v_{\theta_1,\nu_1}}\frac{v_{\sigma, \theta_1} v_{{\zeta_1 ,\xi_2}}}{|v_{\nu_1, \nu_2}|} \frac{|v_{\nu_1, \nu_2}|}{|v_{\nu_2,\nu_3}|v_{\zeta_1, \nu_3}}\frac{v_{\nu_1, \nu_2}}{v_{\nu_2,\nu_3}} \times\\
		&\times \rho^{\rm t}_{\nu_1}\rho_{\xi_2} f_{\xi_2}(R_{\zeta_1, \gamma}^t \partial_\gamma \rho_\gamma \rho^{\rm t}_{\nu_3} \rho_{\nu_3} f_{\nu_3}-\rho^{\rm t}_{\zeta_1} \rho_{\zeta_1} f_{\zeta_1} R^t_{\nu_3, \gamma} \partial_\gamma \rho_\gamma)\int \dd k k^2  V^2_0\hat{\varphi}(k)\hat{\varphi} \left( \frac{v_{\nu_1, \nu_2}}{v_{\nu_2,\nu_3}}k \right) +\\
		&+ 2\pi \int (\delta_{\sigma, \nu_1}\delta_{\eta ,\nu_2}+\delta_{\eta, \nu_1}\delta_{\sigma, \nu_2})(\delta_{\nu_1, \zeta_1}\delta_{\theta_1 ,\xi_2}+\delta_{\theta_1,\zeta_1}\delta_{\nu_1 ,\xi_2})\frac{T^{\rm dr}_{\sigma, \eta} T^{\rm dr}_{\zeta_1, \xi_2}}{\rho^{\rm t}_\eta \rho^{\rm t}_{\zeta_1}} \frac{1}{v_{\theta_1,\eta}}\frac{v_{\sigma, \eta} v_{{\zeta_1, \xi_2}}}{|v_{\nu_1, \theta_1}|} \frac{|v_{\nu_1, \theta_1}|}{|v_{\nu_1,\nu_2}|v_{\zeta_1, \nu_2}}\frac{v_{\nu_1, \theta_1}}{v_{\nu_1,\nu_2}} \times\\
		&\times \rho^{\rm t}_{\eta}\rho_{\xi_2} f_{\xi_2}(R_{\zeta_1, \gamma}^t \partial_\gamma \rho_\gamma \rho^{\rm t}_{\nu_2} \rho_{\nu_2} f_{\nu_2}-\rho^{\rm t}_{\zeta_1} \rho_{\zeta_1} f_{\zeta_1} R^t_{\nu_2, \gamma} \partial_\gamma \rho_\gamma) \int \dd k k^2  V_0^2\hat{\varphi}(k)\hat{\varphi} \left( \frac{v_{\nu_1 ,\theta_1}}{v_{\nu_1,\nu_2}}k \right)\,.
	\end{aligned}
\end{equation}
In order to write the former formula we used the Fourier transform property $FT[f(ax)](k)=(1/|a|)FT[f(x)](k/a)$.
This expression is composed by two different types of 
contributions: proportional to $T^{\rm dr}_{\theta_1,\bullet}T^{\rm dr}_{\theta_!,\bullet}$
or $T^{\rm dr}_{\theta_1,\bullet}T^{\rm dr}_{\bullet,\bullet}$.
More explicitly they can be written as 
\begin{equation}\label{eq:Ioffdiagonal_SM}
	\begin{aligned}
		&\mathcal{I}^{\rm self}_{\theta_1}=  2\pi\int  d \zeta d \sigma \int {\dd}k k^2 V^2_0\hat{\varphi}(k)\rho^{\rm t}_{\sigma}T^{\rm dr}_{\theta_1 \zeta} \frac{v_{\zeta, \theta_1}}{v_{\theta_,1\sigma}} \left[\hat{\varphi}\left(\frac{v_{\zeta, \theta_1 }}{v_{ \theta_1, \sigma}}k \right)\frac{\rho^{\rm t}_{\zeta} }{\rho^{\rm t}_{\theta_1}} T^{\rm dr}_{\sigma, \theta_1} \frac{1}{|v_{\theta_1, \sigma}|v_{\sigma, \zeta }}-2\hat{\varphi}\left(\frac{v_{\zeta, \theta_1 }}{v_{  \sigma ,\zeta }}k \right) T^{\rm dr}_{\zeta, \sigma} \frac{1}{v_{\theta_1, \sigma} |v_{\zeta ,\sigma}|}\right]  \mathcal{B}(\theta_1,\zeta,\sigma).
	\end{aligned}
\end{equation}
This concludes the derivation of the generalised Landau equation.

\section{Effective diffusion in the small time limit for two-point functions}
In this section we show how, in the small time limit $t\to0^+$, i.e., close to a GGE state, the $1/\xi$ correction to the convective spreading of two-point function is diffusive. We also show that the diffusion matrix is the one associated with the local equilibrium state around
which we are linearising.
For the sake of simplicity, we do this derivation in the small $V_0$ case, i.e. when the solution evolution equation for three-point function is given by~\eqref{eq: integrated_g3_1}, otherwise we conjecture that the result is valid beyond this approximation.
Hence, we consider the limit $t\to0^+$ of the three-point function solution~\eqref{eq: t_evolution_g_reg}
\begin{equation}
	\label{eq: t_evolution_g_reg_t0}
	\begin{aligned}
		&\lim_{t\to0^+}g^{(3), \text{reg}}_{\theta_1, \theta_2, \theta_3}(0^+,z';0^+) = \frac{1}{2}R^{-t}_{{\theta_1},{\gamma_1}} R^{-t}_{{\theta_2},{\gamma_2}}\frac{1}{|v_{\gamma_1, \gamma_2}|} R^t_{{\gamma_1},{\mu_1}}R^t_{{\gamma_2},{\mu_2}}M_{\mu_1, \mu_2}^\beta \partial_1g^{(2)}_{\beta,\theta_3} (
		-z';0^+) \\
		&+\frac{1}{2}R^{-t}_{{\theta_1},{\gamma_1}} R^{-t}_{{\theta_2},{\gamma_2}}R^{-t}_{{\theta_3},\gamma_3}\frac{1}{v_{\gamma_1 ,\gamma_3}}\left[\text{sgn}(z')+\text{sgn}(v_{\gamma_1,\gamma_3})\right]R^t_{{\gamma_1},{\mu_1}}R^t_{{\gamma_2},{\mu_2}}R^t_{{\gamma_3},{\mu_3}}M_{\mu_1 ,\mu_3}^\beta \partial_1 g^{(2)}_{\beta,\mu_2} \left( \frac{v_{\gamma_1 ,\gamma_2}}{v_{\gamma_1, \gamma_3}}z';0^+- \frac{z'}{v_{\gamma_1, \gamma_3}}\right) \\
		&+\frac{1}{2}R^{-t}_{{\theta_1},{\gamma_1}} R^{-t}_{{\theta_2},{\gamma_2}}R^{-t}_{{\theta_3},\gamma_3} \frac{1}{v_{\gamma_2, \gamma_3}}\left[\text{sgn}(z')+ \text{sgn}(v_{\gamma_2,\gamma_3})\right] R^t_{{\gamma_1},{\mu_1}}R^t_{{\gamma_2},{\mu_2}}R^t_{{\gamma_3},{\mu_3}} M_{\mu_2, \mu_3}^\beta \partial_1 g^{(2)}_{\beta,\mu_1} \left( \frac{v_{\gamma_2 ,\gamma_1}}{v_{\gamma_2, \gamma_3}}z';0^+- \frac{z'}{v_{\gamma_2, \gamma_3}}\right)\,.
	\end{aligned}
\end{equation}
Here, we stress that, at $t=0^+$, few point correlations in the system are not identically null, yet small. Crucially, imposing the positivity of time dependence in the argument of the two-point function in the last two lines, we observe that they are vanishing. More precisely, considering the second line, we have $z'/v_{\gamma_1,\gamma_3}<0$. This imposes the identity $\text{sgn}(z')+ \text{sgn}(v_{\gamma_1,\gamma_3})=0$. For analogous reasons, the third line is also vanishing.
Hence, we find 
\begin{equation}
	\label{eq: t_evolution_g_reg_t0_fin}
	\lim_{t\to0^+}g^{(3), \text{reg}}_{\theta_1, \theta_2, \theta_3}(x_1^+,x_1^-,x_2;0^+) = -\frac{1}{2}R^{-t}_{{\theta_1},{\gamma_1}} R^{-t}_{{\theta_2},{\gamma_2}}\frac{1}{|v_{\gamma_1, \gamma_2}|} R^t_{{\gamma_1},{\mu_1}}R^t_{{\gamma_2},{\mu_2}}M_{\mu_1, \mu_2}^\beta \partial_{x_1}g^{(2)}_{\beta,\theta_3} (
	x_1,x_2;0^+)\,,
\end{equation}
where we have expressed the derivative in terms of the space coordinate.
Using this result in the equation for two-point function, we get the effective contribution
\begin{equation}
	\label{eq: small_time_diffusion_app}  
	\begin{split}
		\lim_{t\to0^+}\partial_{x_1}H_{\theta_1}^{\gamma,\eta} g^{(3),\rm reg}_{ \gamma,\eta,\theta_2}(x_1^+,x_1^-,x_2;t)&=-\frac{1}{2}\partial_{x_1}H_{\theta_1}^{\gamma,\eta}R^{-t}_{{\gamma},{\gamma_1}} R^{-t}_{{\eta},{\gamma_2}}\frac{1}{|v_{\gamma_1, \gamma_2}|} R^t_{{\gamma_1},{\mu_1}}R^t_{{\gamma_2},{\mu_2}}M_{\mu_1, \mu_2}^\beta \partial_{x_1}g^{(2)}_{\beta,\theta_2} (
		x_1,x_2;0^+)
		\\&
		=-\frac{1}{2}\partial_{x_1}\mathfrak{D}_{\theta_1}^{\gamma}\partial_{x_1}g^{(2)}_{\gamma,\theta_2} (
		x_1,x_2;0^+)
	\end{split}
\end{equation}
where the identity used in the second line has been proven in \cite{Urilyon2025}.

\section{gBBGKY hierarchy for a system of many tubes}
In this section we extend the gBBGKY hierarchy introduced in the main text to the physically relevant case of a system composed by many one dimensional tubes interacting through long-range interactions. More precisely, we consider systems of $N_t$ tubes evolving according to the global Hamiltonian  
\begin{equation}
	\label{eq:HPerturb_Ntubes}
	\hat{H} = \sum_{t=1}^{N_t}\Big[\hat{H}_{a,t}^{\rm int} + \frac{1}{2} \sum_{t'=1}^{N_t}\int \!\!\dd{x}\,\dd{x'} \; V_{(t,t')}(x - x') \,\hat{q}^{(t)}_0(x)\,\hat{q}^{(t')}_0(x') \Big]\,,
\end{equation}
where $H_{a,t}$ is the unperturbed integrable Hamiltonian associated to the $t$-th tube, $\hat{q}^{(t)}_i$ its $i$-th charge density, and $V_{(t,t')}$ the interaction potential between $t$-th and $t'$-th tubes. In particular $V_{(t,t)}$ represent the intra tube interaction.
Importantly, charge density operators associated to different tubes always commute 
\begin{equation}
	[\hat{q}^{(t)}_i,\hat{q}^{(t')}_j]=0 \quad{\rm for\,\, any}\quad t'\neq t\,.
\end{equation}
In this section we repeat the main steps for the derivation of the hierarchy in analogy with single tube case.
The first step is to compute the quantum generalisation of the classical Liouville's equation 
\begin{equation} 
	\begin{aligned}
		&\partial_t \hat q^{(t)}_i(x)  +\partial_x  \hat j^{(t)}_i(x)  =-\sum_{t'=1}^{N_t}\int {\rm d}x' V'_{(t,t')}(x-x')\hat j^{(t)}_{i,0}(x)\hat q^{(t')}_0(x') ,
	\end{aligned}
	\label{eq:chargedyn_Ntubes}
\end{equation}
where we have $i[\hat{H}_{a,t}^{\rm int},\hat q^{(t)}_i]=\partial_x j^{(t)}_i$. It is already interesting to observe that the evolution equation is completely analogous to the single tube case.  The evolution equation for the expectation value of the product of single charges produces the hierarchical structure. For the two function it reads
\begin{equation}
	\begin{aligned}
		&\partial_t \langle q^{(t_1)}_{i_1}(x_1)q^{(t_2)}_{i_2}(x_2) \rangle = -\Big[\partial_{x_1} \langle j^{(t_1)}_{i_1}(x_1) q^{(t_2)}_{i_2}(x_2) \rangle-\sum_{t'=1}^{N_t}\int {\rm d}y V'_{(t_1,t')}(x_1-y)\langle j^{(t_1)}_{i_1,0}(x_1)q^{(t_2)}_{i_2}(x_2)q^{(t')}_0(y)  \rangle\Big]_{(1,2)}.
	\end{aligned}
\end{equation}
It is now important to specify that, because of the interactions between different tubes, the system will develop long-range correlations between them. Hence, in order to introduce them, we perform the average over the ensemble
\begin{equation}
	\begin{aligned}
		\label{eq: longrange_gge_state_Ntubes}
		\rho  = {\rm exp} \Big( &- \sum_{t_1=1}^{N_t}\int \dd{x} \beta_{(t_1)}^i(x) q^{(t_1)}_i(x) - \sum_{t_1,t_2=1}^{N_t}\int {\dd}\vec{x}  \beta_{(t_1,t_2)}^{i,j}(x_1,x_2) [\delta q^{(t_1)}_{i}(x_1)\delta q^{(t_2)}_{j}(x_2) ]^{\rm reg} \\&- \sum_{t_1,t_2,t_3=1}^{N_t}\int {\dd}\vec{x} \beta_{(t_1,t_2,t_3)}^{i,j,k}(x_1,x_2,x_3) [\delta q^{(t_1)}_{i}(x_1)\delta q^{(t_2)}_{j}(x_2)\delta q^{(t_3)}_{k}(x_3)]^{\rm reg}+ \ldots \Big)\,.
	\end{aligned}
\end{equation}
Here, $\beta^{i_1,\ldots,i_n}_{(t_1,\ldots,t_n)}$ is the Lagrange multiplier associated to the $n$- point function between tubes $t_1,\ldots,t_n$.
In addition, we observe that, since the tubes are spatially separated, the connected correlations between them do not exhibit any ultralocal component proportional to delta function. Hence, for example for two and three-pointcorrelator we have
\begin{equation}
	\label{eq: def_2pointfunc_delta_Ntubes}
	\langle q^{(t_1)}_{i_1}(x_1)q^{(t_2)}_{i_2}(x_2) \rangle ^c = \delta_{t_1,t_2}\delta(x_1-x_2)C^{(t_1)}_{i_1 i_2}(x_1)+g^{(t_1,t_2)}_{i_1i_2}(x_1,x_2)
\end{equation}
\begin{equation}
	\begin{split}
		\label{eq: def_3pointfunc_delta_Ntubes}
		\langle &q^{(t_1)}_{i_1}q^{(t_2)}_{i_2}q^{(t_3)}_{i_3}\rangle ^c = \delta_{t_1,t_2}\delta_{t_1,t_3}\delta(x_1-x_2)\delta(x_1-x_3)C^{(t_1)}_{i_1i_2i_3}+[\delta_{t_1,t_2}\delta(x_1-x_2)C^{(t_1)}_{i_1i_2j}C^{jk}_{(t_1)}g^{(t_1,t_3)}_{k,i_3}]_{(1,2,3)}+g^{(t_1,t_2,t_3)}_{i_1i_2i_3}\,.
	\end{split}
\end{equation}
Straightforwardly, if the correlations functions are computed in the same tube, this relations coincide with the ones already introduced in the main text. The state assumption~\eqref{eq: longrange_gge_state_Ntubes}, together with~\eqref{eq:chargedyn_Ntubes},~\eqref{eq: def_2pointfunc_delta_Ntubes} and~\eqref{eq: def_3pointfunc_delta_Ntubes} is enough to derive the full hierarchy of equations and express it in terms of quasiparticles occupations and correlations functions.
In order to avoid confusions, we stress that in this context we changed notation: the connected regular few point functions now, instead of an upper index showing the number of points, have a string of indexes indicating the string of tubes connected by the function.

\subsection{Example: Two tubes system}
In this section we consider the simple case of two coupled one dimensional gases evolving through the Hamiltonian 
\begin{equation}
	\label{eq:HPerturb_2tubes}
	\hat{H} = \hat{H}_{a,1}^{\rm int} +\hat{H}_{a,2}^{\rm int} + \int \dd{x}\,\dd{x'} \; V_{(1,2)}(x - x') \,\hat{q}^{(1)}_0(x)\,\hat{q}^{(2)}_0(x') \,.
\end{equation}
We observe that, differently from the single tube case, here the potential doesn't have the symmetric factor $1/2$ in front since the integration in $x$ and $x'$ in not double counting particles. Also, we stress that, in this specific example, the particles are not interacting within the same tube. The aim of this section is to write the first layers of the hierarchy and derive the associated Landau equation.
The single charge density time evolution reads
\begin{equation} 
	\begin{aligned}
		&\partial_t \hat q^{(1)}_i(x)  +\partial_x  \hat j^{(1)}_i(x)  =-\int {\rm d}x' V'_{(1,2)}(x-x')\hat j^{(1)}_{i,0}(x)\hat q^{(2)}_0(x') ,
	\end{aligned}
	\label{eq:chargedyn_2tubes}
\end{equation}
Taking the coarse grained ensemble average, for a homogeneous system, we find 
\begin{equation} 
	\begin{aligned}
		&\partial_t  q^{(1)}_i(x)  =-\int {\rm d}x' V'_{(1,2)}(x-x') A_{i,0}^{(1),k}\langle q^{(1)}_{k}(x) q^{(2)}_0(x') \rangle^c.
	\end{aligned}
	\label{eq:chargedyn_2tubes}
\end{equation}
Hence, the inter tube two-pointcorrelation evolves according to 
\begin{equation}
	\begin{split}
		\partial_t \langle{q}_{i_1}^{(1)}(x_1){q}_{i_2}^{(2)}(x_2)\rangle^c
		&+\Big[\partial_{x_1}\langle{j}_{i_1}^{(1)}(x_1){q}_{i_2}^{(2)}(x_2)\rangle^c\Big]_{(1,2)}=\\&=-\Big[\int\dd{y}_{(1,2)}'(x_1-y){j}_{i_1,0}^{(1)}(x_1)\langle{q}_0^{(2)}(y){q}_{i_2}^{(2)}(x_2)\rangle^c+\langle{j}_{i_1,0}^{(1)}(x_1){q}_0^{(2)}(y){q}_{i_2}^{(2)}(x_2)\rangle^c\Big]_{(1,2)}\,.
	\end{split}    
\end{equation}
We here observe that the dynamics of this object depends on the intra tube correlation function, that evolves according to a similar equation that, for the sake of brevity, will not be written here. 
% \begin{equation}
	% \begin{split}
		%     \partial_t \langle{q}_{i_1}^{(1)}(x_1){q}_{i_2}^{(1)}(x_2)\rangle^c
		%     &+\Big[\partial_{x_1}\langle{j}_{i_1}^{(1)}(x_1){q}_{i_2}^{(1)}(x_2)\rangle^c\Big]_{(1,2)}=\\&=-\int\dd{y}V_{\rm 1,2}'(x_1-y)\Big[\langle{j}_{i_1,0}^{(1)}(x_1)\rangle\langle{q}_0^{(2)}(y){q}_{i_2}^{(1)}(x_2)\rangle^c+\langle{j}_{i_1,0}^{(1)}(x_1){q}_0^{(2)}(y){q}_{i_2}^{(1)}(x_2)\rangle^c\Big]
		%     \\&-\int\dd{y}V_{\rm 1,2}'(x_2-y)\Big[\langle{j}_{i_2,0}^{(1)}(x_2)\rangle\langle{q}_0^{(2)}(y){q}_{i_1}^{(1)}(x_1)\rangle^c+\langle{j}_{i_2,0}^{(1)}(x_2){q}_0^{(2)}(y){q}_{i_1}^{(1)}(x_1)\rangle^c\Big]\,.
		% \end{split}    
	% \end{equation}
We now extract the singular part of the connected correlations and perform the space and time rescaling with $\xi$, finding 
\begin{equation} 
	\begin{aligned}
		&\partial_t  q^{(1)}_i(x)  =-\int {\rm d}x' V'_{(1,2)}(x-x') A_{i,0}^{(1),k}g^{(1,2)}_{k,0}(x,x').
	\end{aligned}
	\label{eq:chargedyn_2tubes_2}
\end{equation}

\begin{equation}
	\begin{split}
		\partial_tg^{(1,2)}_{i_1,i_2}(x_1,x_2)&+\Big[\partial_{x_1}A_{i_1}^{(1),k}g^{(1,2)}_{k,i_2}(x_1,x_2)+\frac{1}{2}\partial_{x_1} H_{i_1}^{(1),jk}g^{(1,1,2),\rm reg}_{jki_2}(x_1,x_1,x_2)\Big]_{(1,2)}=\\&=-\Big[V_{(1,2)}'(x_1-x_2)A^{(1),k}_{i_1,0}q_{k}^{(1)}C^{(2)}_{0i_2}+A^{(1),k}_{i_1,0}g^{(1,2)}_{km}(x_1,x_2)C^{(2),mn}C^{(2)}_{n ,0,i_2}+\\&-\int\dd{y}V_{(1,2)}'(x_1-y)A^{(1),k}_{i_1,0}q_{k}^{(1)}g^{(2,2)}_{0,i_2}(y,x_2)+A^{(1),k}_{i_1,0}g^{(1,2,2)}_{k,0,i_2}(x_1,y,x_2)\Big]_{(1,2)}\,.
	\end{split}    
\end{equation}
\begin{equation}
	\label{eq: evo_eq_explicit_deltas_3_2N}
	\begin{aligned}
		&\partial_t g^{(1,1,2)}_{i_1i_2i_3}(x_1,x_2,x_3) + \partial_{x_1} A_{i_1}^{(1),k}g^{(1,1,2)}_{ki_2i_3}+\partial_{x_2} A_{i_2}^{(1),k}g^{(1,1,2)}_{i_1ki_3}+\partial_{x_3} A_{i_3}^{(2),k}g^{(1,1,2)}_{i_1i_2k}=
		\\&
		=-\delta(x_1-x_2)\Big(  A_{i_1}^{(1),k}
		C^{(1)}_{k,i_2,m} C^{(1),mn}- 
		C^{(1)}_{i_1,i_2,k} C^{(1),km}A_m^{(1),n}+ H_{i_1}^{(1),mn}C_{m,i_2}^{(1)}\Big)\partial_{x_1}g^{(1,2)}_{n,i_3}(x_1,x_3)+{\rm h.o.t}\,.
	\end{aligned}
\end{equation}
Here, we crucially observe that, differently from what found for the single tube case, the term proportional to Dirac delta function of RHS of Eq.~\eqref{eq: evo_eq_explicit_deltas_3_2N} is not symmetric with respect to permutation of indexes, the consequence of this will be that absence of self contributions in the generalised Landau equation for two coupled tubes. We now proceed writing these equations in quasiparticle picture using the same machinery introduced in the main text, finding 
\begin{equation}\label{eq:BBGKY_quasi_rho_2T}
	\partial_{t}\rho^{(1)}_{\theta_1}=\int\dd{x_2}\dd{\theta_2}V'_{(1,2)}(x_1-x_2)\partial_{\theta_1}g^{(1,2)}_{\theta_1,\theta_2}(x_1,x_2)
\end{equation}
\begin{equation}
	\label{eq:BBGKY_quasi_g2_2T}
	\begin{split}
		\partial_t g^{(1,2)}_{\theta_1, \theta_2}(x_1,x_2)&+ \Big[\partial_{x_1}\big(A_{\theta_1}^{(1)\gamma} g^{(1,2)}_{\gamma, \theta_2}(x_1,x_2)\big) +\frac{1}{2}\partial_{x_1} \big(H_{\theta_1}^{(1),\gamma\gamma'} g^{(1,1,2),\rm reg}_{ \gamma,\gamma',\theta_2}(x_1^+,x_1^-,x_2)\big)\Big]_{(1,2)}=
		\\&
		=\Big[V'_{(1,2)}(x_1-x_2) \int \dd{\theta_3} \partial_{\theta_1}\Big( \rho^{(1)}_{\theta_1}C^{(2)}_{\theta_2\theta_3}+ C_{\theta_2,\theta_3}^{(1),\gamma}g^{(1,2)}_{\theta_1,\gamma}(x_1,x_2)\Big)
		\\&+
		\int {\dd}x_3 \dd{\theta_3} V''_{(1,2)}(x_1-x_3) \partial_{\theta_1}\Big( \rho^{(1)}_{ \theta_1}g^{(1,2)}_{\theta_2,\theta_3}(x_2,x_3)+ g^{(1,2,2)}_{\theta_1,\theta_2,\theta_3}(x_1,x_2,x_3)\Big)\Big]_{(1,2)}
	\end{split}
\end{equation}
\begin{equation}\label{eq:BBGKY_quasi_g3_2T}
	\begin{split}
		\partial_{t}g^{(1,1,2)}_{\theta_1,\theta_2,\theta_3}(x_1,x_2,x_3)+\partial_{x_1}A_{\theta_1}^{(1),\gamma}g^{(1,1,2)}_{\gamma\theta_2\theta_3}(x_1,x_2,x_3)&+\partial_{x_2}A_{\theta_2}^{(2),\gamma}g^{(1,1,2)}_{\theta_1\gamma\theta_3}(x_1,x_2,x_3)+\partial_{x_3}A_{\theta_3}^{(3),\gamma}g^{(1,1,2)}_{\theta_1,\theta_2\gamma}(x_1,x_2,x_3)=\\&=\delta(x_1-x_2)M_{\theta_1,\theta_2}^{\gamma}\partial_{x_1}g^{(1,2)}_{\gamma,\theta_3}(x_1,x_3)+{\rm h.o.t}\,.
	\end{split}
\end{equation}
We now proceed by computing the Landau equation in this specific setup. The main difference from the single tube case is given by the time integration of three-point function, that here reads
\begin{equation}
	\begin{aligned}
		&g^{(1,1,2), \text{reg}}_{\theta_1, \theta_2, \theta_3}(0^+,z';t) = -\frac{1}{2}R^{(1),-t}_{{\theta_1},{\gamma_1}} R^{(1),-t}_{{\theta_2},{\gamma_2}}\frac{1}{|v^{(1,1)}_{\gamma_1, \gamma_2}|} R^{(1),t}_{{\gamma_1},{\mu_1}}R^{(1),t}_{{\gamma_2},{\mu_2}}M_{\mu_1, \mu_2}^{(1),\beta} \partial_{z'}g^{(1,2),1}_{\beta,\theta_3} \left(
		-z';t\right)
	\end{aligned}
\end{equation}
where we introduced $v^{(1,1)}_{\gamma_1 \gamma_2}\equiv v^{(1)}_{\gamma_1}-v^{(1)}_{\gamma_2}$. As compared to~\eqref{eq: t_evolution_g_reg}, this solution is simpler, since it only contains contributions local in time.
The late time solution for $g^{(1,2),1}$ also reads
\begin{equation}\label{eq:g21solution_2SM_2T}
	\lim_{t\to\infty}g^{(1,2),1}_{\theta_1, \theta_2}(z,t) =V'_{(1,2)}(z')R_{\theta_1,\zeta_1}^{(1),-t} R_{\theta_2,\zeta_2}^{(2),-t}\frac{1}{v^{(1,2)}_{\zeta_1,\zeta_2}}R^{(1),t}_{\zeta_1,\eta_1} R^{(2),t}_{\zeta_2,\eta_2}  \Big(\partial_{\eta_1} \rho^{(1)}_{\eta_1} C^{(2)}_{\gamma,\eta_2}-C^{(1)}_{\eta_1,\gamma }\partial_{\eta_2} \rho^{(2)}_{\eta_2}\Big)\,.
\end{equation}
Now, the procedure to derive a scattering integral is exactly analogous to the one introduced in Sec.\ref{sec:Landau_Eq_SM}.
Defining the coupled kinetic equations as
\begin{equation}
	\partial_t \rho^{(1)}_{\theta} = \partial_{\theta}\mathcal{I}^{(1)}[\rho^{(1)},\rho^{(2)}]_{\theta}\quad,\quad \partial_t \rho^{(2)}_{\theta} = \partial_{\theta}\mathcal{I}^{(2)}[\rho^{(1)},\rho^{(2)}]_{\theta}\,,
\end{equation}
we write the scattering integral as
\begin{equation}
	\label{eq:BBGKY_quasi_rho_SM_2_2T}
	\begin{split}
		\mathcal{I}^{(1)}_{\theta_1}&=2\pi\int\dd{k}k^2\hat{V}^2_{(1,2)}(z)\times\\&\times\bigg[\int
		(\delta_{\theta_1,\mu_1}-\delta_{\theta_1,\mu_2}) (T^{(1)\rm dr}_{\mu_2,\mu_1})^2\frac{\rho^{(1)}_{\mu_2} f^{(1)}_{\mu_2}\rho^{(2),\rm tot}_{\zeta} }{(\rho^{(1)\rm tot}_{\mu_1})^2}\frac{|v^{(1,1)}_{\mu_2, \mu_1}|}{v^{(1,2)}_{\mu_1, \zeta}v^{(1,2)}_{\theta_1,\zeta}} (R_{\mu_1, \eta}^{(1),t} \partial_\eta \rho^{(1)}_\eta\rho^{(2),\rm tot}_{\zeta} \rho^{(2)}_{\zeta} f^{(2)}_{\zeta} -\rho^{(2),\rm tot}_{\mu_1} \rho^{(2)}_{\mu_1} f^{(2)}_{\mu_1} R^{(2),t}_{\zeta, \eta} \partial_\eta \rho^{(2)}_\eta)
		\\&+
		\int 
		(\delta_{\zeta,\mu_1}-\delta_{\zeta,\mu_2}) (T^{(2)\rm dr}_{\mu_2,\mu_1})^2
		\frac{\rho^{(2)}_{\mu_2} f^{(2)}_{\mu_2}\rho^{(2)\rm tot}_{\zeta}}{(\rho^{(2),\rm tot}_{\mu_1})^2} \frac{|v^{(2,2)}_{\mu_2, \mu_1}|}{v^{(1,2)}_{\theta_1, \mu_1}v^{(1,2)}_{\theta_1,\zeta}} (R_{\theta_1, \eta}^{(1),t} \partial_\eta \rho^{(1)}_\eta\rho^{(2),\rm tot}_{\mu_1} \rho^{(2)}_{\mu_1} f^{(2)}_{\mu_1} -\rho^{(1),\rm tot}_{\theta_1} \rho^{(1)}_{\theta_1} f^{(1)}_{\theta_1} R^{(2),t}_{\mu_1, \eta} \partial_\eta \rho^{(2)}_\eta)\bigg]\,.
	\end{split}
\end{equation}
Straightforwardly, the scattering integral $\mathcal{I}^{(2)}$ is equivalent up to exchange of tube indexes $1\longleftrightarrow2$. As already mentioned, we stress that in this case there is no cross terms proportional to $T^{(1)}\times T^{(2)}$ appear in the equation. This crucially represent the main qualitative difference between the two tubes late time dynamics, and the single tube case.

\section{Linearized generalised Landau equation}
In this section we consider the generalised Landau equation
\begin{equation}
	\partial_t \rho_\theta= \mathcal{I}_\theta[\rho]
\end{equation}
close to stationary, thermal state. We thus write $\partial_t \delta \rho_\theta = \left(\frac{\delta \mathcal{I}}{\delta \rho}\right)_{\theta \gamma} \delta \rho_\gamma$.
Our goal is to change the degrees of freedom from $\delta \rho$ to perturbations of expectation values of conserved charges $\delta q_n$ defined as $\delta q_n = \int \dd \theta h_n(\theta) \delta\rho(\theta)$.

Firstly, we note that in general it is more natural to work with perturbations of bare pseudoenergy $\delta \epsilon_0$, instead $\delta \rho$. We have the following relations (see SM of~\cite{Lebek2024})
\begin{equation}
	\delta \rho_\theta = -C_{\theta\gamma}(\delta \epsilon_0)_\gamma
\end{equation}
and we also introduce
\begin{equation}
	\left(\frac{\delta \mathcal{I}}{\delta \rho}\right)_{\theta \gamma} \delta \rho_\gamma =  \Gamma_{\theta \gamma}(\delta \epsilon_0)_\gamma.
\end{equation}
In what follows we introduce hydrodynamic inner product defined as
\begin{equation}
	(h|g) = h_\theta C_{\theta\theta'}g_{\theta'}
\end{equation}
and expand 
\begin{equation}
	(\delta \epsilon_0)_\theta = \sum_m (h_m| \delta \epsilon_0) (h_m)_\theta
\end{equation}
where $h_m$ form an orthonormal set with respect to $( \cdot | \cdot)$, constructed with Gram-Schmidt algorithm performed on e.g. set of ultra-local charges $g_m(\theta) =\theta^m/m!$. We note that it follows from the formulas above that $\delta q_n = -(h_n|\delta \epsilon_0)$. We multiply [] by $h_n$ and integrate over $\dd \theta$ finally getting
\begin{equation}
	\partial_t \delta q_n = \Gamma_{nm} \delta q_m, \qquad \Gamma_{nm} = (h_n)_\theta \Gamma_{\theta \theta'}(h_m)_{\theta'}.
\end{equation}
In what follows we determine the matrix $\Gamma_{nm}$. We start with expressions 
\begin{equation}\label{eq:Idiagonal}
	\begin{aligned}
		&\mathcal{I}^{\rm cross}_0[\rho](\theta_1)= 2 \pi V_0^2 \int  \dd k k^2 \hat{\varphi}^2(k)  \int \dd \theta_2 \dd \theta_3   \left[\frac{(\rho^{\rm t}_{\theta_3})^2}{\rho^{\rm t}_{\theta_1}} (T^{\rm dr}_{\theta_2, \theta_1})^2 \frac{|v_{\theta_2, \theta_1}|}{v_{\theta_3, \theta_2} v_{ \theta_1, \theta_3}^2} -\rho^{\rm t}_{\theta_1}(T^{\rm dr}_{\theta_3, \theta_2})^2 \frac{|v_{\theta_3, \theta_2}|}{v_{\theta_1, \theta_3} v_{\theta_2, \theta_1}^2}\right]\mathcal{B}_{\theta_1,\theta_2,\theta_3},
	\end{aligned}
\end{equation}

\begin{equation}\label{eq:Ioffdiagonal}
	\begin{aligned}
		\mathcal{I}^{\rm self}_0[\rho](\theta_1)= 2 \pi  V_0^2 \int  \dd \theta_2 \dd \theta_3 \int \dd k k^2  \hat{\varphi}(k)\rho^{\rm t}_{\theta_3}T^{\rm dr}_{\theta_1, \theta_2} \frac{v_{\theta_2, \theta_1}}{v_{\theta_1,\theta_3}} \bigg[&\hat{\varphi}\left(\frac{v_{\theta_2, \theta_1 }}{v_{ \theta_1, \theta_3}}k \right)\frac{\rho^{\rm t}_{\theta_2} }{\rho^{\rm t}_{\theta_1}}  \frac{T^{\rm dr}_{\theta_3, \theta_1}}{|v_{\theta_1, \theta_3}|v_{\theta_3, \theta_2 }}-\\&-2\hat{\varphi}\left(\frac{v_{\theta_2, \theta_1 }}{v_{  \theta_3, \theta_2 }}k \right) \frac{T^{\rm dr}_{\theta_2, \theta_3}}{v_{\theta_1, \theta_3} |v_{\theta_2, \theta_3}|}\bigg]  \mathcal{B}_{\theta_1,\theta_2,\theta_3},
	\end{aligned}
\end{equation}
introduced in the main text. As a first step, it is useful to rewrite $\mathcal{B}$ factor a bit using
\begin{equation}
	R^{t}_{\theta\gamma} \partial_\gamma \rho_\gamma= - \rho_\theta f_\theta \partial_\theta \epsilon_\theta
\end{equation}
where $\epsilon$ is the pseudoenergy of the state. This identity follows from standard TBA calculations and is derived in SM of~\cite{Lebek2024}. It leads to
\begin{equation}
	\mathcal{B}(\theta,\zeta,\sigma)= - \rho_\theta f_\theta \rho_\sigma f_\sigma \rho_\zeta f_\zeta \left[ \frac{v_{\sigma, \zeta}}{\rho^{\rm t}_\theta} \partial_\theta \epsilon_\theta+\frac{v_{\zeta, \theta}}{\rho^{\rm t}_\sigma} \partial_\sigma \epsilon_\sigma+\frac{v_{\theta, \sigma}}{\rho^{\rm t}_\zeta} \partial_\zeta \epsilon_\zeta \right].
\end{equation}
We note that for thermal state $\epsilon_\theta=\beta(\varepsilon_\theta-\mu)$. Moreover, $k'_\theta=2 \pi \rho^{\rm t}_\theta$ and $v_\theta=\varepsilon_\theta'/k'_\theta$. Thus for thermal states
\begin{equation}
	\mathcal{B}(\theta,\zeta,\sigma)= -2 \pi \beta \rho_\theta f_\theta \rho_\sigma f_\sigma \rho_\zeta f_\zeta \left[ v_{\sigma, \zeta} v_\theta +v_{\zeta, \theta} v_\sigma + v_{\theta, \sigma}v_\zeta \right]=0
\end{equation}
as expected. We thus consider $\epsilon \to \epsilon_{\rm th} +\delta \epsilon_0$ and linearize (to see why large Dressing appears in the expression below see SM of~\cite{Lebek2024}, where similar calculation was performed)
\begin{equation}\label{eq:Gammadiagonal}
	\begin{aligned}
		(\Gamma_0^{\rm cross}\delta \epsilon_0)_\theta&= - 2 \pi \int\dd k  k^2 \hat{V}^2(k)  \int \dd \sigma \dd \zeta   \left[\rho^{\rm t}_\theta(T^{\rm dr}_{\sigma, \zeta})^2 \frac{|v_{\sigma, \zeta}|}{v_{\theta, \sigma} v_{\zeta, \theta}^2}-\frac{(\rho^{\rm t}_\sigma)^2}{\rho^{\rm t}_\theta} (T^{\rm dr}_{\zeta, \theta})^2 \frac{|v_{\zeta, \theta}|}{v_{\sigma, \zeta} v_{ \theta, \sigma}^2} \right] \rho_\theta f_\theta \rho_\sigma f_\sigma \rho_\zeta f_\zeta \times \\
		&\left[ \frac{v_{\sigma, \zeta}}{\rho^{\rm t}_\theta} \partial_\theta ( \delta\epsilon_0)^{\rm Dr}_\theta+\frac{v_{\zeta, \theta}}{\rho^{\rm t}_\sigma} \partial_\sigma ( \delta\epsilon_0)^{\rm Dr}_\sigma+\frac{v_{\theta, \sigma}}{\rho^{\rm t}_\zeta} \partial_\zeta ( \delta\epsilon_0)^{\rm Dr}_\zeta \right],
	\end{aligned}
\end{equation}
where Dressing operation is defined as 
\begin{equation}
	f^{\rm Dr}_\mu = f_\mu-\int {\dd} \mu  F_{\mu,\lambda} n_\mu f'_\mu.
\end{equation}

\begin{equation}\label{eq:Gammaoffdiagonal}
	\begin{aligned}
		(\Gamma^{\rm self}_0 \delta \epsilon_0)_\theta&=  -\int  \dd \zeta \dd \sigma \int \frac{\dd k}{2 \pi} k^2 \hat{V}(k)\rho^{\rm t}_{\sigma}T^{\rm dr}_{\theta \zeta} \frac{v_{\zeta \theta}}{v_{\theta\sigma}} \left[\hat{V}\left(\frac{v_{\zeta \theta }}{v_{ \theta \sigma}}k \right)\frac{\rho^{\rm t}_{\zeta} }{\rho^{\rm t}_\theta} T^{\rm dr}_{\sigma \theta} \frac{1}{|v_{\theta \sigma}|v_{\sigma \zeta }}-2\hat{V}\left(\frac{v_{\zeta \theta }}{v_{  \sigma \zeta }}k \right) T^{\rm dr}_{\zeta \sigma} \frac{1}{v_{\theta \sigma} |v_{\zeta \sigma}|}\right] \times\\
		&\rho_\theta f_\theta \rho_\sigma f_\sigma \rho_\zeta f_\zeta\left[ \frac{v_{\sigma \zeta}}{\rho^{\rm t}_\theta} \partial_\theta ( \delta\epsilon_0)^{\rm Dr}_\theta+\frac{v_{\zeta \theta}}{\rho^{\rm t}_\sigma} \partial_\sigma ( \delta\epsilon_0)^{\rm Dr}_\sigma+\frac{v_{\theta \sigma}}{\rho^{\rm t}_\zeta} \partial_\zeta ( \delta\epsilon_0)^{\rm Dr}_\zeta \right],
	\end{aligned}
\end{equation}

To simplify these expressions we use that under integral we can replace
\begin{equation}
	2 \frac{|v_{\sigma, \zeta}|}{v_{\theta, \sigma}v_{\zeta, \theta}^2} \to -\frac{|v_{\sigma, \zeta}|v_{\sigma, \zeta}}{v_{\theta, \sigma}^2v_{\zeta, \theta}^2} 
\end{equation}
We get

\begin{equation}
	\begin{aligned}
		\Gamma^{\rm cross}_{n,m}&= \pi \int \dd k  k^2 \hat{V}^2(k) \int \dd \theta \dd \sigma \dd \zeta (\rho^{\rm t}_\theta)^2(T^{\rm dr}_{\sigma, \zeta})^2 \frac{|v_{\sigma, \zeta}|}{v_{\theta ,\sigma}^2v_{\theta, \zeta}^2} \rho_\theta f_\theta \rho_\sigma f_\sigma \rho_\zeta f_\zeta \times \\    
		&\left( \frac{v_{\sigma, \zeta}}{\rho^{\rm t}_\theta}  ( h_n')^{\rm dr}_\theta+\frac{v_{\zeta, \theta}}{\rho^{\rm t}_\sigma}  ( h_n')^{\rm dr}_\sigma+\frac{v_{\theta, \sigma}}{\rho^{\rm t}_\zeta}  (h_n' )^{\rm dr}_\zeta \right)\left( \frac{v_{\sigma, \zeta}}{\rho^{\rm t}_\theta}  ( h_m')^{\rm dr}_\theta+\frac{v_{\zeta, \theta}}{\rho^{\rm t}_\sigma}  ( h_m')^{\rm dr}_\sigma+\frac{v_{\theta, \sigma}}{\rho^{\rm t}_\zeta}  (h_m' )^{\rm dr}_\zeta \right),
	\end{aligned}
\end{equation}
and
\begin{equation}
	\begin{aligned}
		\Gamma^{\rm self}_{n,m}&= 2 \pi \int \dd k  k^2 \hat{V}(k) \int \dd \theta \dd \sigma \dd \zeta \rho^{\rm t}_\zeta \rho^{\rm t}_\sigma T^{\rm dr}_{\theta, \zeta} T^{\rm dr}_{\theta, \sigma} \hat{V} \left( \frac{v_{\theta, \zeta}}{ v_{\sigma, \theta}}k \right)  \frac{v_{\theta,\zeta }}{|v_{\sigma, \theta}| v_{\zeta, \sigma}^2 v_{\sigma, \theta}} \rho_\theta f_\theta \rho_\sigma f_\sigma \rho_\zeta f_\zeta \times \\    
		&\left( \frac{v_{\sigma, \zeta}}{\rho^{\rm t}_\theta}  ( h_n')^{\rm dr}_\theta+\frac{v_{\zeta, \theta}}{\rho^{\rm t}_\sigma}  ( h_n')^{\rm dr}_\sigma+\frac{v_{\theta, \sigma}}{\rho^{\rm t}_\zeta}  (h_n' )^{\rm dr}_\zeta \right)\left( \frac{v_{\sigma, \zeta}}{\rho^{\rm t}_\theta}  ( h_m')^{\rm dr}_\theta+\frac{v_{\zeta, \theta}}{\rho^{\rm t}_\sigma}  ( h_m')^{\rm dr}_\sigma+\frac{v_{\theta, \sigma}}{\rho^{\rm t}_\zeta}  (h_m' )^{\rm dr}_\zeta \right).
	\end{aligned}
\end{equation}
\section{Small-momentum limit of Fermi's Golden rule collision integral}
\label{app:small_mom}
We start with the bare collision integral focusing on the $m=3$ case
\begin{align}\label{eq:Istructure1_SM}
	%\begin{aligned}
	\mathcal{I}_0^{(3)}[\rho_{\rm p}](\theta)=\int {\rm d} \mathbf{p} {\rm d} \mathbf{h} \delta(\theta, \mathbf{p})  B( \mathbf{p} \to \mathbf{h})[\rho_{\rm h}(\mathbf{p}) \rho_{\rm p} (\mathbf{h})-\rho_{\rm h}( \mathbf{h})\rho_{\rm p}( \mathbf{p})]\,,
	%\end{aligned}
\end{align}
where $\delta(\theta,\mathbf{p})= \sum_{i=1}^3\delta(\theta-p_i)$ and other quantities are defined as before. We introduce $\lambda_i =(p_i+h_i)/2$ and $q_i=k(p_i)-k(h_i)$. Assuming that $\alpha_i=p_i-h_i$ are small for all $i$ we can approximate $q_i \approx k'(\lambda_i) \alpha_i$ leading to $p_i=\lambda_i+\frac{q_i}{2 k'(\lambda_i)}$ and $h_i=\lambda_i-\frac{q_i}{2 k'(\lambda_i)}$, from which follow the expansion of delta term
\begin{equation}\label{eq:delta_expansion}
	\delta(\theta, \mathbf{p}) =\delta(\theta,\mathbf{\lambda}) + \sum_{i=1}^3 \frac{q_i}{2k'(\lambda_i)} \delta'(\theta-\lambda_i) + \mathcal{O}(q_i^2)\,,
\end{equation}
and difference of the density factors
\begin{equation}
	%\begin{aligned}
	\rho_{\rm h}(\mathbf{p}) \rho_{\rm p} (\mathbf{h})-\rho_{\rm h}( \mathbf{h})\rho_{\rm p}( \mathbf{p})= \rho_{\rm h}( \mathbf{h})\rho_{\rm p}( \mathbf{p}) 
	\left(\sum_{i=1}^3 q_i \frac{\epsilon'(\lambda_i)}{k'(\lambda_i)} +\mathcal{O}(q_i^3) \right)\,,
	%\end{aligned}
\end{equation}
where we have introduced pseudoenergy $\epsilon(\lambda)=\log \left(\rho_h(\lambda)/\rho_p (\lambda)\right)$. The conservation laws inside $B$ factors become $\delta(q_1+q_2+q_3)$ for the momentum and $\delta\left(v_1q_1+v_2q_2+v_3q_3 \right)$ for the energy. From the particle-hole symmetry of form factors the first term in \eqref{eq:delta_expansion} vanishes. We change  integration variables getting \begin{equation}\label{eq:Istructure_smallmom}
	\begin{aligned}
		\mathcal{I}^{(3)}_0[\rho](\theta) &= \frac{1}{3!} \frac{1}{(2\pi)^3} \partial_\theta\int {\rm d} \boldsymbol{\lambda} {\rm d}\mathbf{q}\,  B(\boldsymbol{\lambda}, \mathbf{q})  
		\left(\sum_{i=1}^3 \frac{q_i}{2k'_{\lambda_i}} \delta(\theta - \lambda_i)\right) \left(\sum_{i=1}^3 q_i \frac{\varepsilon'_{\lambda_i}}{k'_{\lambda_i}} \right).
	\end{aligned}
\end{equation}

\AddToHook{enddocument/afteraux}{
\immediate\write18{
cp output.aux SM.aux
}
}

\end{document}